\documentclass{aa}   

\usepackage{graphicx}
\usepackage{lscape}
\usepackage{natbib} 
\usepackage{siunitx}
\usepackage{url}
\usepackage{color}
\usepackage[varg]{txfonts}
\bibpunct{(}{)}{;}{a}{}{,} % citation like A&A style

\begin{document}

\title{Observational properties of massive black 
hole binary progenitors}

   \author{R. Hainich\inst{1}
          \and L. M. Oskinova\inst{1}
          \and T. Shenar\inst{1}
          \and P. Marchant\inst{2,3}
          \and J.\,J. Eldridge\inst{4}
          \and A.\,A.\,C. Sander\inst{1}
          \and W.-R. Hamann\inst{1}
          \and N. Langer\inst{2}
          \and H. Todt\inst{1}
          }

   \institute{Institut f\"ur Physik und Astronomie,
              Universit\"at Potsdam,
              Karl-Liebknecht-Str. 24/25, D-14476 Potsdam, Germany \\
              \email{rhainich@astro.physik.uni-potsdam.de}
              \and 
              Argelander-Institut für Astronomie, 
              Universität Bonn, 
              Auf dem Hügel 71, 53121 Bonn, Germany 
              \and
              Center for Interdisciplinary Exploration 
              and Research in Astrophysics (CIERA) 
              and Department of Physics and Astronomy, 
              Northwestern University, 
              2145 Sheridan Road, Evanston, IL 60208, USA 
              \and
              Department of Physics, 
              University of Auckland, 
              Private Bag 92019, Auckland, New Zealand 
              }
   \date{Received <date> / Accepted <date>}

%-------------------  Abstract --------------------

\abstract
%
% context
%
{The first directly detected gravitational waves (GW\,150914) were emitted
    by two coalescing black holes (BHs) with masses of $\approx 36\,M_\odot$
    and $\approx 29\,M_\odot$.
    Several scenarios have been proposed to put this
    detection into an astrophysical context. The evolution of an isolated
    massive binary system is among commonly considered models.
}
%
% aims heading (mandatory)
%
{Various groups have performed detailed binary-evolution calculations that
    lead to BH merger events. However, the
    question remains open as to whether binary systems with the predicted properties
    really exist. 
    The aim of this paper is to help observers to close this gap by
    providing spectral characteristics of massive binary BH progenitors
    during a phase where at least one of the companions is still non-degenerate.
}
%
% methods heading (mandatory)
%
{
    Stellar evolution models predict fundamental stellar parameters. 
    Using these as
    input for our stellar atmosphere code (PoWR), we compute a set of models
    for selected evolutionary stages of massive merging BH progenitors at
    different metallicities.
}
%
% results heading (mandatory)
%
{
    The synthetic spectra obtained from our atmosphere calculations reveal
    that progenitors
    of massive BH merger events start their lives as O2-3V stars that evolve
    to early-type blue supergiants before they undergo core-collapse during 
    the Wolf-Rayet phase. When the primary has
    collapsed, the remaining system will appear as a wind-fed high-mass X-ray
    binary.
    Based on our atmosphere models, we provide feedback parameters, broad band
    magnitudes, and spectral templates that should help to identify such
    binaries in the future.
}
%
% conclusions
%
{
    While the predicted parameter space for massive BH binary progenitors is partly realized in nature,  none of the known
    massive binaries match our synthetic spectra of massive BH binary progenitors exactly.
    Comparisons of empirically determined mass-loss rates with those assumed by evolution calculations reveal significant differences. The consideration of the empirical mass-loss rates in evolution calculations will possibly entail a shift of the maximum in the predicted binary-BH merger rate to higher metallicities, that is, more candidates should be expected in our cosmic neighborhood than previously assumed.
}


\maketitle

%________________________________________________________________
\section{Introduction}
\label{sect:intro}

The first direct detection of gravitational waves (GW), GW\,150914, proved to
be the  echo of two coalescing black holes (BH) with unexpectedly
high  masses  of $\approx 29\,M_\odot$ and $\approx 36\,M_\odot$
\citep{Abbott2016a}.  The luminosity distance of the event,
$\approx 400$\,Mpc, showed that such massive BHs are present in our cosmic
neighborhood. The most recent GW event GW170104 was due to merging BHs with
masses of $\approx 32\,M_\odot$ and $\approx 19\,M_\odot$  
at a luminosity distance of $\approx 900$\,Mpc \citep{Abbott2017}. Yet,
the masses of accreting BHs in X-ray binaries
in ours and neighboring galaxies are much smaller \citep{Ozel2010} --
they are compatible to those measured in the second
reported GW event, GW\,151226 with $\approx 14\,M_\odot$ and $\approx 8\,M_\odot$
\citep{Abbott2016c}. At present, it is unclear if the massive BHs detected
by GW observatories are remnants of massive stars analogous to the BHs
residing in stellar X-ray binaries.

Hence, among the  most urgent questions in the new field of gravitational wave astrophysics
is to establish whether massive stars, as we presently know and understand
them, could be the progenitors of massive BHs. Answering this question could help to discriminating among the families of models put forward to explain
the first GW detections.

A large  theoretical work is underway to explain GW observations. The majority of
the proposed models invoke massive star evolution. Fast evolutionary
channels leading to massive BH
mergers are related to massive star binaries and could explain
either of the detected GW events
\citep[e.g.,][]{Abbott2016b,Belczynski2016,Belczynski2016a,Marchant2016,Mandel2016,Stevenson2017}. In this
case, the evolutionary time scale from the stellar binary formation until the
double BH merger is about one billion years.
If GW events are the result of this fast channel, potential
progenitors of massive
BHs must be present among observed stellar populations.

Slow evolutionary channels require a time period one order of magnitude longer for
the BH merger. These channels also commonly invoke massive star progenitors
of BHs, such as wide and eccentric massive star binaries \citep{Eldridge2016}
or dynamic interactions in dense star clusters \citep[e.g.,][]{Rodriguez2015}.
In this case one can also expect that the progenitor stars could, in principle,
be identified among present-day massive stars.

Alternative models include Population\,III stars \citep{Inayoshi2017}
and primordial BHs \citep[e.g.,][]{Bird2016}. These models do not require the
existence of Population I massive stars capable of collapsing into massive BHs.

The prevalence of stellar versus cosmological origin could be
distinguished using conventional observations. We need to search and find
massive BH progenitors or rule out their current presence.

The primary goal of this paper is to predict what the potential massive BH and
GW event progenitor stars should look like and how they could be identified.
On this basis, the expected massive BH progenitor properties can be compared
to those of known massive stars. 
If actual stars with the predicted stellar parameters are found,
this will provide strong empirical validation for the 
corresponding evolutionary channel. 

To achieve this goal, we use the theoretical predictions of the progenitor
properties as input for state-of-the-art stellar atmosphere models 
and compute synthetic spectra. We provide photometric fluxes, 
spectral types, stellar and feedback parameters
that can be used to identify such objects
and to evaluate their impact on the Galactic ecology. 
Based on our templates, we discuss whether stars with such properties are already
known.
In this paper, we focus on progenitors of GW\,150914-like events. 

The paper is structured as follows: In Sects.\,\ref{sect:models}
and \ref{sect:evo}, we describe the stellar atmosphere
models, the stellar evolution calculations used in this study, and the
calculated sets of synthetic spectra.
Section\,\ref{sec:owr} constitutes a thorough comparison of the parameters
and templates obtained in this study with observations. The X-ray
properties
of the binary BH progenitors are discussed in Sect.\,\ref{sec:xray}, while
the
feedback provided by those objects is described in Sect.\,\ref{sec:feed}.
The summary and the conclusions are given in Sect.\,\ref{sect:s-and-c}.
Additional tables and synthetic spectra are presented in
Appendices\,\ref{sec:addtables} and \ref{sect:spectra}, respectively.

%________________________________________________________________
\section{The stellar atmosphere models}
\label{sect:models}

The synthetic spectra presented in this paper are calculated with the 
state-of-the-art Potsdam Wolf-Rayet (PoWR) code for expanding stellar 
atmospheres\footnote{Comprehensive grids of stellar atmosphere models for 
O-type and WR stars, calculated for different metallicities, can be found at 
the PoWR website: \url{www.astro.physik.uni-potsdam.de/PoWR}}. This code 
assumes a star with a spherically stationary symmetric outflow 
and fully accounts for deviations from the 
local thermodynamic equilibrium (non-LTE), wind inhomogeneities, and iron line 
blanketing. The code solves the radiative transfer equation in the co-moving 
frame consistently with the rate equations for statistical equilibrium, 
while ensuring energy conservation. Detailed information on 
the assumptions and numerical methods used in the code can be found in 
\citet{Graefener2002}, \citet{Hamann2003,Hamann2004}, and \citet{Sander2015}.

The stellar radius $R_\ast$ is the input parameter that sets the inner boundary 
of the model. It is defined as the radius where the Rosseland continuum 
optical depth reaches $\tau_\mathrm{ross} = 20$. 
This radius is -- to a good approximation -- coincident with the radius of the star
in hydrostatic equilibrium, enabling us to use the output from evolution models 
as input in our atmosphere models.
The outer atmosphere boundary is set to $R_\mathrm{max} = 100\,R_*$ for 
O-type models 
and $1000\,R_*$ for WR models, which proved to be sufficient. The stellar 
temperature $T_\ast$ is the effective temperature, corresponding to $R_\ast$ 
and the luminosity $L$ via the Stefan-Boltzmann law
\begin{equation}
\label{eq:sblaw}
L = 4 \pi \sigma_\mathrm{SB} R_\ast^2 T_\ast^4~.
\end{equation}  

In the main model iteration, we account for the random motion within the 
stellar atmosphere by assuming Gaussian line profiles with a Doppler width of 
$30\,\mathrm{km}\,\mathrm{s}^{-1}$ and $100\,\mathrm{km}\,\mathrm{s}^{-1}$ for the O and WR stars, 
respectively. This velocity is disassembled in its components, a depth-dependent
thermal motion and a micro turbulence velocity $\xi$, during the final calculation of the
emergent spectrum.
The latter is assumed to grow with the wind velocity up to a value of 
$\xi(R_\mathrm{max}) = 0.1\,v_\infty$, while it is set to $\xi(R_*) = 20\,
\mathrm{km}\,\mathrm{s}^{-1}$ and $\xi(R_*) = 100\,\mathrm{km}\,\mathrm{s}^{-1}$ at the inner boundary for 
the models that resemble O and WR stars, respectively. 
 
In O-type models, the quasi-hydrostatic part of the atmosphere is calculated 
self-consistently to fulfill the hydrostatic equation \citep{Sander2015}. In 
the supersonic part, corresponding to the stellar wind, $v(r)$ is prescribed by 
the so-called $\beta$-law \citep{Castor1979,Pauldrach1986}. For O-type stars, the 
finite disk corrected CAK theory \citep{Castor1975,Pauldrach1986} 
predicts a $\beta$ exponent of about 0.8 \citep[e.g.,\ ][]{Kudritzki1989}, which is 
consistent with the results obtained from the propagation of clumps in O-type 
star winds \citep[e.g.,\ ][]{Eversberg1998}. This value is therefore assumed in 
the calculation of the O-type star models. For WR stars, different estimates of 
the $\beta$ exponent are available in the literature. While empirical studies 
suggest a velocity law with $\beta = 1.0$ for hydrogen rich WR stars and a 
$\beta$ in excess of four for more evolved WR stars 
\citep{Lepine1999,Dessart2005}, theoretical predictions based on hydrodynamic 
consistent stellar atmosphere models point to a more complex velocity law in 
the form of a double-$\beta$ law \citep{Hillier1999,Graefener2005}. Therefore, 
models that resemble hydrogen-rich WN stars are calculated with $\beta = 1.0$, 
while models of more evolved WR stars are calculated with a double-$\beta$ law 
in the form as presented in \citet{Todt2015}.

PoWR model atmospheres account for wind inhomogeneities in the form 
of the so-called ``microclumping'' approximation, assuming optically thin clumps 
that fill a volume fraction $f_\mathrm{V}$, with a void interclump medium 
\citep{Hillier1991,Hamann1998}. The clumping factor $D = {f_\mathrm{V}}^{-1}$ 
describes the density contrast between the clumps and a homogeneous model with 
the same mass-loss rate $\dot{M}$. The clumping factor can have a radial 
dependency, that is, $D(r)$. For all models presented in this paper, we assume that 
clumping starts at the sonic point, increases outwards, and reaches its 
maximum with $D = 10$ ($f_{\rm V}=0.1$) at a radius of 
$10\,R_\ast$ \citep{Runacres2002}. This choice of 
the maximum clumping factor is motivated by theoretical simulations 
\citep[e.g.,\ ][]{Feldmeier1997} as well as spectral analyses \citep[e.g.,\ 
][]{Hainich2014,Hainich2015}, although larger values, especially for O-type 
stars, are sometimes claimed in the literature  \citep[e.g.,][]{Bouret2012}. 

For the non-LTE calculations, detailed model atoms of \element{H}, 
\element{He}, \element{C}, \element{N}, \element{O}, \element{Mg}, 
\element{Si}, \element{P}, and \element{S} are used (see Table\,\ref{table:model_atoms} 
for details). For the iron group elements \element{Fe}, 
\element{Sc}, \element{Ti}, \element{V}, \element{Cr}, \element{Mn}, \element{Co}, and 
\element{Ni}, a ``superlevel approach'' is 
applied where thousands of levels and millions of line transitions are grouped 
into superlevels with pre-calculated, complex functions for the transition 
cross-sections \citep[see\ ][]{Graefener2002}. 

One output of binary evolution models are tracks on the 
Hertzsprung-Russell Diagram (HRD) that describe the changes in 
fundamental stellar parameters, $T_\ast$, $R_\ast$, and $L$, 
in the course of stellar evolution. These parameters are used as an input for our PoWR 
atmosphere calculations. The mass-loss rates and chemical abundances for each element are also 
adopted from the stellar evolution models. Since the BPASS models do not 
consider the elements \element{Mg}, \element{Si}, \element{P}, and \element{S},
the corresponding solar abundances \citep{Asplund2009} are used 
and scaled according to the metallicity of the stellar evolution tracks. 

Model spectra of the binary systems are computed by adding up the absolute line
spectra of the individual binary components. Afterwards, this combined spectrum is normalized with the composite continuum.
We neglect all asymmetries that can occur
in close binary systems or during Roche lobe overflow (RLOF), such as wind-wind collision zones or accretion flows.

%%%%%%%%%%%%%%%%%%%  Fig.1 %%%%%%%%%%%%%%%%%%
%---------------------------------------------------------------
\begin{figure}[tbp]
    \centering
    \includegraphics[width=\hsize]{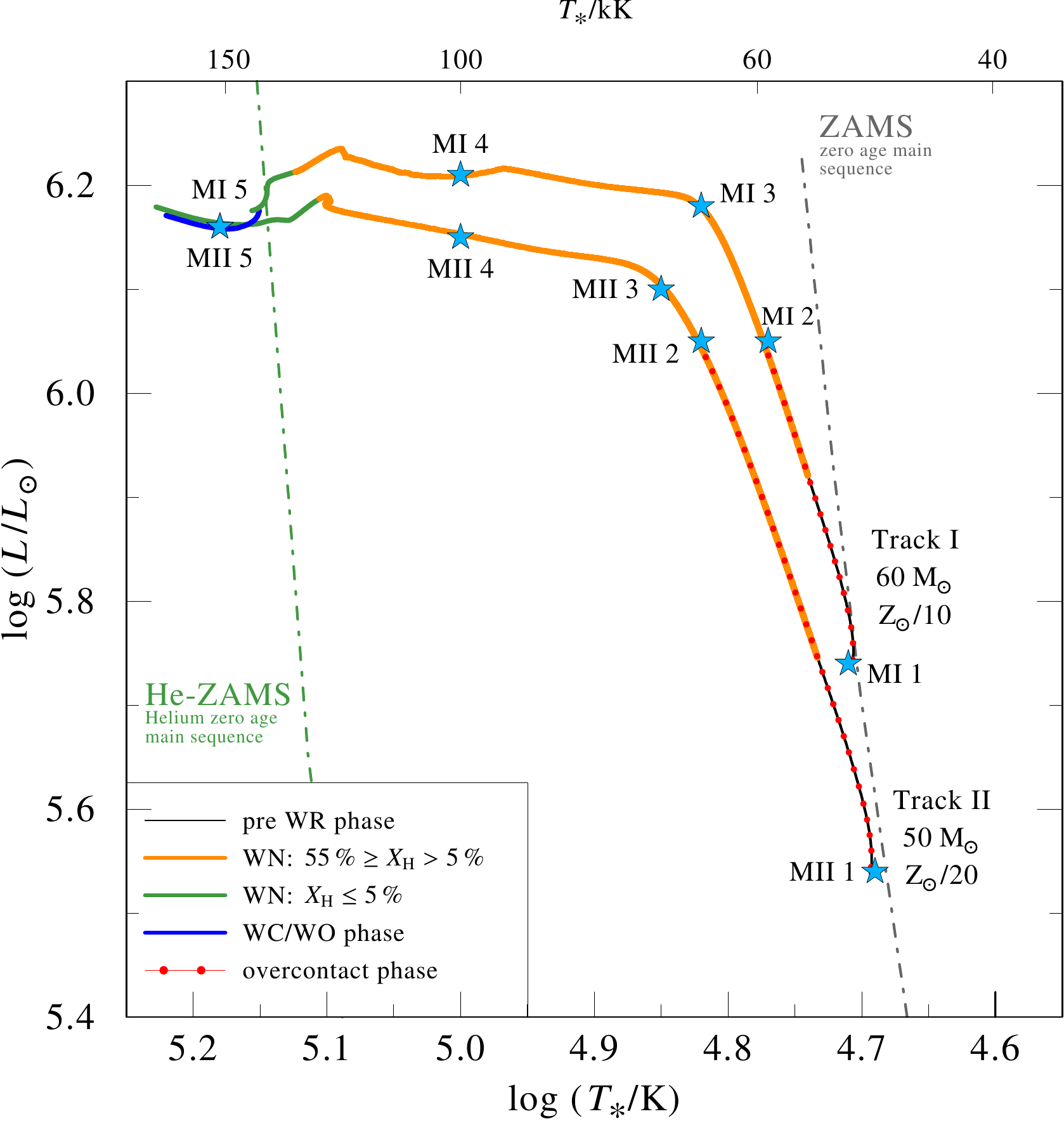}
    \caption{Hertzsprung-Russell diagram showing the two stellar evolution tracks 
        calculated by \citet{Marchant2016} that are used in this paper. Only 
        the evolution path of the primary is shown, since the evolution of the secondary
        is virtually identical after the initial mass transfer. The blue filled asterisks 
        mark the positions at which stellar atmosphere models where calculated. 
        The labels near these asterisks are the atmosphere-model identifiers used throughout 
        the text and Tables\,\ref{table:parameters-pablo-01Zsun} and \ref{table:parameters-pablo-005Zsun}.
        The highlighted and color-coded parts of the tracks refer to different WR phases 
        and the period where the Roche lobe volume is filled, respectively (see inlet). 
        } 
    \label{fig:hrd_pablo}
\end{figure}
%---------------------------------------------------------------     

%%%%%%%%%%%%%%%%%%%%%%%%%%%%%%%%%%%%%%%%%%%%%%%%%%%%%%%%%%%%%%%%%%%%%%%%%%%%%%%%%%%%%
%%%%%%%% table parameters
%%%%%%%%%%%%%%%%%%%%%%%%%%%%%%%%%%%%%%%%%%%%%%%%%%%%%%%%%%%%%%%%%%%%%%%%%%%%%%%%%%%%%
\begin{table*}[th!]
\caption{Parameters of the stellar atmosphere models calculated for the $60\,M_\odot$ track with $Z = 1/10\,Z_\odot$ \citep[][see track I in Fig.\,\ref{fig:hrd_pablo}]{Marchant2016}.}
\label{table:parameters-pablo-01Zsun}
\centering  
\begin{tabular}{lllcccccc}
\hline \hline \rule[0mm]{0mm}{4.0mm}   %-------------------------------------------------
 Model  & MI\,1 & MI\,2 & \multicolumn{2}{c}{MI\,3} & \multicolumn{2}{c}{MI\,4} & \multicolumn{2}{c}{MI\,5}  \\
           &    &    & \multicolumn{1}{c}{a} & \multicolumn{1}{c}{b} & \multicolumn{1}{c}{a} & \multicolumn{1}{c}{b} & \multicolumn{1}{c}{a} & \multicolumn{1}{c}{b}                \\
\hline  %--------------------------------------------------------
 Spectral type                                        & O3\,V((f*))z & O2-3\,If* & \multicolumn{2}{c}{WN2}   & \multicolumn{2}{c}{WN2}   & \multicolumn{2}{c}{WO1}   \rule[0mm]{0mm}{4.0mm}  \\
 age $[10^{6}\,\mathrm{yr}]$                          & 0.09         & 3.45      & \multicolumn{2}{c}{4.39}  & \multicolumn{2}{c}{4.65}  & \multicolumn{2}{c}{4.95} \\
 $T_*\, [\mathrm{kK}]$                                & 52.2         & 59.4      & \multicolumn{2}{c}{66.0}  & \multicolumn{2}{c}{100.0} & \multicolumn{2}{c}{150.0} \\
 $\log L\, [L_{\odot}]$                               & 5.73         & 6.05      & \multicolumn{2}{c}{6.18}  & \multicolumn{2}{c}{6.2}   & \multicolumn{2}{c}{6.16}  \\
 $M\, [M_{\odot}]$                                    & 63.1         & 59.7      & \multicolumn{2}{c}{52.8}  & \multicolumn{2}{c}{47.5}  & \multicolumn{2}{c}{34.8}  \\
 $R_*\, [R_{\odot}]$                                  & 9.0          & 10.0  & \multicolumn{2}{c}{9.4}   & \multicolumn{2}{c}{4.2}   & \multicolumn{2}{c}{1.8}   \\
 $\log \dot M\, [M_{\odot}/\mathrm{yr}]$              & -6.0        & -5.6     & -4.8  & -5.1   & -3.2\tablefootmark{a}  &  -3.5\tablefootmark{a}   & -4.5 & -4.9  \\
 $\varv_{\infty}\, [\mathrm{km/s}]$                   & 3155       & 1600  & \multicolumn{2}{c}{1600}  & \multicolumn{2}{c}{2400}  & \multicolumn{2}{c}{3000}  \\
 $\varv_\mathrm{rot}\, [\mathrm{km/s}]$               & 381          & 348       & \multicolumn{2}{c}{221}   & \multicolumn{2}{c}{253}   & \multicolumn{2}{c}{110}   \\
 $\varv_\mathrm{orbit}\, [\mathrm{km/s}]$             & 493          & 463       & \multicolumn{2}{c}{391}   & \multicolumn{2}{c}{342}   & \multicolumn{2}{c}{246}   \\                                                          
 $X_{\element{H}}$                                    & 0.75         & 0.37      & \multicolumn{2}{c}{0.14}  & \multicolumn{2}{c}{0.08}  & \multicolumn{2}{c}{0.0}   \\
 $X_{\element{C}}\, [10^{-3}]$                        & 0.29       & 0.02  & \multicolumn{2}{c}{0.02}  & \multicolumn{2}{c}{0.02}  & \multicolumn{2}{c}{450.0} \\
 $X_{\element{N}}\, [10^{-3}]$                        & 0.09      & 1.08    & \multicolumn{2}{c}{1.08}  & \multicolumn{2}{c}{0.11}  & \multicolumn{2}{c}{0.0}   \\
 $X_{\element{O}}\, [10^{-3}]$                        & 0.8        & 0.02  & \multicolumn{2}{c}{0.02}  & \multicolumn{2}{c}{0.02}  & \multicolumn{2}{c}{147.0}   \\
\hline  %--------------------------------------------------------
 $M_\mathrm{U}\, [\mathrm{mag}]$                      & -6.8        & -7.0      & -7.0 & -7.0  & -9.0                   & -8.4                   & -6.4   & -5.8  \rule[0mm]{0mm}{4.0mm}   \\
 $M_\mathrm{B}\, [\mathrm{mag}]$                      & -5.5        & -5.8      & -5.8     & -5.7  & -8.2                    & -7.7                   & -5.2     & -4.6     \\
 $M_\mathrm{V}\, [\mathrm{mag}]$                      & -5.2        & -5.4      & -5.4     & -5.4  & -8.0               & -7.3                   & -4.7     & -3.9     \\
 $M_\mathrm{J}\, [\mathrm{mag}]$\tablefootmark{b}     & -4.4        & -4.6      & -4.9     & -4.7  & -8.1                   & -7.4                   & -4.7     & -3.7     \\
 $M_\mathrm{H}\, [\mathrm{mag}]$\tablefootmark{b}     & -4.3        & -4.6      & -4.9     & -4.7  & -8.4                   & -7.7                   & -5.0     & -4.0     \\
 $M_\mathrm{K}\, [\mathrm{mag}]$\tablefootmark{b}     & -4.0        & -4.4      & -5.1     & -4.6  & -8.7                   & -8.3                   & -5.3     & -4.3     \\
 $\log Q_\mathrm{\element{H}}\, [\mathrm{s^{-1}}]$    & 49.5        & 49.9      & 50.1     & 50.1  & 49.8  & 51.2  & 49.9     & 49.9      \\
 $T_{\mathrm{Zanstra}, \element{H}}\, [\mathrm{kK}]$  & 52.0         & 65.9       & 68.8      & 72.7   & 30.6   & 77.7   & 81.5      & 109.3     \\
 $\log Q_\mathrm{\ion{He}{i}}\, [\mathrm{s^{-1}}]$    & 49.1       & 49.5      & 49.7     & 49.7  & 49.3  & 51.2  & 49.8 & 49.8     \\
 $\log Q_\mathrm{\ion{He}{ii}}\, [\mathrm{s^{-1}}]$   & 45.1        & 46.5      & 41.9     & 46.0  & 37.4\tablefootmark{c}  & 41.5\tablefootmark{c}  & 48.9     & 49.1     \\
 $T_{\mathrm{Zanstra}, \ion{He}{ii}}\, [\mathrm{kK}]$ & 40.6         & 50.8       & 26.8      & 46.0   & 17.4\tablefootmark{c}   & 23.7\tablefootmark{c}   & 98.8      & 125.0     \\
\hline  %--------------------------------------------------------
 $P$ [d]                                              & 1.2          & 1.4       & \multicolumn{2}{c}{2.1}   & \multicolumn{2}{c}{2.9}   & \multicolumn{2}{c}{5.6}   \rule[0mm]{0mm}{4.0mm} \\
 orbital separation $[R_{\odot}]$           & 23.8         & 26.5      & \multicolumn{2}{c}{33.0}  & \multicolumn{2}{c}{38.6}  & \multicolumn{2}{c}{54.7}  \\
\hline  \rule[0mm]{0mm}{4.0mm} 
%---------------------------------------------------------------------------------------------------------
Model atom set\tablefootmark{d} & I & I & \multicolumn{2}{c}{I} & \multicolumn{2}{c}{II} & \multicolumn{2}{c}{III} \\
\hline   %-----------------------------------------------------------------------
\multicolumn{9}{c}{Wind accretion rates and accretion X-ray luminosity\tablefootmark{e}} \rule[0mm]{0mm}{4.0mm} \\
\hline  %-----------------------------------------------------------------------
\rule[0mm]{0mm}{4.0mm}  
$\log S_{\rm accr}\, [M_{\odot}/\mathrm{yr}]$ &  &  & -7.0 & -7.3 & -6.2 & -6.5 
& -8.2  & -8.6 \\
$s_{\rm accr}$\tablefootmark{f}               &  &  &  1.1  &  0.5  & 6.3 & 3.1 & 0.06  &   
0.03 \\
$L_{\rm X}$ [erg\,s$^{-1}$]\tablefootmark{g} &  &  & $6\times 10^{38}$ & 
$3\times 10^{38}$ & $3\times 10^{39}$  & $2\times 10^{39}$  &         $4\times 
10^{37}$  & $2\times 10^{37}$ \\       
\hline  \rule[0mm]{0mm}{4.0mm}
\end{tabular}
\tablefoot{The WN models were calculated with two different mass-loss rates. The ``a'' models use the same mass-loss rate as given in the tracks, assuming a metallicity scaling of $Z^{0.85}$ as it is observed 
for O-type stars. The mass-loss rates of the ``b'' models were scaled according to the mass-loss metallicity relation presented by \citet{Hainich2015}. 
\tablefoottext{a}{The high mass-loss rate of this model is a result of the high angular velocity, leading to a rotational enhancement of $\dot{M}$ compared to a non-rotating model. For details, we refer the reader to \citet{Marchant2016}.}
\tablefoottext{b}{Monochromatic magnitudes at $1.26\,\mu\mathrm{m}$, $1.60\,\mu\mathrm{m}$, and $2.22\,\mu\mathrm{m}$, respectively.}
\tablefoottext{c}{Only lower limits, since the \ion{He}{ii} ionization edge is optically thick at the outer boundary of the respective stellar atmosphere models.}
\tablefoottext{d}{Model atom set used in the corresponding stellar atmosphere calculations (see Table\,\ref{table:model_atoms}).}
\tablefoottext{e}{The evolution model predicts BH formation only if the secondary is already in its final evolutionary stage (see text for details). Accretion onto a  $36\,M_\odot$ BH  is assumed.}
\tablefoottext{f}{Accretion rate normalized to the Eddington accretion rate: $s_{\rm accr}\equiv 
\frac{S_{\rm accr}}{S_{\rm Edd}} \approx 1.5\times 10^{7}\cdot
 \frac{M_{\rm BH} \dot{M}}{v_{8}^4 a_{\rm BH}^2}$ (see Sect.\,\ref{subsect:hmxb} for details).}
\tablefoottext{g}{Upper limit since an accretion efficiency of 0.1 is assumed.}
}
\end{table*}
%%%%%%%%%%%%%%%%%%%%%%%%%%%%%%%%%%%%%%%%%%%%%%%%%%%%%%%%%%%%%%%%%%%%%%%%%%%%%%%%%%%%%

%________________________________________________________________
\section{Binary evolution models}
\label{sect:evo}

In this paper, we make use of the 
binary-evolution tracks computed by two independent groups and codes,
\citet{Marchant2016} and \citet{Eldridge2016}. Both models predict
evolutionary paths leading to binary BHs that would merge producing  an event similar
to GW\,150914. The choice of these two models is primarily motivated by
their detailed predictions of the fundamental stellar parameters at different
evolutionary stages, which are required for the computations of synthetic spectra and
comparisons with observed stellar populations.

Any massive star evolution-model depends on a number of input parameters,
some of the most important being the initial metallicity $Z$ and stellar wind
mass-loss rate $\dot{M}(Z)$ at the various phases of stellar evolution.
The mass loss largely determines whether the star will maintain its fast rotation and 
undergo chemical mixing, when and how its core will collapse, how the orbit evolves, 
and what the final BH mass will be. 

Fast rotating massive stars experience strong internal mixing
that can lead to quasi-chemically homogeneous evolution \citep[QCHE,\ ][]{Heger2000,Yoon2005}, but see \citet{Vink2017}.
Such stars do not significantly expand during their evolution. Even in close binary
systems, they typically do not fill their Roche lobes and avoid mass
transfer onto their companions. Thus, the components may remain very
massive and have a small orbital separation at the moment of collapse.

Hot star winds are mainly driven by scattering of UV photons in metal lines
\citep{Lucy1970,Castor1975}. Hence, stellar winds are generally weaker at
low metallicity environments, and remove less mass and
angular momentum. Thus, a metal-poor star will
undergo its core collapse at a higher mass and at faster rotation
than a star with initially the same mass but higher metallicity.

For a given initial stellar mass, the mass of the
immediate BH progenitor is a function of the mass-loss rate. 
Both sets of evolution models considered in this study rely on recipes 
prescribing $\dot{M}(Z)$ at different evolutionary stages. 
\citet{Marchant2016} adopted the
scaling of the mass-loss rate with metallicity as \mbox{$\dot{M}(Z)\propto
    (Z/Z_\odot)^{0.85}$} for stars at all evolutionary stages.
\citet{Eldridge2016} use $\dot{M}(Z)\propto (Z/Z_\odot)^{0.69}$ for
hydrogen-rich \mbox{O-type} stars and $(Z/Z_\odot)^{0.5}$ for more
evolved, hydrogen-poor stars.
For more details on the mass-loss recipes applied in the individual evolution calculations, we refer the reader to \citet{Marchant2016}, \citet{Eldridge2016}, and the references therein.

Recent theoretical work and empirical measurements of mass-loss rates from
massive stars show that the \mbox{$\dot{M}(Z)\propto (Z/Z_\odot)^{0.7-0.9}$}
scaling adopted in evolution calculations is reasonably accurate
for O-type stars  \citep{Mokiem2007, Lucy2012}.
On the other hand, for the more evolved Wolf-Rayet (WR) stars, which
represent massive stars during a late evolutionary stage 
prior to core collapse, the metallicity
dependence of the mass-loss rate is steeper than assumed in evolution models.
For example, \citet{Hainich2015} showed that the mass-loss rates of WR stars of the nitrogen sequence (WN stars) scale with metallicity as $\dot{M} \propto Z^{1.2}$. This
scaling was later confirmed by \citet{Tramper2016}.

In this work, we calculate synthetic stellar spectra of O stars adopting
the
same $\dot{M}(Z)$ and other wind parameters as used in the evolution
calculations. For WN stars, however, we use two alternative prescriptions for the
mass-loss rate metallicity scaling -- the one adopted in the evolution
calculations, and the empirical one. 

\subsection{Progenitors of GW\,150914-like systems in the quasi-chemically 
homogeneous evolution channel}
\label{sec:qhe}
\label{sec:mar16}

The stellar evolution calculations by \citet{Marchant2016} focus on the QCHE in 
close binary systems. This scenario involves a very tight binary consisting 
of two massive stars that remain fully mixed during the course of their evolution 
as a result of their tidally induced high spin. 
The models account for detailed effects of tidal interactions and differential
rotation.

The binary components start their lives with a mass ratio $q$ so close to  
unity that exchange of mass during core hydrogen burning leads to a quick 
equalization of the component masses. After this,
the components stay in contact with each other for some time, before 
the system subsequently evolves as a detached binary.
These tracks are shown in Fig.\,\ref{fig:hrd_pablo} for two metallicities. The 
tracks reflect the evolutionary paths of the primaries only. The secondaries 
have virtually identical tracks because of the mass equalization early  
in their evolution \citep[see Fig.\,1 of ][]{Marchant2016}. 

One of the two tracks displayed in Fig.\,\ref{fig:hrd_pablo} is calculated 
for a star with an initial mass of $60\,M_\odot$ at a
metallicity\footnote{\citet{Marchant2016} define solar metallicity is 
$Z_\odot=0.017$, while \citet{Eldridge2016} adopt $Z_\odot=0.02$.} of 
$0.1\,Z_\odot$, while the other one shows the evolution of a star with an 
initial mass of $50\,M_\odot$ at a lower metallicity ($Z = 0.05\,Z_\odot$). 
These metallicities roughly correspond to dwarf galaxies in the Local Group, 
like Leo\,I, IC\,1613, Phoenix, WLM, or Sextans A and B 
\citep{Kniazev2005,Leaman2013,Ross2015}.

After leaving the zero-age main sequence (ZAMS), the tracks always stay in the blue part of the HRD,
given their high surface temperatures in excess of 50\,kK. This is a consequence
of the rotationally induced QCHE. While constantly evolving 
to higher luminosities and higher temperatures, the stellar mass drops, mainly 
because of wind-mass loss, to about $34\,M_\odot$ and $35\,M_\odot$ shortly before
core collapse for the $Z = 0.1\,Z_\odot$ and $Z = 0.05\,Z_\odot$ models, respectively.

For both tracks shown in Fig.\,\ref{fig:hrd_pablo}, stellar atmosphere models 
and corresponding synthetic spectra have been calculated at five selected 
evolutionary stages. For each track, the first atmosphere model (marked by No.\,1 in 
Fig.\,\ref{fig:hrd_pablo}) is computed at the evolutionary stage
shortly after the mass ratio of the binary components has reached unity following 
the initial RLOF phase. The next atmosphere model is calculated for the point No.\,2  
when the binary components are well detached from each other and continue to 
evolve quasi-homogeneously. When the track turns to the left (No.\,3),     
that is,\ the star evolves to significantly higher surface temperatures, the third 
atmosphere model is calculated. An atmosphere model corresponding to the
evolutionary stage characterized by a reduced hydrogen abundance, as 
observed in WN stars, is calculated next  
(No.\,4). The final model (No.\,5) is calculated at the stage when the hydrogen 
abundance has dropped below 5\,\% (mass fraction) -- roughly corresponding to 
the detection limit for hydrogen in the atmospheres of WR stars. 
In the track for the higher metallicity, this evolutionary stage corresponds to the WR 
stars of carbon/oxygen spectral type (WC/WO stars). The model parameters and the 
results of our calculations for $Z=0.1\,Z_\odot$ and $0.05\,Z_\odot$ are summarized in 
Tables\,\ref{table:parameters-pablo-01Zsun} and \ref{table:parameters-pablo-005Zsun}.

\subsection{The GW-event progenitors as predicted by BPASS stellar evolution models}
\label{sec:bpass}
\label{sec:bpass-track}

%%%%%%%%%%%%%%%%%%%%%% Fig. 4 %%%%%%%%%%%%%%%%%%%%%%%%%%%%%%%%%%
%---------------------------------------------------------------
\begin{figure}[tbp]
    \centering
    \includegraphics[width=\hsize]{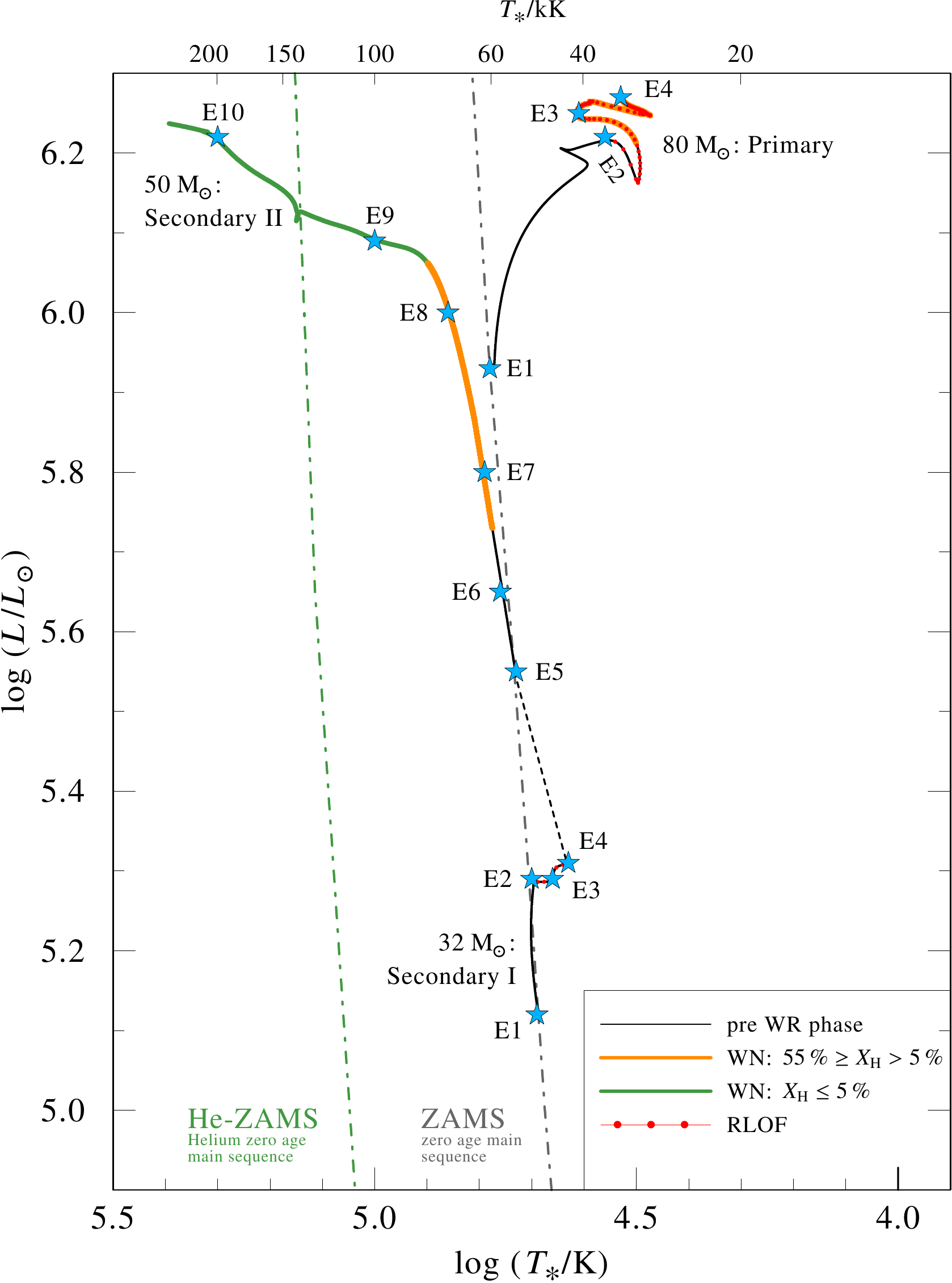}
    \caption{
        Same as Fig.\,\ref{fig:hrd_pablo} but showing the BPASS binary evolution 
        tracks. Since the BPASS binary-calculations are stopped after the first core collapse event,
        the track of the secondary is split in two parts, leaving a gap between the positions E4 and E5 (dashed line, see text for details).
        } 
    \label{fig:hrd_JJ}
\end{figure}
%---------------------------------------------------------------

The \emph{Binary Population and Spectral Synthesis} (BPASS) models\footnote{\url{bpass.auckland.ac.nz}}
consider a general population of binaries with a distribution of masses, mass ratios, 
and orbits. The code follows the evolution of both the primary and the secondary 
star \citep{Eldridge2008}. 
In contrast to the \citet{Marchant2016} models 
discussed above, the evolution channels presented by \citet{Eldridge2016} 
do not require initially very tight binaries and quick equalization 
of binary component masses to produce massive BH binaries.
The BPASS models in principle also account for common-envelope phases. However,  
this poorly understood evolutionary stage, where a nearly arbitrary amount of mass
could be removed from the system, is avoided in the models presented in this paper.

The population synthesis conducted by \citet{Eldridge2016} shows that the formation 
of GW progenitor binaries is most likely at low metallicities. 
According to this study, the occurrence of the GW progenitor 
systems is highest at \mbox{$Z=0.005\,Z_\odot$}, while mergers similar to GW\,150914 are predicted
up to metallicities of $Z = 0.5\,Z_\odot$. At a metallicity of $Z = 10^{-4}\,Z_\odot$, \citet{Eldridge2016} predict that more than a quarter of all expected BH mergers would have masses similar to those measured in the GW\,150914 event. 

Tracks for both components of a binary evolution model
leading to the formation of a GW\,150914-like binary are shown in  Fig.\,\ref{fig:hrd_JJ}.
In contrast to the scenarios developed by \citet{Marchant2016}, 
in the BPASS models the binary components do not experience mass 
transfer before core hydrogen exhaustion. Both components evolve 
like single stars until the mass transfer sets in, which then lasts for about 
$3 \times 10^{5}$ years.  At the end of this phase, the primary explodes 
as a supernova (SN), and leaves a remnant BH with a mass of about 
$35\,M_\odot$. A probability is assigned to the SN kick and for the system 
to remain bound after a SN. 

Due to the mass gain, the secondary increases its mass from $32\,M_\odot$ on 
the ZAMS to about $50\,M_\odot$ at the end of the mass transfer. The track 
of the secondary star in Fig.\,\ref{fig:hrd_JJ} has a gap between positions E\,4
and E\,5 (dashed line). This gap is because of the approximate 
treatment by the BPASS model of the secondary evolution during the 
fast RLOF mass-accretion stage.   

Initially, the evolution of the secondary star is conventional. 
However, the secondary is significantly spun up by the mass and 
angular momentum transfer during the primary RLOF phase. The newly acquired 
high rotational velocity makes the secondary evolve quasi-homogeneously 
after the first BH formation. Therefore, the secondary star 
remains compact during its subsequent evolution. Hence, a common 
envelope phase between the secondary and the BH, which could potentially lead to a BH 
in-spiral, is avoided. 

During all phases of the binary evolution prior to the primary collapse, the total 
luminosity of the system is dominated by the primary -- the contribution 
of the secondary to the overall flux is only a few percent. 

Stellar atmosphere models were calculated at characteristic points on 
the BPASS tracks as highlighted by the numbers in Fig.\,\ref{fig:hrd_JJ}. 
The key difference compared to the approach used in 
Sect.\,\ref{sec:mar16} is that we now have to compute two 
separate sets of synthetic spectra, one for the primary and one for the secondary star.  

%---------------------------------------------------------------
\begin{figure*}[htbp]
    \includegraphics[width=\hsize]{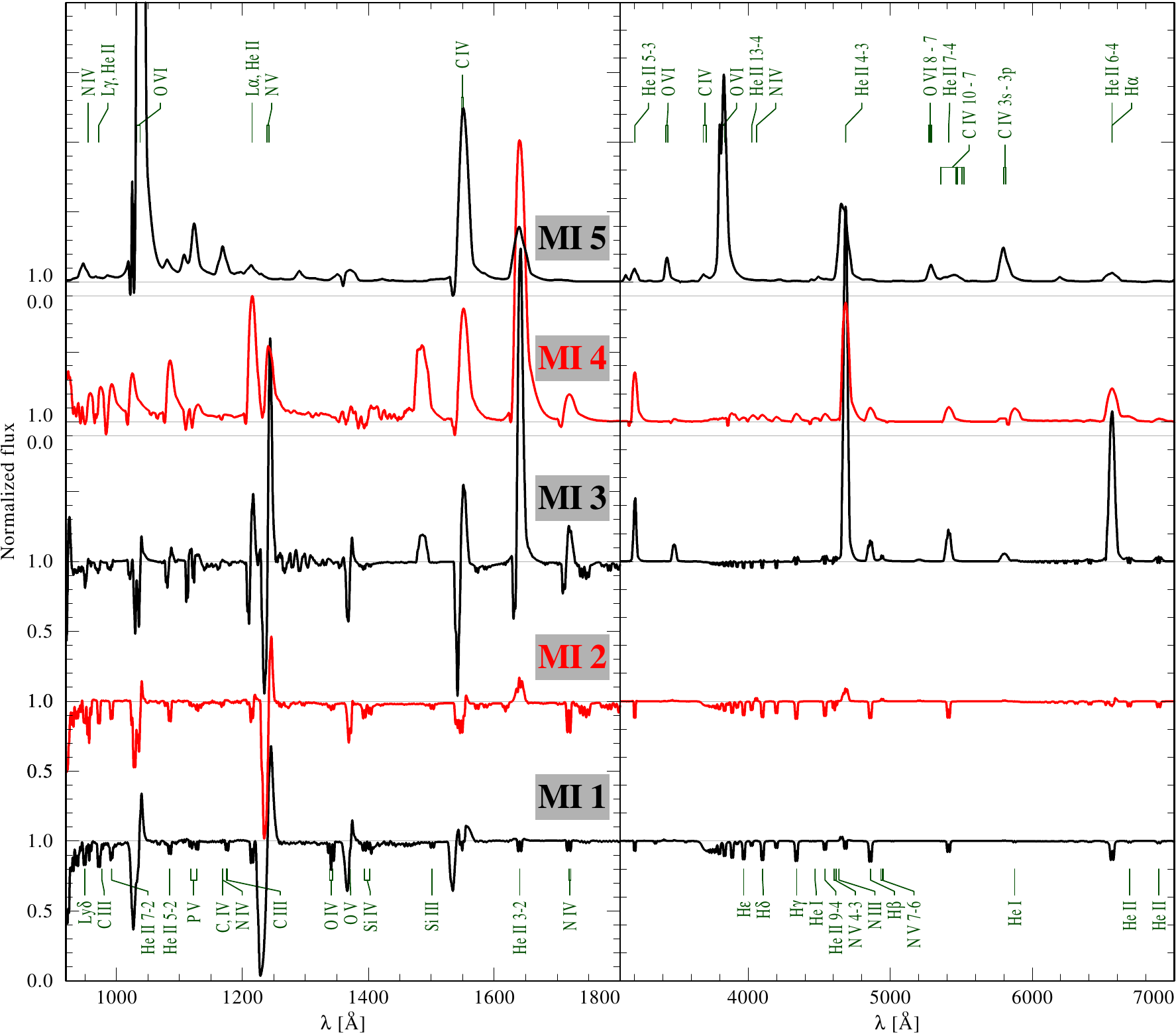}
    \caption{
        Sequence of synthetic spectra at five different 
        evolutionary stages (marked by the
        identifiers also used in Fig.\,\ref{fig:hrd_pablo} and Table\,\ref{table:parameters-pablo-01Zsun})
        of a massive star with initial mass $60\,M_\odot$ evolving quasi-homogeneously 
        at $Z=0.1\,Z_\odot$ (Marchant track I). 
        The continuum levels and the zero lines are indicated by horizontal thin gray lines.
        The time series starts at the bottom with the spectrum of the 
        model at the ZAMS. 
        To facilitate a 
        comparison with observed spectra, the model spectra are convolved with a Gaussian 
        profile with a FWHM of 1.0\,\AA.
        } 
    \label{fig:spec_comb_01}
\end{figure*}
%---------------------------------------------------------------
%%%%%%%%%%%%%%%%%%%%%%%%%%%%%%%%%%%%%%%%%%%%%%%%%%%%%%%%%%%%%%%%%%

The first two spectral models correspond to the phase when both binary 
components are on the ZAMS (E\,1 in Fig.\,\ref{fig:hrd_JJ}). 
The next stage at which synthetic spectra are computed is immediately 
before the mass-transfer starts (E\,2).
The following atmosphere models refer to a phase during which the primary looses its outer hydrogen-rich envelope and contracts (E\,3). The last atmosphere model for the primary was calculated for the short phase between the end of mass transfer and the core collapse of the primary (E\,4).

After the primary has collapsed, we compute spectral models only for the secondary 
star. The secondary now evolves quasi-homogeneously 
as a single star (E\,5, 6, and 7 in Fig.\,\ref{fig:hrd_JJ}).
We note that the BPASS models account
for rotational mixing in a somewhat simplified way. Furthermore, it is
assumed that the fast rotation required for the QCHE is
the result of mass transfer only.
Throughout the QCHE, the temperature and the luminosity increase, while the
hydrogen abundance drops. Eventually, the star starts to display WR-type 
abundances (E\,8). The star ends its 
evolution with a low surface hydrogen abundance (E\,9 and 10), 
leaving the second BH remnant upon core collapse.   
The BH merger is expected within 10\,Gyr.

\subsection{Predicted spectral sequence of potential GW-progenitors}
\label{sec:spmar16}

\begin{figure*}[tbp]
    \centering
    \includegraphics[width=\hsize]{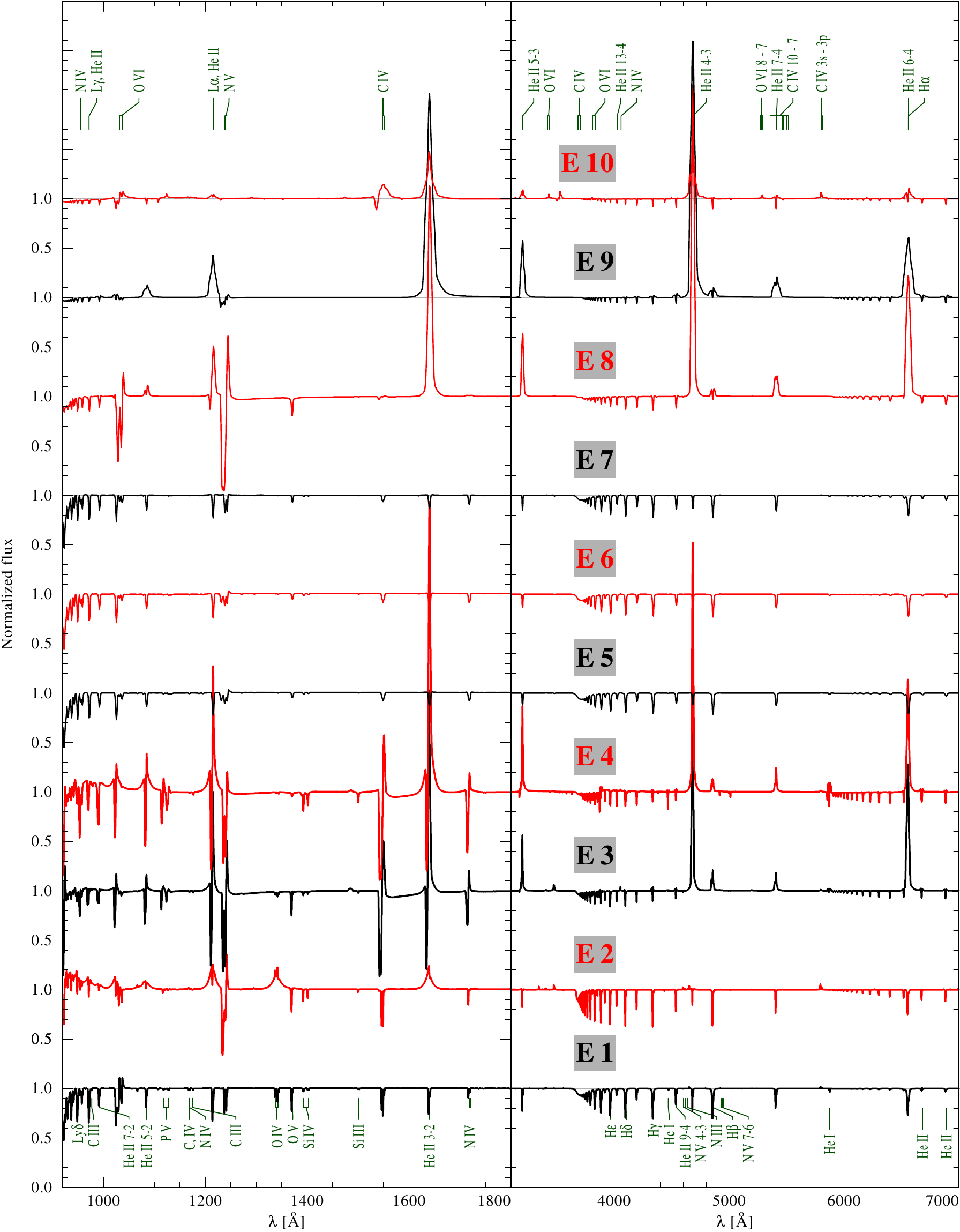}
    \caption{
             Same as Fig.\,\ref{fig:spec_comb_01} but for the BPASS 
             stellar evolution tracks and a metallicity of $0.005\,Z_\odot$.
             }
    \label{fig:spec_comb_JJ}
\end{figure*}

An exemplary sequence of synthetic UV and optical spectra at different evolutionary stages is 
shown in Fig.\,\ref{fig:spec_comb_01}. These Figures illustrate the 
stellar atmosphere models calculated for the $60\,M_\odot$ track ($Z=0.1\,Z_\odot$) published by 
\citet{Marchant2016}. Corresponding Fures for the $50\,M_\odot$ track ($Z=0.05\,Z_\odot$) can 
be found in Appendix\,\ref{sect:spectra} and spectra of a binary system evolving according to the BPASS 
tracks \citep{Eldridge2008} are shown in Fig.\,\ref{fig:spec_comb_JJ}.

In each of those plots, the time series starts at the bottom of each Figure with the spectrum of the 
model at the ZAMS. The spectra of the subsequent 
evolution steps are shifted upwards by one flux unit each for clarity. 

For illustration, the spectra of the binary components are shifted according to the maximum 
velocity amplitude of the orbital motion, assuming the statistically most probable inclination of $i \approx 57\deg$.
The spectra are also convolved with a rotational profile accounting for the rotational velocity predicted by the evolution models.
To facilitate a comparison with observations, the model spectra are degraded to a medium resolution by convolving them with a Gaussian profile with a FWHM of 1.0\,\AA. 
Moreover, the spectra 
account for a macroturbulence velocity of $20\,\mathrm{km}\,\mathrm{s}^{-1}$. In 
Appendix\,\ref{sect:spectra}, we also present plots for the different stages in 
the evolution of the investigated binary system, showing the complete normalized spectrum 
together with the spectral energy distribution (SED) for the different evolution phases.
These spectral templates can be used to identify such binaries in low-metallicity stellar 
populations.

We derived spectral types for all synthetic spectra presented in this 
work (see Tables\,\ref{table:parameters-pablo-01Zsun}, \ref{table:parameters-pablo-005Zsun}, and \ref{table:parameters-JJ}) using  
the classification criteria published by 
\citet{Crowther1998}, \citet{vanderHucht2001}, \citet{Walborn2002}, 
\citet{Evans2004}, and \citet{Sota2011}. 
As one would expect for homogeneously evolving stars, the derived spectral types reflect the high 
surface temperatures, resulting in early-type spectra throughout their evolution. In contrast, the 
BPASS models \citep{Eldridge2008} also exhibit later spectral types, which is especially evident 
during the WR phase of the primary, permitting a WN9 classification. This highlights that binary
BH progenitors can come in very different shapes, while eventually leading to similar massive BH
systems. 

The SED from the extreme UV to the infrared (IR) 
spectral range and for the different evolution phases are shown in Figs.\,\ref{fig:sed_005},
\ref{fig:sed_01}, and \ref{fig:sed_jj}, revealing significant differences in the stellar fluxes
depending on $T_\ast$, $\dot{M}$, and $Z$. This has a strong influence on observable
properties like broad band magnitudes as well as feedback parameters (see Sect.\,\ref{sec:feed} 
for details).

Comparing the evolutionary sequence of spectral types at different metallicities as
shown in Fig.\,\ref{fig:spec_comb_01} and \ref{fig:spec_comb_005} (see also Tables\,\ref{table:parameters-pablo-01Zsun} 
and \ref{table:parameters-pablo-005Zsun}), one can notice the differences arising 
from the reduction of the wind strength at lower $Z$. At a metallicity of
$Z=0.1\,Z_\odot$, the mass-loss rate is high enough that the star passes 
through the WN phase and is evolving further, exhibiting optical spectra with prominent 
metal lines, representative for WC/WO stars (see e.g.,\ Fig.\,\ref{fig:spec_comb_01}). 
This stage directly precedes the  
gravitational collapse into a BH. On the other hand, at the lower metallicity
$Z=0.05\,Z_\odot$, the stellar wind is not sufficiently strong enough to remove a 
significant part of the helium-rich outer envelope. At such low metallicity, 
the model predicts that massive QCHE stars will end their lives as WN-type stars. 

Reflecting a metal-poor chemical composition, the synthetic spectra 
calculated for our exemplary binary system are characterized by
the weakness or even absence of metal lines. 
The spectra of the main 
sequence stars, for example, basically show only hydrogen and helium lines in the optical 
and infrared  range. However, already very early in their evolution,  the QCHE leaves its footprint in a significant 
nitrogen self-enrichment (see 
Table\,\ref{table:parameters-pablo-005Zsun} and 
Table\,\ref{table:parameters-pablo-01Zsun}). This increase in the nitrogen 
abundance results in the appearance of weak but noticeable  nitrogen lines in the
optical spectra. The strength of the nitrogen lines is  
steadily decreasing with the increasing effective temperature (see 
Fig.\,\ref{fig:spec_comb_005}) due to the shift in the ionization balance 
towards higher ions that, in the case of nitrogen, do not have noticeable 
lines in the optical. 

Hence, at low metallicities, a spectral classification and analysis  
can be  troublesome. In particular, spectral classification criteria in 
the optical for Of and WN stars lose significance with reduced metallicities. 
However, even at $Z = 0.005\,Z_\odot$ , some metal lines are present in the near and 
far UV (see Fig. \ref{fig:spec_comb_JJ}). 
Moreover, since the winds of massive stars are significantly decreasing with
    metallicity, observational constraints on the mass-loss rates at low metallicity are scarce. UV spectra often provide the only diagnostics to determine 
    mass-loss rates at these metallicities.
Consequently, the access to the UV 
spectral range is essential for a detailed spectral classification 
and robust spectral analysis of those objects. 
   
A key for identifying GW progenitor systems in our neighborhood is the detection of their binary nature. All of the prototypical GW progenitor systems
presented here exhibit substantial radial velocity (RV) amplitudes throughout their evolution. Assuming an average inclination of $i = \left\langle i \right\rangle =57^\circ$, the models from
\citet{Marchant2016} predict projected RVs with amplitudes in excess of $400\,\mathrm{km}\,\mathrm{s}^{-1}$ in the initial phase of their evolution. The RV amplitudes drop as the systems
evolve, but remain in excess of $\approx 200\,\mathrm{km}\,\mathrm{s}^{-1}$. Since the primary and secondary are essentially equally massive throughout the system's evolution, their RV
amplitudes are virtually identical. Moreover, the stars are almost equally bright, making the spectrum of the system easily detectable as SB2. This applies for both
for the $Z=0.01\,Z_\odot$ as well as the $Z=0.05\,Z_\odot$ scenarios. We conclude that GW progenitor systems undergoing QCHE should be easily detectable as SB2 binaries even
with low spectral resolution ($R \approx 2000$) and a modest signal-to-noise ratio
(S/N).

GW progenitors that do not evolve as very tight binary systems (the slow evolution channel), such as those modeled by the BPASS code, which are generally harder to identify as binaries.
The reason for this is twofold: First, this scenario typically entails systems with mass ratios that significantly differ from unity. Thus the primary, the more massive star, strongly dominates the spectrum. For example, in the prototypical BPASS scenario presented here, the primary is
brighter than the secondary in the optical by a factor of ten or more throughout the evolution of the system. Secondly, the RV amplitudes in those systems
are generally much smaller compared to those in systems undergoing QCHE, since the components are not necessarily as close. In the BPASS model used here, the
projected RV amplitudes are rather large: The primary's RV amplitude ranges from initially $\approx 300\,\mathrm{km}\,\mathrm{s}^{-1}$ to $\approx 200\,\mathrm{km}\,\mathrm{s}^{-1}$ shortly before its
core collapse. The secondary's projected RV amplitude increases in the same time from $\approx 100\,\mathrm{km}\,\mathrm{s}^{-1}$ initially to $\approx 200\,\mathrm{km}\,\mathrm{s}^{-1}$. However, in the framework of the BPASS models \citep{Eldridge2016} one may also expect GW progenitor systems with initially low RV amplitudes that might be identified at best as SB1 binaries. To identify both components and properly analyze such systems, spectra with a S/N of at least $\approx150$ and a resolving power of $R \gtrsim 10000$ would be needed.

The binary systems discussed in this work are hot and very luminous ($\log L/L_\odot = 5.5 - 6.2$), and thus could in principle be observed in galaxies beyond the Local Group \citep[see, e.g.,][]{Kudritzki2014,Kudritzki2016}. However, spectroscopy of those objects with a modern instrument mounted to a 8\,m class telescope would still be restricted to a low spectral resolution on the order of $R \approx 1000$ to obtain a useful S/N. This restriction makes the detection of the investigated systems outside the Local Group very difficult.

%---------------------------------------------------------------
%---------------------------------------------------------------
%%%%%%%%%%%%%%%%%%%%%%%%%%%%%%%%%%%%%%%%%%%%%%%%%%%%%%%%%%%%%%%%

\section{Comparison with observations}
\label{sec:owr}

\subsection{Early O-type stars}

%%%%%%%%%%%%%%%%%%%%%%%%%%%%%%% TAB O3 %%%%%%%%%%%%%%%%%%%%%%%%%%%%%
\begin{table*}
    \caption{Stellar and wind parameters of the GW  
        progenitor models on the main sequence, compared to the earliest-type main sequence 
        stars observed in the SMC and IC\,1613 (see the original publications\tablefootmark{a} for details)
        }
    \label{tab:o3}
    \centering  
    \tabcolsep 0.76ex
    \begin{tabular}{l|ccc|cccc}
        \hline \hline \rule[0mm]{0mm}{4.0mm}   
        %-------------------------------------------------
        & \multicolumn{3}{c|}{Models of BH progenitors } &  
        \multicolumn{4}{c}{Empirical parameters of early low-$Z$ O stars} \\
        \hline
        %-----------------------------------------------------------
        Identifier & MI\,1 & MII\,1 & E\,5 & Sk\,183 & Cl*\,NGC\,MPG\,324 & Cl*\,NGC\,MPG\,355 &
        [BUG2007]\,A\,13 \rule[0mm]{0mm}{4.0mm} \\ 
        & (Table\,\ref{table:parameters-pablo-01Zsun}) & (Table\,\ref{table:parameters-pablo-005Zsun}) & (Table\,\ref{table:parameters-JJ}) & & & & \\
        \hline
        Spectral type                      & O3\,V((f*))& O3\,V((f*))& O3\,V((f*))z & O3\,V((f*)) 
        & O4\,V((f)) & O2\,III(f*) & O3\,V((f)) \rule[0mm]{0mm}{4.0mm} \\
        $T_{\ast}$ [kK]        & 52.2 & 50.4 & 54.1 & 47.5  & 41.5 & 52.5 & 42.5 \\  
        $\log(L/L_\odot)$       & 5.7  & 5.5 & 5.6 & 5.6 & 5.5 & 6.0 & 5.6 \\
        % $R_\ast$ [$R_\odot$]                      & 9    & 6.8    & 10  &         \\
        $v_{\rm rot}$ [km\,s$^{-1}$]  & 381 & 433 & & $60/\sin{i}$ & $70/\sin{i}$ & 
        $110/\sin{i}$ & \\    
        $v_{\rm \infty}$  [km\,s$^{-1}$]    & 3160 & 2790 & 1980 & 3000 &  2300 & 2200 
        & 2800 \\
        $M_\ast$ [$M_\odot$]                      & 63  & 50 & 50 & 38 & 44 & 64 & 
        30 \\      
        $Z/Z_\odot$  & 0.1  & 0.05  & 0.005 & 0.2 & 0.2 & 0.2 & 0.14 \\
        $\log{\dot{M}}$ [$M_\odot\,{\rm yr}^{-1}$] & -6.0 & -6.7 & -7.8 & -7.0  &  -7.0 & -6.7 & 
        -6.6\\
        \hline
    \end{tabular}
    \tablefoot{
        \tablefoottext{a}{Empirically derived stellar parameters for the SMC stars: 
            Sk\,183 \citep{Evans2012}; Cl*\,NGC\,346\,MPG\,324, Cl*\,NGC\,346\,MPG\,355 \citep{Bouret2003}; 
            [BUG2007]\,A\,13 \citep{Garcia2014,Bouret2015}.}
    }
\end{table*}
%%%%%%%%%%%%%%%%%%%%%%%%%%%%%%%%%%%%%%%%%%%%%%%%%%%%%%%%%%%%%%%%%%%%%%%%%%%%%%

In both sets of evolution models considered here, the progenitor stars settle on the 
main sequence as a very early O-type star and evolve through the WN 
spectral type towards a WC/WO type. Stars with similar spectral types are known 
in local low-metallicity galaxies, thus enabling a comparison between 
model predictions and observations.    

In general, very massive O-type stars are scarce. In Table\,\ref{tab:o3}, we 
compare the model predictions and the properties of the earliest O-type stars in 
the Small Magellanic Could (SMC) and IC\,1613. In the whole SMC, only 
four O stars with spectral types earlier than O4 are known \citep{Evans2012}. 
The earliest-type O dwarfs in the SMC occupy positions on the HRD that agree well
with the predictions of those massive BH progenitor models that evolve quasi-homogeneously.
Supporting QCHE, significant nitrogen enrichment compared to the baseline 
SMC value was measured in Sk\,183 \citep{Evans2012}. The enhanced nitrogen 
abundance in the most massive O-stars in the SMC, such as Cl*\,NGC\,346\,MPG\,324, was 
discussed in \citet{Bouret2003}. They noted that standard evolution models 
do not predict a nitrogen surface enrichment during the main-sequence phase, 
but fast rotation induces mixing and can lead to the observed abundance pattern 
\citep{Maeder1987}. Thus, analyses of observations support the idea that 
the stars may indeed evolve quasi-chemically homogeneously. Since the measured projected 
rotational velocity of the earliest O-stars in the SMC is quite significant, it is 
plausible to invoke rotational mixing to explain the abundance 
patterns. 

Another, indirect evidence for quasi-homogeneous evolution of massive stars in 
the SMC is that, except for one object, the red supergiant (RSG) 
stars in the SMC have luminosities below $\log{L/L_\odot}< 
5.8$ \citep{Massey2003a}. This suggests that while the less massive stars 
in the SMC may follow the standard evolution channels, the stars initially 
more massive than $40\,M_\odot$ are evolving quasi-homogeneously and do not 
become RGSs \citep{Hamann2017}. 

The binary status of the earliest metal poor O stars is not well known. The 
spectral analysis of Sk\,183 lead \citet{Evans2012} to suggest that this star 
is a binary with a less massive OB-type companion; binarity of Cl*\,NGC\,346\,MPG\,324 
and MPG\,355 could not be ruled out either \citep{Bouret2003}.

The stellar masses of stars in the SMC and IC\,1613 listed in Table\,\ref{tab:o3} 
are spectroscopically determined. They are lower than those predicted by the evolution models
for massive BH progenitors. However, the typical uncertainty
in spectroscopically derived surface gravities $\log{g}$ is 0.2\,dex, implying an 
error in the spectroscopic masses of about 60\,\%. Systematic discrepancies between 
spectroscopic and evolution masses are notoriously found for yet unknown reasons 
\citep{Herrero1992,Repolust2004,Massey2012}.

O-type stars with stellar parameters similar to those given in Table\,\ref{table:parameters-pablo-01Zsun}, \ref{table:parameters-pablo-005Zsun}, and \ref{table:parameters-JJ} can also be found at metallicities that are significantly higher than the metallicity range explored in this paper. An intriguing example is the LMC star VFTS\,755 \citep{Bestenlehner2014} that closely resembles our model MI\,1, albeit the mass-loss rate and the stellar temperature of VFTS\,755 are a bit lower. The same can be concluded for HD\,93128, a Galactic O3\,V((f)) star \citep{Repolust2004}. Therefore, we state that the parameter range investigated in this work is also partly realized at super-SMC metallicities.

Overall, we conclude that, nonetheless rare, stars with properties similar 
to those expected for a massive BH progenitor do exists in our neighboring 
galaxies. 

\subsection{Metallicity, clumping, and mass-loss rate}
\label{sect:Mdot}

Comparing the empirical and model stellar properties shown in Table\,\ref{tab:o3}, it 
is obvious that the empirical mass-loss rates at SMC metallicity are already
much lower than those adopted in the evolution models for $Z = 0.1\,Z_\odot$. 
Only at even lower metallicities ($Z \lesssim 0.05\,Z_\odot$) are the mass-loss rates assumed 
in the evolution calculations comparable or lower than those empirically derived for 
unevolved early O-type stars in the SMC. 
For example, the empirical mass-loss rate of an O3V star at $Z=0.2\,Z_\odot$ is only six 
times higher than the mass-loss rate assumed for a forty times smaller metallicity 
($Z=0.005\,Z_\odot$), while it is an order of magnitude lower than the one 
assumed for $Z=0.1\,Z_\odot$. 

How justified is the comparison between the GW progenitor models and real massive O 
stars, given the different metallicities adopted in the models and 
the one observed in local dwarf galaxies, like the SMC?
The metallicity requirements in evolution models are set by the model 
parametrization of mass loss via stellar winds. Mass-loss rates 
for O stars may not exceed certain limits $\dot{M} \lesssim 10^{-6}$
to not significantly lose angular momentum and maintain the fast rotation 
necessary for the quasi-homogeneous evolution. These low $\dot{M}$ values are also necessary to limit the orbit widening and to retain a stellar mass high enough 
for the production of a massive BH. As can be seen in Table\,\ref{tab:o3}, the 
evolution models assume that stellar winds are sufficiently weak only at very low 
metallicities. The adopted mass-loss rates in evolution calculations are significantly 
too large for early-type O dwarfs, and consequently the corresponding models underestimate the 
metallicity needed for the production of massive BHs.   

Nevertheless, when comparing theoretical and empirical mass-loss rates, one must be well 
aware of the problems with empirical mass-loss rate determinations. 
One caveat regarding the mass-loss diagnostic based on UV, optical, and IR data is that some of the stellar wind might be in a hot shock-heated phase and therefore not be detectable in the UV and optical. This scenario was considered by \citet{Oskinova2011} and \citet{Huenemoerder2012} as a solution for the so-called weak-wind problem. This phenomenon is based on the finding that the mass-loss rates inferred from spectral analyses of OB-dwarfs with low luminosity
$(\log L_\mathrm{bol}/L_\odot < 5.2)$ 
are significantly lower than predicted by standard mass-loss recipes \citep{Bouret2003,Martins2005,Marcolino2009}.       

Arguably, 
the most serious problem affecting mass-loss diagnostics is stellar wind 
clumping (see \citealt{Hamann2008} for an overview). Depending on the assumption on clumping 
properties, the empirically determined mass-loss rates are drastically 
different. 

Wind clumping on small scales (micro clumping) reduces mass-loss rates empirically measured from 
fitting recombination lines in the stellar spectrum (usually the H$\alpha$ line)
by a factor $\sqrt{f_{\rm V}} = 1 / \sqrt{D}$ \citep{Fullerton2006}.  Mass-loss rates measured 
by modeling of resonance lines in the UV spectra are affected by wind clumping in the opposite 
way -- namely neglecting for optically thick wind clumping in spectral modeling can lead
to an underestimation of mass-loss rates \citep{Oskinova2007}. The question of real mass-loss 
rates is not yet firmly settled. The spectral analyses of Galactic early O-type 
stars show that the standard mass-loss recipe \citep{Vink2001} most likely gives overly large values, 
while the true mass-loss rates are a factor of between 1.3 and 3 lower 
\citep{Sundqvist2011,Bouret2012,Surlan2013,Shenar2015}.   

Yet, while the existence of stellar wind inhomogeneities is indisputable, 
robust clumping diagnostics are scarce. Clear demonstrations of
wind clumping were provided by the detection of stochastic variability in the
He\,{\sc ii}\,$\lambda$4686\,\AA\ emission line in the spectrum of an O
supergiant \citep{Eversberg1998}. \citet{Markova2005} concluded that the
H$\alpha$-line variability observed for a large sample of O-type supergiants
could be explained by a structured wind consisting of shell fragments.
Using spectral diagnostics, \citet{Prinja2010} showed that the winds of 
B-type supergiants are clumped. In a recent study, \citet{Martins2015} found 
that spectral lines formed in the winds of all OB supergiants in their 
sample are variable on various time scales. \citet{Lepine2008} 
monitored the line-profile variations in a sample of O stars and 
explained their observations using a phenomenological model that depicts winds
as being made up of a large number of clumps. \citet{Brown2000} and 
\citet{Davies2007} modeled polarimetric variability arising in a clumped wind.
All these studies are largely restricted to supergiants  
\citep[see, e.g.,][]{Puls2006,Najarro2011}. The knowledge of 
clumping in the winds of O dwarfs is still limited; its improvement 
requires quite sophisticated approaches, such as three-dimensional (3D) wind simulations 
\citep{Surlan2013}.

Studies confirm that massive star winds are also clumped at low metallicities 
\citep{Marchenko2007}. This implies that empirically derived mass-loss rates 
corrected for clumping would be lower than those obtained using the standard recipe. 
Moreover, clumping may not be the sole reason for the low mass-loss rates 
empirically measured from O stars in low-metallicity galaxies. Recent models using a
different approach than \citet{Vink2001} predict lower mass-loss rates 
at low metallicities \citep{Lucy2015}. For an extensive discussion, we refer the 
reader to \citet[][and references therein]{Bouret2015}.

\begin{figure}
    \centering
    \includegraphics[width=\hsize]{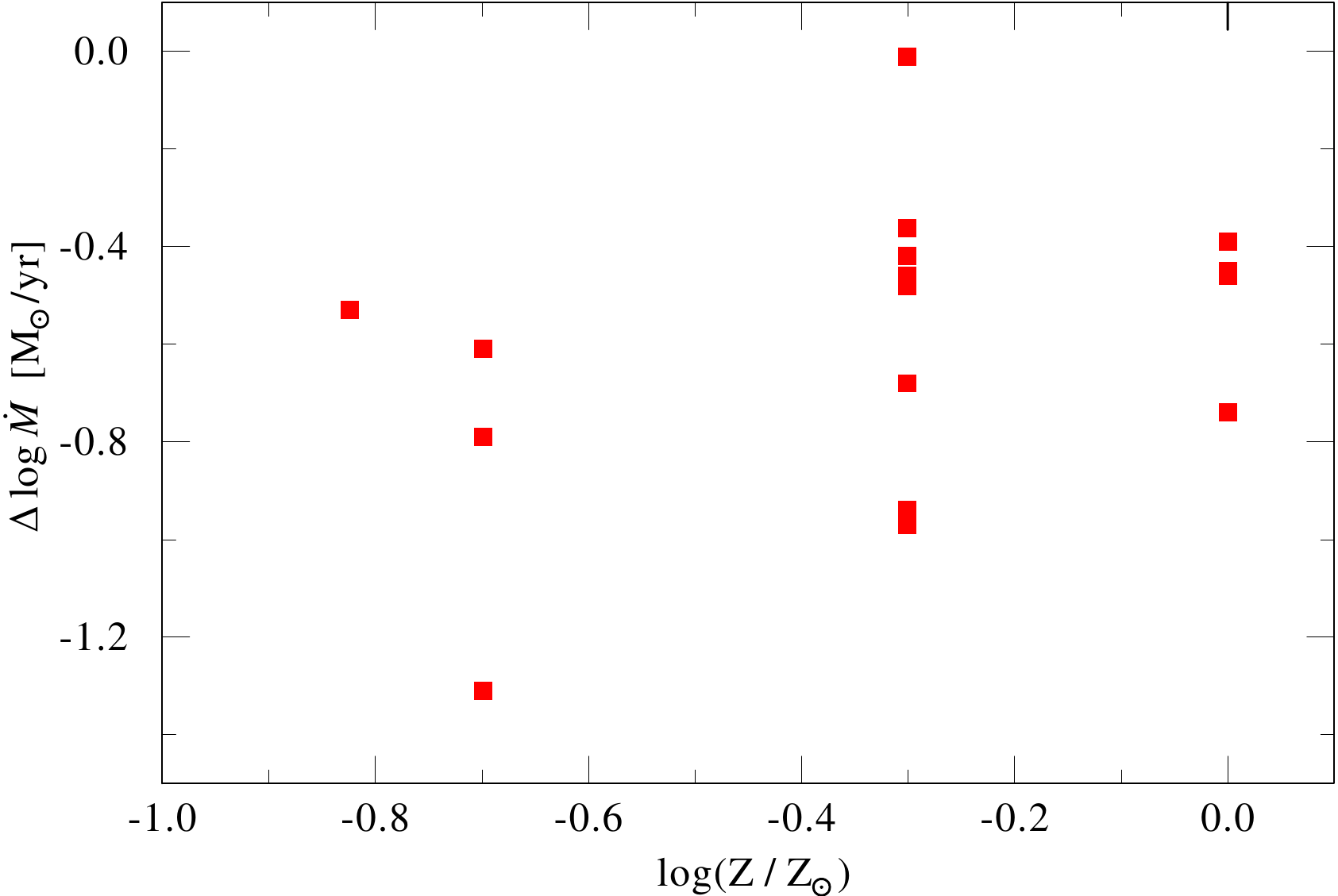}
    \caption{
        Difference between empirical mass-loss rates and those predicted by the mass-loss recipe proposed by \citet{Vink2001}. 
        The sample consists of 14 early O-dwarfs with spectral types 
        earlier than O4V (see text for the references and discussion) in the Galaxy 
        ($Z=1\,Z_\odot$), the LMC ($Z=0.5\,Z_\odot$), the SMC ($Z=0.2\,Z_\odot$), and IC\,1613 ($Z=0.14\,Z_\odot$).
    } 
    \label{fig:O3mdot}
\end{figure}

The lower mass-loss rates are corroborated by the trends seen in 
Fig.\,\ref{fig:O3mdot}, where we compile the ratio between empirically determined 
mass-loss rates and those predicted by the standard mass-loss recipe derived by \citet{Vink2001}
for a sample of 16 massive dwarfs with spectral types earlier 
than O4V, residing in four galaxies with different metallicities. 

%---------------------------------------------------------------
%---------------------------------------------------------------

Among the stars in the sample shown in Fig.\,\ref{fig:O3mdot} are 
four Galactic early O stars,  HD\,93128 (O3V((f))), HD\,93250 (O3V((f))),
and HD\,303308 (O4V((f$^+$))), that  were analyzed by \citet{Repolust2004}. The 
mass-loss rates from their Table\,4 were reduced by a factor of three to account for 
the effect of wind clumping adopting $f_{\rm V}=0.1$. The clumped ($f_{\rm 
V}=0.1$) mass-loss rate of HD\,46223 (O4V((f))) is from \citet{Martins2012}.  
The mass-loss rates for the eight early LMC O-stars are from the analyses by   
\citet{Bestenlehner2014} and \citet{Ramachandran2017}, who also adopted $f_{\rm V}=0.1$. We note
that the metallicity of the stars in the LMC sub-sample is likely slightly 
higher than $Z=0.5\,Z_\odot$. Concerning the sub-sample of the SMC stars, the 
clumped mass-loss rates of Cl*\,NGC\,346\,MPG\,324 (O4\,V((f)), $f_{\rm V}=0.1$) and
Cl*\,NGC\,346\,MPG\,368 
(O4-5\,V((f)), $f_{\rm V}=0.05$) are from \citet{Bouret2003}. The unclumped mass-loss 
rate for Sk\,183 (O3((f))) from \citet{Evans2012} was corrected for clumping with 
$f_{\rm V}=0.1$.  The clumped ($f_{\rm V}=0.03$) mass-loss rate of [BUG2007]\,A\,13
(O3V((f))) is from \citet{Bouret2015}, who notes that a higher, SMC-like, 
metallicity cannot be ruled out for this star in IC\,1613.  

\begin{figure}[htbp]
    \centering
    \includegraphics[width=\hsize]{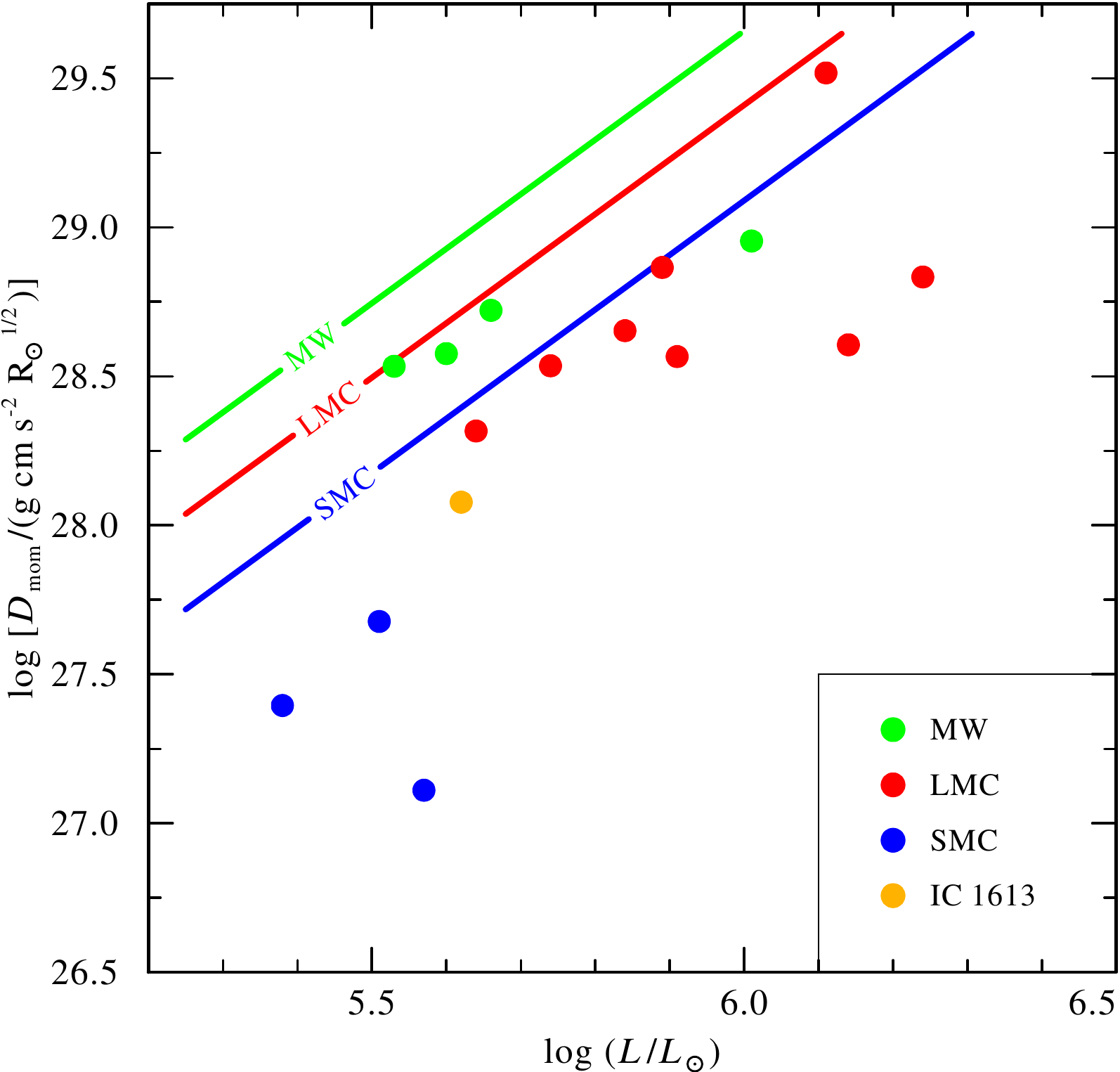}
    \caption{
        Modified wind momentum ($D_\mathrm{mom} = \dot{M} v_\infty R_*^{1/2}$) of the 
        early O-type stars discussed in Sect.\,\ref{eq:mdot_ostars} plotted over their
        luminosity. The straight lines illustrate the predictions by the standard mass-loss recipe derived by \citet{Vink2001} for solar (green), LMC (red), SMC (blue), and sub SMC (orange) metallicity. 
    } 
    \label{fig:dmom}
\end{figure}

In Fig.\,\ref{fig:dmom}, we show the modified wind momentum of these stars as a function of their luminosity. The modified wind momentum is a measure for the strength of the stellar wind \citep{Kudritzki1995,Puls1996,Kudritzki1999} that is defined as $D_\mathrm{mom} = \dot{M} v_\infty R_*^{1/2}$. The line-driven wind theory predicts a distinct relation of the form $D_\mathrm{mom} \propto L^{\alpha}$, the so-called wind-momentum luminosity relation \citep[WLR,\ ][]{Kudritzki1999}, which is expected to show a metallicity dependence. To illustrate the deviations between empirical derived values and what is predicted by a  mass-loss recipe applied in most stellar evolution calculations, we also plot the WLRs predicted by \citet{Vink2001} in Fig.\,\ref{fig:dmom}. 
Only for the primary in the binary system N206-FS\,180 \citep{Ramachandran2017} is the mass-loss rate  predicted by the standard mass-loss recipe of \citet{Vink2001} in agreement with the empirically derived value (see also Fig. \ref{fig:O3mdot}). However, the spectral analysis of this object is complicated by its binarity.
Figures\,\ref{fig:O3mdot} and \ref{fig:dmom} demonstrate that the standard mass-loss recipe severely overestimates the mass-loss rates of massive stars at early stages of their evolution.

%---------------------------------------------------------------
%---------------------------------------------------------------

The available sample of low-metallicity massive stars with empirically
established mass-loss rates is currently very limited, but nevertheless
some trends are apparent (see Fig.\,\ref{fig:O3mdot}). Based on this small
sample of very early-type O dwarfs (with $T_{\rm eff} > 40\,\mathrm{kK}$ 
and $\log{L_\ast/L_\odot} > 5.5$), we suggest an empirically calibrated
$\dot{M}$ relation of the form
\begin{equation}
\label{eq:mdot_ostars}
\dot{M}\approx 10^{-6.1} \cdot (Z/Z_\odot)^{0.85}$\,$M_\odot$\,yr$^{-1}~.
\end{equation}
The uncertainty in this correlation is on the order of a factor 2--3. 
A thorough investigation of the mass-loss rates of very early O-stars, including
the derivation of a detailed mass-loss recipe, will be the subject of a forthcoming paper.

\subsection{Wolf-Rayet stars}

%%%%%%%%%%%%%%%%%%%%%%%%%%%%%%%%%%%%%%%%%%%%%%%%%%%%%%%%%%%%%%%%%%%%%%%%%%%%%
%%%%%%%%%%%%%%%%%%%%%%%%%%%%%%%%%%%%%%%%%%%%%%%%%%%%%%%%%%%%%%%%%%%%%%%%%%%%%
\begin{table*}
    \caption{Comparison between the stellar and wind parameters of the GW 
        model progenitors during the WR phase and the earliest WR stars observed at low 
        metallicity (see the original 
        publications\tablefootmark{a} for details)
        }
    \label{tab:wr}
    \centering  
    \tabcolsep 1.0ex
    \begin{tabular}{l|ccc|ccccc}
        \hline \hline \rule[0mm]{0mm}{4.0mm}   
        %-------------------------------------------------
        & \multicolumn{3}{c|}{BH progenitor models} &  
        \multicolumn{5}{c}{Empirical parameters of low-$Z$ WR stars} \\ 
        \hline
        %-----------------------------------------------------------
        Model & MI\,4 & MII\,3 & MI\,5 & SMC AB\,1 &SMC AB\,9 &SMC\,AB12 & SMC\,AB8 & DR\,1 in IC\,1613 \rule[0mm]{0mm}{4.0mm}\\
        &(Table\,\ref{table:parameters-pablo-01Zsun}) 
        &(Table\,\ref{table:parameters-pablo-005Zsun}) 
        &(Table\,\ref{table:parameters-pablo-01Zsun}) & & & &  \\ 
        \hline
        Spectral type     & WN2 & WN2.5 & WO1 & WN3ha & WN3ha & WN3ha & WO4 (+O4V) & WO3 \rule[0mm]{0mm}{4.0mm}\\
        $T_{\ast}$ [kK]   & 100 & 70.3  & 150 & 79    & 100   & 112   & 141 &  150 \\  
        $\log(L/L_\odot)$ & 6.2  & 6.1  & 6.2 & 6.1   & 5.9   & 6.2   & 6.2 & 5.7 \\
        $v_{\rm rot}$ [km\,s$^{-1}$] & 250  & 270 & 110 &  & & & & \\    
        $v_{\rm \infty}$  [km\,s$^{-1}$] & 2400 & 1600 &3000 & 1700 & 1800 & 1800 & 
        3700 & 2750 \\
        $M_\ast$ [$M_\odot$]               & 47.5 & 44.3 & 35 & 75   & 62 & 42 & 20 & 
        20 (?) \\ 
        
        $Z/Z_\odot$  & 0.1  & 0.005 & 0.1 & 0.2 & 0.2 & 0.2 & 0.2 & 0.14 \\
        $\log{\dot{M}}$ [$M_\odot\,{\rm yr}^{-1}$] & -4.8   & -5 & -4.5 & -5.6  &  -5.7 
        & -5.8 & -4.8 & -4.7 \\
        \hline
    \end{tabular}
    \tablefoot{
        \tablefoottext{a}{Empirically derived stellar parameters for the SMC stars are from 
            \citet{Hainich2015}, \citet{Shenar2016}, \citet{Tramper2013}}
    }
\end{table*}
%%%%%%%%%%%%%%%%%%%%%%%%%%%%%%%%%%%%%%%%%%%%%%%%%%%%%%%%%%%%%%%%%%%%%%%%%%%%%%
%%%%%%%%%%%%%%%%%%%%%%%%%%%%%%%%%%%%%%%%%%%%%%%%%%%%%%%%%%%%%%%%%%%%%%%%%%%%%%

The comprehensive empirical analyses of the Galactic WR star population 
show that WN stars form two distinct sub-classes \citep{Hamann2006}. 
The descendants of stars with initial masses in the range 
$20-40\,M_\odot$ become WN early (WNE) stars, that are typically hydrogen-free. 
Such WNE stars are hotter and less luminous than the stars of the WN late (WNL) 
type. The latter stem directly from O-stars with masses exceeding $\sim 
40\,M_\odot$. WNL stars have high luminosities ($\log L/L_\odot > 6$), are 
somewhat cooler than WNEs, and usually contain a significant amount of 
hydrogen in their atmospheres.

This dichotomy, however, does not hold for the WN populations in the LMC \citep{Hainich2014} 
and SMC \citep{Hainich2015}. For the latter,
\citet{Hainich2015} found that 
the evolutionary origin of the SMC WN stars can be explained by QCHE. 
In this case, even those WN stars that originate from very massive O stars 
remain compact and hot during their evolution, while still containing hydrogen 
in their atmosphere.
However, \citet{Hainich2015} noticed that lower mass-loss rates during the WN phase need 
to be used in the QCHE models to bring the predicted hydrogen surface abundance 
in SMC WN stars in accordance with observations.

These suggestions are in line with the predictions of the GW progenitor 
models investigated in this paper. Let us first consider the spectral evolutionary 
sequence in the case of quasi-homogeneous evolution as presented by \citet{Marchant2016}
and as illustrated in 
Fig.\,\ref{fig:hrd_pablo} and Table\,\ref{table:parameters-pablo-01Zsun}. As 
one can see, the \mbox{O-star} stage is directly followed by a very early  
WN-type stage. The star contracts and becomes hotter, while retaining hydrogen in 
its atmosphere.      

The BPASS models \citep{Eldridge2016} predict a different evolution of the spectral 
types (see Fig.\,\ref{fig:spec_comb_JJ} and Table\,\ref{table:parameters-JJ}). 
In this case, the 
primary does not evolve quasi-homogeneously. Therefore, following the \mbox{O-star} 
stage and directly preceding its core collapse, the star becomes a very 
luminous and rather cool WNL star of a late subtype with a sizable amount
of hydrogen in its atmosphere.

On the other hand, the secondary, spun up after the primary's SN 
explosion, evolves quasi-homogeneously and becomes a much hotter but still very 
luminous WN star of an early sub-type. Such stars can evolve further and become 
early-type carbon/oxygen rich (WO/WC) stars that collapse into a BH.   

The population of WR stars in the SMC is considered to be fully known. 
The complete sample of SMC WR stars was analyzed in \citet{Martins2009}, 
\citet{Hainich2015} and \citet{Shenar2016}. The former two papers presented 
support for QCHE of the most massive SMC single stars. The 
latter, concentrating on the WR binaries, found no evidence of QCHE in those systems,  
which might avoid full mixing because their rotation is slowed down 
by tidal interactions in wide binaries. 

As shown in Table\,\ref{tab:wr}, the observed WR stars in local galaxies have 
properties similar to the model predictions. 
Two of these stars, namely SMC\,AB\,1 \& 9, closely resemble  some of the
stellar atmosphere models presented in this paper (model MII\,3 and model 
MI\,4, respectively) in terms of stellar temperature, 
luminosity, and mass. This shows that the stellar parameter range predicted for 
the massive BH progenitors is, to our current knowledge, quite valid.
 
Interestingly, the empirical masses 
of WN stars are even higher than expected for the GW progenitor, with the
exception being the WO-type stars that obviously cannot produce a BH with more than
$\approx 20\,M_\odot$. 
SMC\,AB\,1 \& 9 have current masses in a range 
where pair-instability supernovae (PISNe) are expected to occur.
If these stars continue to evolve quasi-homogeneously and if their mass-loss rates
do not significantly change during the rest of their evolution, the cores of 
these objects might become susceptible to PISNe \citep{Heger2002,Chatzopoulos2012},
leading to the
total disruption of the stars rather than the formation of BHs. 
However, the uncertainty of empirical WR masses are 
high because they are solely based on luminosities and mass-luminosity relations. Unfortunately,
no direct observational handle on the mass or surface gravity is available
for those stars \citep[see discussion in][]{Shenar2016}. 

As in the case of O stars, the evolution models adopt overly large mass-loss rates
during the WN evolutionary phase and, consequently, underestimate the metallicity domain
where massive BH binaries can be found. In comparison to O-type stars, an even steeper relationship between the mass-loss rate and the metallicity is observed for WN stars. 
\citet{Hainich2015} found a relation of the form $\dot{M} \approx 10^{-4.4}\, Z/Z_\odot^{1.2}\,M_\odot\,\mathrm{yr}^{-1}$.
For a more thorough derivation and discussion of this dependency, we refer the reader to \citet{Hainich2015}.

Among the binary WR stars in the SMC is AB\,8 (see Table\,\ref{tab:wr}), the only one 
that contains a WO type star known in this galaxy (while no WC-type stars 
are identified in the SMC). This system (WO4 +O4I, $20+60\,M_\odot$) is in a
relatively tight orbit, $P=16.6$\,d, only slightly larger than the orbits expected for 
a GW progenitor system. However, this is already sufficient to reduce the spin
because of tidal interactions hampering QCHE and preventing the formation of a tight BH binary. 
SMC\,AB\,8 can thus be considered as a binary BH progenitor, but most likely
it would not follow the fast evolution channel.  
Similar conclusions can be drawn for SMC\,AB\,5 (HD\,5980), which is a 
hierarchical quadrupole system containing a WN6h+WN6-7 ($61+66\,M_\odot$) 
binary with a 19.3\,d orbital period.

All of the most massive O and WR binaries observed  so far in the SMC have wider orbits 
than required for either of the GW progenitor models. It remains to be seen 
whether sufficiently close and massive binary systems will be identified in 
future surveys of low-metallicity galaxies. 

\subsection{Implications}
\label{sect:impli}

The above findings have some implications for the metallicity dependence of the BH merger rate. \citet{Belczynski2016} predicted massive BH mergers for metallicities up to $Z \le 0.1\,Z_\odot$, similar to what \citet{Marchant2016} obtained. \citet{Eldridge2016} found an upper metallicity limit of $0.5\,Z_\odot$, while predicting that the BH merger rate peaks at metallicities as low as \mbox{$Z \le 0.005\,Z_\odot$}.

However, as shown above, overly large mass-loss rates were adopted in evolution calculations of massive stars at early stages of their evolution and in the WN stage, especially at low metallicities. 
This is one reason why evolution models predict massive BH progenitors preferentially at low metallicities.    
Stellar evolution models that account for realistic mass-loss rates might lead not only to larger BH masses but also less orbital widening in the course of the stellar evolution due to angular momentum loss. Potentially giving rise to a larger fraction of stars evolving quasi homogeneously, this reinforces the importance of the evolution channels discussed in this paper for the formation of binary BHs that will merge within a Hubble time.
Moreover, we suspect that the discussed evolution channels would occur already at higher metallicities, if a steep $\dot{M}(Z)$ relation for the WN phase and lower mass-loss rates for massive main sequence stars (see Sect.\,\ref{sect:Mdot}) are used in stellar evolution calculations.

A population synthesis study evaluating the binary BH merger rates and the rates of expected LIGO events is beyond the scope of this paper. Such studies have been performed by various authors \citep[e.g.,\ ][]{Dominik2015,Belczynski2016,Mandel2016,Marchant2016,Eldridge2016}, but should be repeated with more realistic mass-loss rates as discussed above. We speculate that this will shift the peak of the binary BH merger rate to higher metallicities than predicted hitherto.

Apart from the wind mass-loss rates, further important uncertainties affect the BH masses predicted by evolution models. Episodic mass loss associated with luminous blue variable (LBV) phases might contribute significantly to the total mass lost over the lifetime of a massive star and hence substantially influence its final mass.
The LBV phenomenon is generally attributed to the proximity of these stars to the Eddington limit, at which the surface gravity is balanced by the radiative acceleration, while additional mechanisms were proposed to explain the different variabilities associated with this transient state \citep[see review by][]{Vink2012}.
An LBV-like mass-loss episode might also be related to unstable late nuclear burning phases or to the onset of a pulsational pair-instability supernova \citep{Woosley2007,Chatzopoulos2012}, as was proposed for the supernova imposter SN\,2009ip \citep{Mauerhan2013}.
However, despite intensive research on LBVs, their eruptions and the underlying mechanisms are still not understood.

In recent years, growing evidence has been provided that massive stars in a certain mass range might not explode, but end as a failed SN or directly collapse to form a BH \citep[see, e.g.,][]{OConnor2011,Kochanek2015,Sukhbold2016,Adams2017}. Albeit failed SNe can eject a substantial amount of mass \citep{Nadezhin1980,Lovegrove2013}, the remaining BH masses can be significantly higher compared to those from an energetic SN \citep[see also][]{Woosley2016}.  
The direct collapse scenario also avoids large kicks as expected for SNe that form neutron stars (NSs) or BHs \citep{Fryer2012,Mirabel2017}. This entails that multiple systems where at least one of the components directly collapses to a BH have a lower probability of becoming disrupted compared to those systems where all components exhibit a SN. 
Another consequence of the direct collapse scenario is the increased importance of stellar winds, since, in this scenario, only the winds chemically enrich the surrounding interstellar medium (ISM) and inject kinetic energy into it.  

Nevetheless, we only consider binaries in this work; an interesting possibility to form close pairs is via triple systems. A hierarchical triple system might experience an orbital tightening of the inner pair due to the Lidov-Kozai mechanism \citep[e.g.,\ ][]{Shappee2013}.

\section{X-ray properties of massive binary BH progenitors}
\label{sec:xray}

X-rays provide one of the best observational windows for the identification of binaries among the
general massive star population. Massive binary components drive strong 
supersonic stellar winds that collide, shock, and thereby power X-ray emission. During a 
later evolutionary phase, when one of the binary components has already collapsed 
into a BH, stellar wind accretion leads to characteristically strong X-ray 
emission. Therefore, observations in X-rays are pivotal for finding 
potential binary BH progenitors.    

\subsection{Colliding wind binaries}

Both components in a massive binary drive stellar winds. When the binary
orbital parameters and wind strengths are favorable, the two winds 
collide and power strong X-ray emission \citep[see recent review by][]{Rauw2016}. 
As a result, the X-ray luminosity of the binary system can be 
significantly higher than that of the components. 
Such a situation is realized in the \citet{Marchant2016} models. In these models, 
both binary components have the same spectral type during their evolutionary 
paths, drive strong winds, and are at favorable separations
to facilitate a significant X-ray emission from their colliding winds.
Therefore, X-ray observations are potentially 
effective for identifying binaries, such as those predicted by the \citet{Marchant2016} evolutionary 
scenario.

\citet{Stevens1992} showed that the efficiency of radiative 
cooling in a colliding wind binary can be estimated based on the parameter 
$\chi\approx v^4 d/\dot{M}$, where $v$ is the wind velocity in 
1000\,km\,s$^{-1}$, $\dot{M}$ is the mass-loss rate for each binary component 
in $10^{-7}\,M_\odot\,{\rm yr}^{-1}$, and $d$ is the distance to the shock 
zone in $10^7$\,km.  
In the case of identical components, this distance 
is half of the orbital separation. 
We stress that the above equation was derived for solar abundances and that the cooling efficiency will be slightly different at sub-solar metallicities. Therefore, we only use this equation for an order of magnitude estimate.
When $\chi > 1$, the cooling is adiabatic, 
and for  $\chi \ll 1$ the cooling is radiative. For models shown in
Tables\,\ref{table:parameters-pablo-01Zsun} and \ref{table:parameters-pablo-005Zsun}, 
the cooling is adiabatic only for the models MI\,1, MII\,I, MI\,5b, and MII\,5b. In all other cases, 
the cooling is radiative. The models based on \citet{Eldridge2016} evolutionary tracks 
are characterized by larger orbital separation compared to the \citet{Marchant2016} 
tracks. For all binary parameters shown in Table\,\ref{table:parameters-JJ} the 
cooling in the colliding wind zone is adiabatic.

With the help of equation (10) from \citet{Stevens1992}, we can roughly estimate the X-ray luminosities assuming adiabatic cooling.
For this purpose, we use the model binary separations and wind parameters given in Tables\,\ref{table:parameters-pablo-01Zsun} and \ref{table:parameters-pablo-005Zsun}, but note that these models do not include effects of radiative braking, which may act to reduce the derived X-ray luminosity \citep{Gayley1997}.
Our models show that 
the winds would collide when their velocity is about 75\,\%\ of $v_\infty$, 
that is,\ already very high. For an O3V+O3V binary (model MI\,1 and MII\,1), the expected X-ray 
luminosity is \mbox{$L_\mathrm{X, CWB} \approx 10^{34}$\,erg\,s$^{-1}$}, while for an WO+WO 
binary (model MI\,5b), X-ray luminosities up to $L_\mathrm{X, CWB} \lesssim 10^{36}$\,erg\,s$^{-1}$ might occur. 
Hence, if an apparently single O3\,V star has an X-ray luminosity exceeding $10^{34}$\,erg\,s$^{-1}$, this would be a strong indication for a colliding wind binary. However, if an early-type star is X-ray dim, this does not exclude its binary nature, because, for example, the binary separation might be just too large to facilitate a substantial X-ray emission. 

Binary systems with parameters shown in Table\,\ref{table:parameters-JJ} are expected to generate moderately strong X-ray emission before the primary's core collapse. The 
highest X-ray luminosity is predicted for the WN7-9+O5V systems (models E\,3 and 
E\,4), but does not exceed $\approx 10^{34}$\,erg\,s$^{-1}$
for all models in this table.

In order to estimate the X-ray luminosity for the cases where the shocked plasma cools 
radiatively, we consider the semi-analytical models presented by
\citet{Antokhin2004}. These models predict upper limits to the  
X-ray luminosity of colliding wind binaries. For none of the binary models
considered in this work, do the upper limits exceed
\mbox{$L_{\rm X, CWB}\approx 10^{34}$\,erg\,s$^{-1}$}.   
This upper limit is in agreement with observations. In the Galaxy, 
X-ray luminosities of massive O+O binaries, in general, do not exceed 
$10^{34}$\,erg\,s$^{-1}$, while some WR+O binaries can be an
order of magnitude brighter \citep{Oskinova2005, Naze2011, Gagne2012}.    

The known colliding wind binaries in the SMC are not especially X-ray bright. The X-ray 
brightest system is the quadrupole star AB\,5 (WN6h+WN6-7, HD\,5980) with 
\mbox{$L_{\rm X, CWB}\approx 2\times 10^{34}$\,erg\,s$^{-1}$} \citep{Naze2007, Shenar2016}. 
The putatively single O-type stars have lower X-ray luminosities -- none of them was so 
far detected in X-rays, putting the upper limit on their X-ray emission at about
$10^{33}$\,erg\,s$^{-1}$ \citep{Oskinova2013}. Therefore, it appears safe to 
conclude that, in general, O-type and WR stars with X-ray luminosities exceeding 
$10^{33}$\,erg\,s$^{-1}$ are most likely colliding wind massive binaries.

Yet, it is important to realize that, while X-ray luminosities in the range 
$10^{33}-10^{35}$\,erg\,s$^{-1}$ for O and WR-type stars may be a sufficient 
criterion to identify them as colliding wind binaries, it is not a necessary 
condition. For example, the WO4 + O4 binary AB\,8 in the SMC is not detected in 
X-rays with an upper limit \mbox{$< 5 \times 10^{32}$\,erg\,s$^{-1}$}. Orbital 
geometry, wind structure and opacity, as well as radiative braking are 
among the factors that are capable of significantly reducing the emergent X-ray flux. 
Hence, even very massive binaries may be X-ray faint.         
 
\subsection{High-mass X-ray binaries}
\label{subsect:hmxb}

At some point in the evolution of a massive binary, the 
primary might be already collapsed to a BH, while the secondary is still a 
normal non-degenerate star. The accretion of matter from the secondary into the 
BH will power strong X-ray emission. Such systems are known as high-mass X-ray 
binaries (HMXBs). The GW progenitor models predict the orbital configuration, 
the BH mass, and the parameters of the donor star. With this information at hand,
one can calculate the expected X-ray luminosities of HMXBs during an immediate 
phase before the formation of a binary black hole.   

An upper limit to the luminosity of a BH is set by the Eddington luminosity. 
This is the luminosity at which the radiative acceleration from 
the scattering of photons by electrons equals the inward gravitational 
force, \mbox{$L_{\rm Edd}= 4\pi G M_{\rm BH} c \kappa^{-1}$}, 
where $\kappa$ is the mass-absorption coefficient, and other
notations have their usual meaning. Assuming that the plasma is fully ionized and 
that it only consists of hydrogen and helium, the Eddington luminosity can be 
written as 
\mbox{$L_{\rm Edd}\approx 2.55\times 10^{38}(\frac{M_{\rm BH}}{M_\odot})\,/\, (1+X_{\element{H}})$\,erg\,s$^{-1}$},
with $X_{\element{H}}$ being the hydrogen mass fraction (solar value: \mbox{$X_{\element{H}} = 0.7$}).
For example, a $36\,M_\odot$ mass BH accreting matter with a solar hydrogen abundance, 
this limit becomes $5.4\times 10^{39}$\,erg\,s$^{-1}$. 

This classical Eddington X-ray luminosity is comparable to that of ultra-luminous X-ray 
sources (ULXs).  The latter are usually defined as non-nuclear, point-like 
X-ray sources with an apparent isotropic X-ray luminosity exceeding 
$10^{39}$\,erg\,s$^{-1}$ in the \mbox{0.3--10.0\,keV} band  
\citep[see][and references therein]{Kaaret2017}. 

The X-ray luminosity of an accreting BH is related to the accretion rate, 
$S_{\rm accr}$, via the accretion efficiency constant $\epsilon$ that depends 
on the detailed physics of accretion: $L_\mathrm{X}=\epsilon S_\mathrm{accr} \,c^2$. We 
can define the Eddington accretion rate such that $S_{\rm Edd}\equiv L_{\rm 
Edd} c^{-2}$. Then, the dimensionless accretion rate and luminosity may be 
written as
\begin{equation}
s_{\rm accr}\equiv \frac{S_{\rm accr}}{S_{\rm Edd}}~~~{\rm and}~~~l\equiv \frac{L_{\rm X}}{L_{\rm Edd}}~.
\label{eq:ls}
\end{equation}
Hence, the luminosity of an accreting BH can be expressed in dimensionless 
units as
\begin{equation}
 l=\epsilon s_{\rm accr}
 \label{eq:l}
.\end{equation}
In both sets of evolution models that we consider in this work, the systems 
remain detached after the formation of the first black hole. Therefore, the X-ray 
emission of the BH is powered only by stellar wind accretion.  The 
stellar wind accretion rate can be estimated using the Bondi-Hoyle-Lyttleton 
formalism \citep[e.g.,][]{Davidson1973,Martinez-Nunez2017} as
\begin{equation}
 s_{\rm accr}\approx 1.5\times 10^{7}\cdot
 \frac{M_{\rm BH} \dot{M}}{v_{8}^4 a_{\rm BH}^2}, 
\label{eq:sb}
 \end{equation}
where $M_{\rm BH}$ is in $M_\odot$,  $\dot{M}$ is the 
stellar wind mass-loss rate in units of $M_\odot$\,yr$^{-1}$, $v_8$ is 
the stellar wind velocity in $10^8\,\mathrm{cm}\,\mathrm{s}^{-1}$, and $a_{\rm BH}$ is the orbital 
separation in $R_\odot$. 

The Bondi accretion rate (Eq.\,\ref{eq:sb}) gives an upper limit on the 
accretion rate from stellar winds, since it does not account for radiative 
feedback effects. On the other hand, the Eddington rate $S_{\rm Edd}$ gives 
the maximum accretion rate limited by the radiative pressure feedback close to 
the BH. While detailed numeric simulations are required to obtain robust 
estimates of accretion rates \citep{Park2011}, one can more readily estimate 
the accretion efficiency $\epsilon$. In the standard Shakura-Sunyaev disk 
model \citep{Shakura1973}, accretion occurs via a geometrically thin but optically 
thick disk. For such disks, $\epsilon \approx 0.1$ is a constant 
\citep[e.g.,][]{Shapiro1973}. On the other hand, in case of advection-dominated 
disks or spherical accretion,  $\epsilon \propto s_{\rm accr} \lesssim 10^{-4}$ 
\citep[see Fig.\,2 in][]{Park2001}. 
For $s_{\rm accr} > 1$, the super-Eddington accretion regime is realized and the 
system appears as an ULX characterized by a high X-ray luminosity and powerful 
outflows \citep{Pout2007}.  

The predicted accretion rates based on Eq.\,\ref{eq:sb} and X-ray luminosities for the 
QCHE GW progenitor models are given in Tables\,\ref{table:parameters-pablo-01Zsun} and
\ref{table:parameters-pablo-005Zsun}, assuming $\epsilon \approx 0.1$ and a $36\,M_\odot$ BH. As an example, let us consider 
the progenitor models given in Table\,\ref{table:parameters-pablo-01Zsun}. 
For the given orbital parameters and masses, super-Eddington accretion rates 
are predicted only for the WN stages. Thus, ULXs with a WR-type donor may 
be expected. Based on existing observations, only five WR stars with 
relativistic companions are suspected. The Galactic system Cyg\,X-3 is  strongly obscured, revealing only little information about 
its donor star, even though its identification as a WR star seems to be secure
\citep{Zdziarski2013}. The BH mass in Cyg\,X-3 seems to be quite low, ($\approx 
2\,M_\odot$). 
The extragalactic systems IC\,10 X-1 and NGC\,300 X-1 may host WN-type donors \citep{Crowther2007}. 
The mass of the compact object in \mbox{IC\,10 X-1} was recently reassessed  
and also found to be quite low ($\approx 2\,M_\odot$) \citep{Laycock2015}.    
In the case of NGC\,300 X-1, \citet{Binder2015} showed that a low-mass 
donor star is not excluded and that the BH mass measurements  are not reliable.
The WR-type features observed in M101\,ULX-1 are likely not from the donor star, 
but from a disk wind instead \citep{Soria2016}. The nature of  
CXOUJ123030.3+413853 is still controversial \citep{Esposito2015}.  Hence, 
it appears that no WN HMXB hosting a massive BH (as expected for a GW 
progenitor system) is firmly detected yet.  
 
According to the models by \citet{Marchant2016}, the  first BH is 
formed when the secondary is already an evolved WR-type star. For a
GW\,150914-like progenitor, the first BH is born when the secondary is a WO star 
and thus the lifetime of the HMXB would be very short. At this stage, the 
accretion rate is approximately a few percent of the Eddington rate. 
Such systems would be observed as HMXBs, but would not manifest themselves 
as ULXs. Moreover, WC/WO winds are optically very thick for X-rays 
\citep{Oskinova2009}. As a consequence, the observable X-ray 
luminosity might be significantly lower than the intrinsic one. 
HMXBs with WC/WO-type donors are not known.

The evolutionary models by \citet{Eldridge2016} predict orbital parameters 
only up to the first core collapse (see Table\,\ref{table:parameters-JJ}). To estimate 
the X-ray luminosity of the HMXB formed after the primary's collapse into a 
BH, we assumed that the orbital separation remains the same as in the moment 
of the BH formation, that is, $64\,R_\odot$. The X-ray luminosities powered by 
direct accretion shown in Table\,\ref{table:parameters-JJ} are modest, so no 
ULX formation is predicted for these models.

\section{Feedback parameters of massive BH progenitors}
\label{sec:feed}

The massive black hole progenitor models predict their formation in a low $Z$ 
environment. This requirement is to reduce the amount of matter lost by the
massive star via its wind before collapse.   

Since the stellar winds are weaker at low $Z$, such massive stars inject less 
mechanical energy and momentum into the ISM. The exception 
could be WR stars that exhibit a significant self-enrichment with nuclear 
burning products, like WC and WO stars. However, the $50\,M_\odot$ track 
based on \citet{Marchant2016} models for a metallicity of $0.05\,Z_\odot$ shows 
that self enrichment is quite limited even for the stars evolving 
quasi-homogeneously during a significant fraction of their life. Moreover, 
more massive stars may directly collapse into BHs avoiding a SN explosion, 
further reducing the massive star feedback at low $Z$.   

While mechanical energy input from massive stars at low $Z$ may be reduced 
compared to solar metallicity, the radiative energy feedback is very 
significant. A salient characteristic of the QCHE binary models shown in 
Fig.\,\ref{fig:hrd_pablo} (and partly also those shown in Fig.\,\ref{fig:hrd_JJ}) 
is that these stars evolve to very high stellar 
temperatures even at very low metallicities, in strong contrast to 
the standard non-homogeneous models. Consequently, the radiative feedback
from a homogeneously evolving star of a certain initial mass is significantly 
higher than its counterpart that follows a classical, non homogeneous evolution 
path (see e.g.,\ \citealt{Stanway2016} for the effect of QCHE on stellar
populations). Therefore, just a few very hot and luminous WN stars resulting
from QCHE can, in principle, dominate the entire ionizing radiation budget 
of a low metallicity dwarf galaxy. 

An intriguing example of such a galaxy is I~Zw\,18, which is characterized 
by a metallicity of $Z \approx 1/32\,Z_\odot$ \citep{Vilchez1998}. 
\citet{Kehrig2015} determined the rate of \ion{He}{ii} ionizing photons
necessary to power the observed \ion{He}{ii} nebular emission to be 
$1.33 \times 10^{50}\,\mathrm{s}^{-1}$. For comparison, \mbox{model MII\,5} calculated for the 
$50\,M_\odot$ track ($Z = 0.05\,Z_\odot$)
provided by \citet{Marchant2016} emits a \ion{He}{ii} ionizing photon flux of 
$\log Q_\mathrm{\ion{He}{ii}} = 49.31$ photons\,$\mathrm{s}^{-1}$. Thus, only 
six of these stars (three binary systems) would be able to provide the necessary ionizing 
radiation to explain the observed \ion{He}{ii} nebular emission in I~Zw\,18.

Homogeneously evolving massive and very massive stars at the metallicity
of I~Zw\,18 were studied by \citet{Szecsi2015}. The authors
investigated the radiative feedback provided by their models, assuming blackbody fluxes.
In comparison to those results, our sophisticated stellar atmosphere models give 
similar hydrogen ionizing fluxes, but the predicted \ion{He}{ii} ionizing 
fluxes are an order of magnitude lower. 

As shown above, the mass-loss rates assumed by evolution models seem to be overestimated,
which is partly attributable to the $Z$ scaling (see Sect.\,\ref{sect:evo}). 
It remains to be seen how a steeper $\dot{M}$-$Z$ prescription will 
affect the outcome of the evolution calculations. Here we only explore the
effect of different mass-loss rates in the WN phase on the number of ionizing photons 
and other observable quantities, such as broad-band magnitudes. For this purpose, 
we have calculated synthetic spectra with different $\dot{M}$, according to the two 
$\dot{M}(Z)$ prescriptions given in Sect.\,\ref{sect:evo}. The comparison between these 
two model sets is presented in Tables\,\ref{table:parameters-pablo-01Zsun} and
\ref{table:parameters-pablo-005Zsun}. 

The most severe effect is seen in the atmosphere models with $Z = 0.1\,Z_\odot$. The 
number of \ion{He}{ii} ionizing photons of those models using the steep 
$\dot{M}(Z)$ prescription \citep{Hainich2015} can be several orders of magnitude higher compared 
to the models that follow the shallower dependence of $\dot{M}(Z)$
(e.g.,\ see model MI\,3 in Table\,\ref{table:parameters-pablo-01Zsun}). In contrast, at 
lower metallicities ($Z = 0.05\,Z_\odot$) the number of \ion{He}{ii} ionizing photons is 
slightly higher in those models that assume higher mass-loss rates. 

Most of the potential progenitor systems investigated in this work are characterized by 
high luminosities and high surface temperatures, entailing hard flux distributions 
(see Figs.\,\ref{fig:sed_01}, \ref{fig:sed_005}, and \ref{fig:sed_jj}). Consequently,
those stars are much brighter in the UV and optical in comparison to the IR 
(see Tables\,\ref{table:parameters-pablo-01Zsun}, \ref{table:parameters-pablo-005Zsun}, 
and \ref{table:parameters-JJ}). 
Moreover, the low-metallicity massive stars are significantly fainter in the 
IR compared to similar stars at higher metallicities \citep[see also ][]{Crowther2006b}, 
because of the lower mass-loss rates at low $Z$.
It is informative to compare, for example, model MI\,4 with MII\,4 calculated for the $0.1\,Z_\odot$ 
and $0.05\,Z_\odot$ tracks, respectively (see Tables\,\ref{table:parameters-pablo-01Zsun} and 
\ref{table:parameters-pablo-005Zsun}). In latter case, the K-band magnitude is more 
than 4\,mag lower than in the former case, which  is mainly attributable to the 
significantly lower $\dot{M}$. This hardening of the massive star SED 
at lower metallicities needs to be taken into account in an IR-based census of 
massive stars in low $Z$ galaxies.

%__________________________________________________________________
\section{Summary and conclusions}
\label{sect:s-and-c}

The GW observatories have discovered BHs with large masses, well above 
those previously known from observations of X-ray binaries. Elaborate 
models have been proposed showing that such massive BHs can be a natural 
result of stellar binary evolution. 

In this paper, we consider two independent model sets presented by 
\citet{Marchant2016} and \citet{Eldridge2016}.
Both models provide detailed stellar evolution tracks for massive BH 
progenitors that eventually lead to a BH merger event similar to GW\,150914. 
Quasi-chemically homogeneous evolution is realized in both sets of 
evolution models, while the common envelope evolution is avoided and 
mass removal by stellar winds is an important factor in the evolution towards
the BH binary system.

On the basis of these evolution models and using advanced non-LTE stellar 
atmosphere PoWR models, we compute synthetic stellar spectra of massive BH progenitors 
at key evolutionary stages to provide spectral templates. Their spectral 
classification, photometry, and stellar feedback parameters are  
established. This allows direct comparison between model predictions and 
empirical parameters of massive stars. Our main conclusions are: 

\smallskip\noindent
1) The range of stellar parameters predicted for massive BH progenitors is realized in nature.

\smallskip\noindent
2) The massive BH progenitors evolve from O3V through WN towards WO spectral 
types, while the latter is only reached for the highest metallicity ($Z = 0.1\,Z_\odot$) 
considered in this work. 

\smallskip\noindent
3) Existing evolution models adopt stellar mass-loss rates, which are significantly too large.
Empirically derived mass-loss rates are up to an order of magnitude lower. For early O-type dwarfs (with $T_{\rm eff} > 40\,\mathrm{kK}$ 
and $\log{L_\ast/L_\odot} > 5.5$), empirical measurements
suggest a $\dot{M}$ relation of the form 
\mbox{$\dot{M}\approx 10^{-6.1} \,(Z/Z_\odot)^{0.85}\,M_\odot\,\mathrm{yr}^{-1}$}, resulting in 
significantly lower mass-loss rates compared to standard prescriptions.
The corresponding relation for WN stars is $\dot{M} \approx 10^{-4.4} \, (Z/Z_\odot)^{1.2}\,M_\odot\,\mathrm{yr}^{-1}$.

\smallskip\noindent
4) We suspect that the metallicity at which the massive binary-BH merger rate peaks as predicted by population synthesis studies will be higher, if mass-loss rates in agreement with the empirical values are used in stellar evolution calculations. A significant number of massive BHs may be formed already at SMC-like metallicities \mbox{($\lesssim 0.2\,Z_\odot$)}.

\smallskip\noindent
5) There is no one-to-one correspondence between GW progenitor prediction models and already known objects. This conclusion holds for both known massive binaries and HMXBs.

\smallskip\noindent
6) We provide spectral templates and broad band magnitudes that should help to identify 
progenitors of massive BH merger events.

%__________________________________________________________________
\begin{acknowledgements}

We thank our anonymous referee for their constructive comments.
A.\,A.\,C.\,S. is supported by the Deutsche Forschungsgemeinschaft (DFG) under grant HA 
1455/26. T.\,S. is grateful for financial support from the Leibniz Graduate 
School for Quantitative Spectroscopy in Astrophysics, a joint project of the 
Leibniz Institute for Astrophysics Potsdam (AIP) and the Institute of Physics 
and Astronomy of the University of Potsdam. 
This research has made use of the VizieR catalogue access tool, Strasbourg, France. The original description of the VizieR service was published in A\&AS 143, 23.

\end{acknowledgements}

\bibliographystyle{aa}
\bibliography{paper}

%%%%%%%%%%%%%%%%%%%%%%%%%% ONLINE MATERIAL %%%%%%%%%%%%%%%%%%%%%%%%%%%%%%%%%%%%

\Online
\label{onlinematerial}

\begin{appendix} %first online appendix
\section{Additional tables}
\label{sec:addtables}

%%%%%%%%%%%%%%%%%%%%%%%%%%%%%%%%%%%%%%%%%%%%%%%%%%%%%%%%%%%%%%%%%%%%
%%%%%%%% additional tables
%%%%%%%%%%%%%%%%%%%%%%%%%%%%%%%%%%%%%%%%%%%%%%%%%%%%%%%%%%%%%%%%%%%%

%%%%%%%%%%%%%%%%%%%%%%%%%%%%%%%%%%%%%%%%%%%%%%%%%%%%%%%%%%%%%%%%%%%%
%%%%%%%% model atoms II
%%%%%%%%%%%%%%%%%%%%%%%%%%%%%%%%%%%%%%%%%%%%%%%%%%%%%%%%%%%%%%%%%%%%
\begin{table*}[htbp]
\caption{Atomic models used in the stellar atmosphere calculations} 
\label{table:model_atoms}
\centering  
\tabcolsep 1.1ex
\begin{tabular}{lSSSSSS}
\hline\hline %---------------------------------------
    &
   \multicolumn{2}{c}{Model atom set I} &
   \multicolumn{2}{c}{Model atom set II}\rule[0mm]{0mm}{3.5mm} &
   \multicolumn{2}{c}{Model atom set III}
   \\
%-------
   Ion &
   \multicolumn{1}{c}{Number of levels} &
   \multicolumn{1}{c}{Number of lines\tablefootmark{a}} &
   \multicolumn{1}{c}{Number of levels} &
   \multicolumn{1}{c}{Number of lines\tablefootmark{a}} &
   \multicolumn{1}{c}{Number of levels} &
   \multicolumn{1}{c}{Number of lines\tablefootmark{a}}  \rule[0mm]{0mm}{3.mm}
   \\
\hline  %---------------------------------------------------------------------
  \ion{H}{i}                      &  22    & 231   &  22  &  231   &   0  &    0 \rule[0mm]{0mm}{3.4mm}\\
   \ion{H}{ii}                    &  1     &   0   &   1  &    0   &   0  &    0 \\
   \ion{He}{i}                    &  35    & 271   &  35  &  271   &  35  &  271 \\
   \ion{He}{ii}                   &  26    & 325   &  26  &  325   &  26  &  325 \\
   \ion{He}{iii}                  &   1    &   0   &   1  &    0   &   1  &  0 \\
   \ion{N}{ii}                    &  38    & 201   &   0  &    0   &   0  &  0 \\
   \ion{N}{iii}                   &  36    & 146   &  36  &  146   &  36  &  146 \\
   \ion{N}{iv}                    &  38    & 154   &  38  &  154   &  38  &  154  \\
   \ion{N}{v}                     &  20    & 114   &  20  &  114   &  20  &  114  \\
   \ion{N}{vi}                    &  14    &  48   &  14  &   48   &  14  &  48  \\
   \ion{C}{ii}                    &  32    & 148   &   0  &    0   &   0  &  0 \\
   \ion{C}{iii}                   &  40    & 226   &  40  &  226   &  40  &  226 \\
   \ion{C}{iv}                    &  25    & 230   &  25  &  230   &  25  &  230 \\
   \ion{C}{v}                     &  29    & 120   &  29  &  120   &  29  &  120  \\
   \ion{C}{vi}                    &  15    & 105   &  15  &  105   &  15  &  105  \\
   \ion{O}{ii}                    &  37    & 150   &   0  &    0   &   0  &  0 \\
   \ion{O}{iii}                   &  33    & 121   &  33  &  129   &  33  &  129 \\
   \ion{O}{iv}                    &  29    &  76   &  29  &   76   &  29  &   76  \\
   \ion{O}{v}                     &  36    & 153   &  36  &  153   &  36  &  153  \\
   \ion{O}{vi}                    &  16    & 101   &  16  &  101   &  16  &  101 \\
   \ion{O}{vii}                   &   0    &   0   &  15  &   64   &  15  &   64 \\
   \ion{O}{viii}                  &   0    &   0   &   1  &    0   &   1  &  0 \\
   \ion{S}{iii}                   &  23    &  38   &  23  &   38   &  23  &   38  \\
   \ion{S}{iv}                    &  11    &  13   &  11  &   15   &  11  &   15  \\
   \ion{S}{v}                     &  10    &   8   &  10  &    8   &  10  &    8  \\
   \ion{S}{vi}                    &   1    &   0   &   1  &    0   &   1  &  0 \\
   \ion{Mg}{i}                    &   1    &   0   &   0  &    0   &   0  &  0  \\
   \ion{Mg}{ii}                   &  32    & 120   &  32  &  120   &  32  &  120  \\
   \ion{Mg}{iii}                  &  43    & 158   &  43  &  158   &  43  &  158  \\
   \ion{Mg}{iv}                   &  17    &  27   &  17  &   27   &  17  &   27  \\
   \ion{Mg}{v}                    &   0    &   0   &  20  &   25   &  20  &   25  \\
   \ion{Mg}{vi}                   &   0    &   0   &  21  &   32   &  21  &   32  \\
   \ion{Mg}{vii}                  &   0    &   0   &   1  &    0   &   1  &    0  \\
   \ion{Si}{ii}                   &   1    &   0   &   1  &    0   &   1  &    0  \\
   \ion{Si}{iii}                  &  24    &  68   &  24  &   68   &  24  &   68  \\
   \ion{Si}{iv}                   &  23    &  72   &  23  &   72   &  23  &   72  \\
   \ion{Si}{v}                    &   1    &   0   &   1  &    0   &   1  &    0  \\
   \ion{P}{iv}                    &  12    &  16   &  12  &   16   &  12  &   16  \\
   \ion{P}{v}                     &  11    &  22   &  11  &   22   &  11  &   22  \\
   \ion{P}{vi}                    &   1    &   0   &   1  &    0   &   1  &  0 \\
   \ion{G}{ii}\tablefootmark{b}   &   1    &   0   &   0  &    0   &   0  &  0 \\
   \ion{G}{iii}\tablefootmark{b}  &  13    &  40   &   1  &    0   &   1  &  0 \\
   \ion{G}{iv}\tablefootmark{b}   &  18    &  77   &  18  &   77   &  18  &  77  \\
   \ion{G}{v}\tablefootmark{b}    &  22    & 107   &  22  &  107   &  22  &  107  \\
   \ion{G}{vi}\tablefootmark{b}   &  29    & 194   &  29  &  194   &  29  &  194  \\
   \ion{G}{vii}\tablefootmark{b}  &  19    &  87   &  19  &   87   &  19  &  87  \\
   \ion{G}{viii}\tablefootmark{b} &  14    &  49   &  14  &   49   &  14  &  49 \\
   \ion{G}{ix}\tablefootmark{b}   &  15    &  56   &  15  &   56   &  15  &  56  \\
   \ion{G}{x}\tablefootmark{b}    &   1    &  0    &  28  &  170   &  28  &  170  \\
   \ion{G}{xi}\tablefootmark{b}   &   0    &  0    &  26  &  161   &  26  &  161  \\
   \ion{G}{xii}\tablefootmark{b}  &   0    &  0    &  13  &   37   &  13  &  37  \\
   \ion{G}{xiii}\tablefootmark{b} &   0    &  0    &  15  &   50   &  15  &  50  \\
   \ion{G}{xiv}\tablefootmark{b}  &   0    &  0    &  14  &   49   &  14  &  49  \\
   \ion{G}{xv}\tablefootmark{b}   &   0    &  0    &  10  &   25   &  10  &  25  \\
   \ion{G}{xvi}\tablefootmark{b}  &   0    &  0    &   9  &   20   &   9  &  20  \\
   \ion{G}{xvii}\tablefootmark{b} &   0    &  0    &   1  &    0   &   1  &  0  \\
\hline %---------------------------------------------------------------------
\end{tabular}
\tablefoot{
    \tablefoottext{a}{Not counted are transitions with a negligible oscillator strength.}
    \tablefoottext{b}{G denotes a generic atom that incorporates the following 
iron group elements \element{Sc}, \element{Ti}, \element{V}, \element{Cr}, 
\element{Mn}, \element{Fe}, \element{Co}, and \element{Ni}. The corresponding ions are 
treated by means of a superlevel approach \citep[][]{Graefener2002}.}
}
\end{table*}
%%%%%%%%%%%%%%%%%%%%%%%%%%%%%%%%%%%%%%%%%%%%%%%%%%%%%%%%%%%%%%%%%%%%

%%%%%%%%%%%%%%%%%%%%%%%%%%%%%%%%%%%%%%%%%%%%%%%%%%%%%%%%%%%%%%%%%%%%%%%%%%%%%%%%%%%%%
%%%%%%%% table parameters
%%%%%%%%%%%%%%%%%%%%%%%%%%%%%%%%%%%%%%%%%%%%%%%%%%%%%%%%%%%%%%%%%%%%%%%%%%%%%%%%%%%%%
\begin{table*}
\caption{Parameters of the stellar atmosphere models calculated for the $50\,M_\odot$ track with $Z = 0.05\,Z_\odot$ \citep{Marchant2016}.}
\label{table:parameters-pablo-005Zsun}
\centering  
\begin{tabular}{lllcccccc}
\hline \hline \rule[0mm]{0mm}{4.0mm}   %-------------------------------------------------
 Model     & MII\,1 & MII\,2 & \multicolumn{2}{c}{MII\,3} & \multicolumn{2}{c}{MII\,4} & \multicolumn{2}{c}{MII\,5}  \\
           &    &    & \multicolumn{1}{c}{a} & \multicolumn{1}{c}{b} & \multicolumn{1}{c}{a} & \multicolumn{1}{c}{b} & \multicolumn{1}{c}{a} & \multicolumn{1}{c}{b}                \\
\hline   %--------------------------------------------------------
 Spectral type                                        & O3\,V((f*)) & Of/WN & \multicolumn{2}{c}{WN2.5} & \multicolumn{2}{c}{WN2}   & \multicolumn{2}{c}{WN2}  \rule[0mm]{0mm}{4.0mm} \\
 age $[10^{6}\,\mathrm{yr}]$                          & 0.11        & 5.07  & \multicolumn{2}{c}{5.36}  & \multicolumn{2}{c}{5.54}  & \multicolumn{2}{c}{5.83} \\
 $T_*\, [\mathrm{kK}]$                                & 50.4        & 66.5  & \multicolumn{2}{c}{70.3}  & \multicolumn{2}{c}{100.0} & \multicolumn{2}{c}{150.0} \\
 $\log L\, [L_{\odot}]$                               & 5.53        & 6.05  & \multicolumn{2}{c}{6.10}  & \multicolumn{2}{c}{6.15}  & \multicolumn{2}{c}{6.16}  \\
 $M\, [M_{\odot}]$                                    & 50.1        & 46.4  & \multicolumn{2}{c}{44.3}  & \multicolumn{2}{c}{42.2}  & \multicolumn{2}{c}{35.4}  \\
 $R_*\, [R_{\odot}]$                                  & 7.7         & 8.0   & \multicolumn{2}{c}{7.6}   & \multicolumn{2}{c}{4.0}   & \multicolumn{2}{c}{1.8}  \\
 $\log \dot M\, [M_{\odot}/\mathrm{yr}]$              & -6.7        & -5.3 & -5.0  & -5.5  & -4.7 &  -5.2  & -4.7   &  -5.1  \\
 $\varv_{\infty}\, [\mathrm{km/s}]$                   & 2790        & 1600  & \multicolumn{2}{c}{1600}  & \multicolumn{2}{c}{2400}  & \multicolumn{2}{c}{3000}  \\
 $\varv_\mathrm{rot}\, [\mathrm{km/s}]$               & 433         & 335   & \multicolumn{2}{c}{270}   & \multicolumn{2}{c}{285}   & \multicolumn{2}{c}{253}   \\
 $\varv_\mathrm{orbit}\, [\mathrm{km/s}]$             & 513         & 453   & \multicolumn{2}{c}{423}   & \multicolumn{2}{c}{394}   & \multicolumn{2}{c}{322}   \\
 $X_{\element{H}}$                                    & 0.75        & 0.17  & \multicolumn{2}{c}{0.1}   & \multicolumn{2}{c}{0.05}  & \multicolumn{2}{c}{0.0}   \\
 $X_{\element{C}}\, [10^{-3}]$                        & 0.15        & 0.01  & \multicolumn{2}{c}{0.01}  & \multicolumn{2}{c}{0.01}  & \multicolumn{2}{c}{2.12}  \\
 $X_{\element{N}}\, [10^{-3}]$                        & 0.04        & 0.54  & \multicolumn{2}{c}{0.54}  & \multicolumn{2}{c}{0.54}  & \multicolumn{2}{c}{1.22}  \\
 $X_{\element{O}}\, [10^{-3}]$                        & 0.4         & 0.01  & \multicolumn{2}{c}{0.01}  & \multicolumn{2}{c}{0.01}  & \multicolumn{2}{c}{0.06}  \\
\hline   %--------------------------------------------------------
 $M_\mathrm{U}\, [\mathrm{mag}]$                      & -6.2       & -6.6 & -6.5 & -6.6  & -5.8  & -5.7  & -5.0 &  -4.6 \rule[0mm]{0mm}{4.0mm} \\
 $M_\mathrm{B}\, [\mathrm{mag}]$                      & -5.0       & -5.4 & -5.3 & -5.3  & -4.6  & -4.4  & -3.9 &  -3.3 \\
 $M_\mathrm{V}\, [\mathrm{mag}]$                      & -4.7       & -5.0 & -5.0 & -5.0  & -4.3  & -4.1  & -3.7 &  -3.0 \\
 $M_\mathrm{J}\, [\mathrm{mag}]$\tablefootmark{a}     & -3.9       & -4.3 & -4.4 & -4.2  & -4.0  & -3.3  & -3.8 &  -2.4 \\
 $M_\mathrm{H}\, [\mathrm{mag}]$\tablefootmark{a}     & -3.8       & -4.3 & -4.4 & -4.2  & -4.2  & -3.2  & -4.1 &  -2.4 \\
 $M_\mathrm{K}\, [\mathrm{mag}]$\tablefootmark{a}     & -3.6       & -4.3 & -4.5 & -4.0  & -4.5  & -3.1  & -4.5 &  -2.5 \\
 $\log Q_\mathrm{\element{H}}\, [\mathrm{s^{-1}}]$    & 49.3       & 49.9 & 50.0 &  50.0 &  50.0 & 50.0  & 49.9 & 49.9  \\
 $T_{\mathrm{Zanstra}, \element{H}}\, [\mathrm{kK}]$  & 53.7       & 72.5 & 76.6 &  81.1 & 95.7  &  113.5 & 105.9 & 149.7  \\
 $\log Q_\mathrm{\ion{He}{i}}\, [\mathrm{s^{-1}}]$    & 48.9       & 49.6 & 49.7 &  49.7 & 49.8  &  49.8  & 49.8 & 49.8  \\
 $\log Q_\mathrm{\ion{He}{ii}}\, [\mathrm{s^{-1}}]$   & 45.9       & 47.1 & 47.5 & 47.2  & 48.7  &  48.4  & 49.3 & 49.3  \\
 $T_{\mathrm{Zanstra}, \element{He}}\, [\mathrm{kK}]$ & 48.0       & 58.9 & 63.4 & 60.8  & 97.4  &   92.5 & 135.9 & 159.3 \\
\hline   %--------------------------------------------------------
 $P$ [d]                                              & 0.9        & 1.2   & \multicolumn{2}{c}{1.4}   & \multicolumn{2}{c}{1.7}   & \multicolumn{2}{c}{5.1}  \rule[0mm]{0mm}{4.0mm} \\
 orbital separation $[R_{\odot}]$                     & 18.1        & 21.5  & \multicolumn{2}{c}{23.6}  & \multicolumn{2}{c}{26.0}  & \multicolumn{2}{c}{32.6}  \\
\hline  \rule[0mm]{0mm}{4.0mm}  
%---------------------------------------------------------------------------------------------------------
Model atom set\tablefootmark{b} & I & II & \multicolumn{2}{c}{II} & \multicolumn{2}{c}{II} & \multicolumn{2}{c}{III} \\
\hline   %-----------------------------------------------------------------------
\multicolumn{9}{c}{Wind accretion rates and accretion X-ray luminosity\tablefootmark{c}} \rule[0mm]{0mm}{4.0mm}\\
\hline  %-----------------------------------------------------------------------
\rule[0mm]{0mm}{4.0mm}  
$\log S_{\rm accr}\, [M_{\odot}/\mathrm{yr}]$ &  &  & -6.9 & -7.4 & -7.4 & -7.9 
& -8.0  & -8.4 \\
$s_{\rm accr}$\tablefootmark{d}               &  &  &  1.3  &  0.4  & 0.4 & 0.1 & 0.1  &   
0.04 \\
$L_{\rm X}$  [erg\,s$^{-1}$]\tablefootmark{f}                  &  &  & $7\times 10^{38}$ & 
$2\times 10^{38}$ & $2\times 10^{38}$  & $7\times 10^{37}$  & $6\times 10^{37}$  & $2\times 10^{37}$ \\   
\hline  \rule[0mm]{0mm}{4.0mm}  %---------------------------------------------------------------------------------------------------------
\end{tabular}
\tablefoot{The WN models were calculated with two different mass-loss rates. The ``a'' models use the same mass-loss rate as given in the tracks, assuming a metallicity scaling of $Z^{0.85}$ as is observed 
for O-type stars. The mass-loss rate of the ``b'' models was scaled according to the mass-loss metallicity relation presented by \citet{Hainich2015}. 
\tablefoottext{a}{Monochromatic magnitudes at $1.26\,\mu\mathrm{m}$, $1.60\,\mu\mathrm{m}$, and $2.22\,\mu\mathrm{m}$, respectively.}
\tablefoottext{b}{Model atom set used in the corresponding stellar atmosphere calculations (see Table\,\ref{table:model_atoms}).}
\tablefoottext{c}{The evolution model predicts BH formation only if the secondary is already in its final evolutionary stage (see text for details). Accretion onto a  $36\,M_\odot$ BH  is assumed.}
\tablefoottext{d}{Accretion rate normalized to the Eddington accretion rate: $s_{\rm accr}\equiv 
\frac{S_{\rm accr}}{S_{\rm Edd}} \approx 1.5\times 10^{7}\cdot
 \frac{M_{\rm BH} \dot{M}}{v_{8}^4 a_{\rm BH}^2}$ (see Sect.\,\ref{subsect:hmxb} for details).}
\tablefoottext{f}{Upper limit since an accretion efficiency of 0.1 is assumed.}
}
\end{table*}
%%%%%%%%%%%%%%%%%%%%%%%%%%%%%%%%%%%%%%%%%%%%%%%%%%%%%%%%%%%%%%%%%%%%%%%%%%%%%%%%%%%%%

%%%%%%%%%%%%%%%%%%%%%%%%%%%%%%%%%%%%%%%%%%%%%%%%%%%%%%%%%%%%%%%%%%%%%%%%%%%%%%%%%%%%%
%%%%%%%% table parameters
%%%%%%%%%%%%%%%%%%%%%%%%%%%%%%%%%%%%%%%%%%%%%%%%%%%%%%%%%%%%%%%%%%%%%%%%%%%%%%%%%%%%%
\begin{sidewaystable*}
\caption{Parameters of the stellar atmosphere models calculated for the BPASS stellar evolution tracks.}
\label{table:parameters-JJ}
\small
\centering  
\begin{tabular}{lllllllllllllll}
\hline \hline \rule[0mm]{0mm}{4.0mm}   %--------------------------------------------------------------------------------------------------
 Model  & \multicolumn{2}{c}{E\,1} & \multicolumn{2}{c}{E\,2} & \multicolumn{2}{c}{E\,3} & \multicolumn{2}{c}{E\,4} & E\,5  & E\,6 & E\,7 & E\,8 & E\,9 & E\,10 \\
           & \multicolumn{1}{c}{A} & \multicolumn{1}{c}{B} & \multicolumn{1}{c}{A} & \multicolumn{1}{c}{B} & \multicolumn{1}{c}{A} & \multicolumn{1}{c}{B} & \multicolumn{1}{c}{A} & \multicolumn{1}{c}{B} &  B  &  B  &  B  &  B  &  B  &  B  \\
\hline  %---------------------------------------------------------------------------------------------------------
 Spectral type                                    & O2-3\,V & O4\,V & O6\,III & O4\,V & WN7-8? & O5\,V & WN9? & O5\,V  & O3-2\,V\,((f*))z  & O3-2\,V\,((f*))z  & O2\,V\,((f*)) & WN2.5 & WN2 & WN2/WC4   \rule[0mm]{0mm}{4.0mm}   \\
 age $[10^{6}\,\mathrm{yr}]$                      & \multicolumn{2}{c}{0.01} & \multicolumn{2}{c}{3.32} & \multicolumn{2}{c}{3.46} & \multicolumn{2}{c}{3.66} & 3.75 & 5.05 & 6.59 & 8.29 & 8.98 & 9.32 \\
 $T_*\, [\mathrm{kK}]$                            & 59.6   & 48.6   & 36.6  & 49.6 & 41.1 & 45.6 & 33.6 & 42.8 & 54.1  & 57.0  & 62.1 & 72.0 & 100.0 & 200.0 \\
 $\log L\, [L_{\odot}]$                           & 5.93   & 5.12   & 6.22  & 5.29 & 6.25 & 5.29 & 6.27 & 5.31 & 5.55  & 5.65  & 5.8  & 6.0  & 6.09  & 6.22   \\
 $M\, [M_{\odot}]$                                & 80.0   & 32.0   & 79.5  & 32.0 & 54.1 & 41.7 & 40.7 & 50.3 & 50.0  & 50.0  & 50.0 & 45.3 & 38.3  & 34.5   \\
 $R_*\, [R_{\odot}]$                              & 8.7    & 5.1    & 32.0  & 5.9  & 26.3  & 7.1 & 40.1 & 8.2  & 6.8   & 6.9   & 6.9  & 6.4  & 3.7   & 1.1 \\
 $\log \dot M\, [M_{\odot}/\mathrm{yr}]$          & -7.8   & -8.1   & -6.1  & -7.8 & -4.7 & -7.9 & -4.4 & -7.9 & -7.8  & -7.9  & -9.0 & -5.2 & -4.8  & -5.4   \\
 $\varv_{\infty}\, [\mathrm{km/s}]$               & 2210   & 1827   & 1150  & 1692 & 1000 & 1766 & 1000 & 1807 & 1979  & 1966  & 1600 & 1600 & 2400  & 3000  \\
 $\varv_\mathrm{orbit}\, [\mathrm{km/s}]$         & 159    & 397    & 158   & 394  & 237  & 308  & 287  & 233  & -     & -     & -    & -    & -     & -      \\
 $X_{\element{H}}$                                & \multicolumn{2}{c}{0.75} & \multicolumn{2}{c}{0.75} & 0.35 & 0.75 & 0.15 & 0.75 & 0.75 & 0.64 & 0.47 & 0.18 & 0.002 & 0.0    \\
 $X_{\element{C}}\, [10^{-4}]$                & \multicolumn{2}{c}{0.17}  & \multicolumn{2}{c}{0.17} & 0.01 & 0.17 & 0.01 & 0.17 & 0.01 & 0.01 & 0.01 & 0.01 & 0.02  & 899.0 \\
 $X_{\element{N}}\, [10^{-4}]$                & \multicolumn{2}{c}{0.05}  & \multicolumn{2}{c}{0.05} & 0.63 & 0.05 & 0.66 & 0.05 & 0.58 & 0.65 & 0.65 & 0.65 & 0.65  & 1.52   \\
 $X_{\element{O}}\, [10^{-4}]$                & \multicolumn{2}{c}{0.48}  & \multicolumn{2}{c}{0.48} & 0.04 & 0.48 & 0.01 & 0.48 & 0.10 & 0.01 & 0.01 & 0.01 & 0.006 & 0.41  \\
\hline  %---------------------------------------------------------------------------------------------------------
 $M_\mathrm{U}\, [\mathrm{mag}]$                  & -6.6 & -5.3 & -8.7 & -5.6 & -8.4 & -5.9 & -9.1 & -6.1 & -6.0 & -6.1 & -6.2 & -6.3 & -5.6 & -3.9 \rule[0mm]{0mm}{4.0mm} \\
 $M_\mathrm{B}\, [\mathrm{mag}]$                  & -5.3 & -4.1 & -7.6 & -4.4 & -7.2 & -4.7 & -7.9 & -4.9 & -4.8 & -4.9 & -4.9 & -5.0 & -4.5 & -2.6 \\
 $M_\mathrm{V}\, [\mathrm{mag}]$                  & -5.0 & -3.7 & -7.3 & -4.1 & -6.9 & -4.4 & -7.7 & -4.6 & -4.5 & -4.5 & -4.6 & -4.7 & -4.1 & -2.3 \\
 $M_\mathrm{J}\, [\mathrm{mag}]$\tablefootmark{a} & -4.2 & -3.0 & -6.6 & -3.3 & -6.3 & -3.6 & -7.2 & -3.9 & -3.7 & -3.8 & -3.8 & -4.1 & -3.7 & -2.0 \\
 $M_\mathrm{H}\, [\mathrm{mag}]$\tablefootmark{a} & -4.2 & -2.9 & -6.6 & -3.3 & -6.2 & -3.5 & -7.2 & -3.8 & -3.6 & -3.7 & -3.7 & -4.1 & -3.9 & -2.1 \\
 $M_\mathrm{K}\, [\mathrm{mag}]$\tablefootmark{a} & -4.0 & -2.7 & -6.5 & -3.1 & -6.2 & -3.4 & -7.2 & -3.7 & -3.5 & -3.5 & -3.6 & -4.1 & -4.2 & -2.3 \\
 $\log Q_\mathrm{\element{H}}\, [\mathrm{s^{-1}}]$    & 49.7  & 48.9 & 49.9 & 49.1 & 50.0 & 49.0 & 50.0 & 49.0 & 49.4 & 49.5 & 49.6 & 49.9 & 49.9 & 49.8 \\
 $T_{\mathrm{Zanstra}, \element{H}}\, [\mathrm{kK}]$  & 64.6  & 50.5 & 39.4 & 52.0 & 46.0 & 46.9 & 36.9 & 43.5 & 57.5 & 61.3 & 68.1 & 78.4 & 98.7 & 186.8 \\
 $\log Q_\mathrm{\ion{He}{i}}\, [\mathrm{s^{-1}}]$    & 49.4  & 48.5 & 49.2 & 48.7 & 49.5 & 48.6 & 49.1 & 48.5 & 49.0 & 49.1 & 49.3 & 49.6 & 49.8 & 49.7 \\
 $\log Q_\mathrm{\ion{He}{ii}}\, [\mathrm{s^{-1}}]$   & 45.7  & -    & 45.2 & -    & -    & -    & 40.4 &  -   & -    & 45.0 & -    & 46.8 & 48.6 & 49.5 \\
 $T_{\mathrm{Zanstra}, \element{He}}\, [\mathrm{kK}]$ & 45.0  & -    & 36.5 & -    & -    & -    & 21.6 &  -   & -    & 41.6 & -    & 56.3 & 96.5 & 208.2 \\
\hline  %---------------------------------------------------------------------------------------------------------
 $P$ [d]                                                   & \multicolumn{2}{c}{6.3}  & \multicolumn{2}{c}{6.4}    & \multicolumn{2}{c}{5.7}   & \multicolumn{2}{c}{6.2}   & -  & - & - & - & - & - \rule[0mm]{0mm}{4.0mm} \\
 orbital separation $[R_{\odot}]$               & \multicolumn{2}{c}{69.3} & \multicolumn{2}{c}{69.6}  & \multicolumn{2}{c}{61.4} & \multicolumn{2}{c}{64.0} & -  & - & - & - & - & - \\
\hline  \rule[0mm]{0mm}{3.0mm}  %---------------------------------------------------------------------------------------------------------
Model atom set\tablefootmark{b} & \multicolumn{2}{c}{I} & \multicolumn{2}{c}{I} & \multicolumn{2}{c}{I} & \multicolumn{2}{c}{I} & I & I & I & II & II & III \rule[0mm]{0mm}{3.5mm} \\
\hline   %---------------------------------------------------------------------------------------------------------
\multicolumn{15}{c}{Wind accretion rates and accretion X-ray luminosity\tablefootmark{c}} \rule[0mm]{0mm}{3.5mm}  \\
\hline  %-----------------------------------------------------------------------
\rule[0mm]{0mm}{3.5mm}  
$\log S_{\rm accr}\, [M_{\odot}/\mathrm{yr}]$ &  &  &   &   &   & 
&    &      &  & -11  & -12   & -8   &  -8  &  -9   \\
$s_{\rm accr}$\tablefootmark{d} &  &  &  &  &  &  &  &  & & 0.0001 & 0.00002 & 0.1 & 
0.05  & 0.006 \\ 
$L_{\rm X}$  [erg\,s$^{-1}$]\tablefootmark{f} &  &  &  & &  &  &  & & 
& $10^{35}$ & $10^{34}$ & $10^{38}$ & $10^{37}$ &  $10^{36}$ 
\\       
\hline  
\end{tabular}
\tablefoot{
The capital letters A and B denote the primary and the secondary in the binary systems.
\tablefoottext{a}{Monochromatic magnitudes at $1.26\,\mu\mathrm{m}$, $1.60\,\mu\mathrm{m}$, and $2.22\,\mu\mathrm{m}$, respectively.}
\tablefoottext{b}{Model atom set used in the corresponding stellar atmosphere calculations (see Table\,\ref{table:model_atoms})}
\tablefoottext{c}{The evolution model predict BH formation when the secondary has the spectral type O3-2V (see text for details). Accretion onto a  $36\,M_\odot$ BH is assumed. The assumed orbital separation is $64\,R_\ast$.}
\tablefoottext{d}{Accretion rate normalized to the Eddington accretion rate: $s_{\rm accr}\equiv 
\frac{S_{\rm accr}}{S_{\rm Edd}} \approx 1.5\times 10^{7}\cdot
 \frac{M_{\rm BH} \dot{M}}{v_{8}^4 a_{\rm BH}^2}$ (see Sect.\,\ref{subsect:hmxb} for details).}
\tablefoottext{f}{Upper limit since an accretion efficiency of 0.1 is assumed.}
}
\end{sidewaystable*}
%%%%%%%%%%%%%%%%%%%%%%%%%%%%%%%%%%%%%%%%%%%%%%%%%%%%%%%%%%%%%%%%%%%%%%%%%%%%%%%%%%%%%

%%%%%%%%%%%%%%%%%%%%%%%%%%%%%%%%%%%%%%%%%%%%%%%%%%%%%%%%%%%%%%%%%%%%
%%%%%%%%%%%%%%%%%%%%%%%%%%%%%%%%%%%%%%%%%%%%%%%%%%%%%%%%%%%%%%%%%%%%

\clearpage

\section{Synthetic spectra}
\label{sect:spectra}

%---------------------------------------------------------------
\begin{figure*}[htbp]
\centering
\includegraphics[width=\hsize]{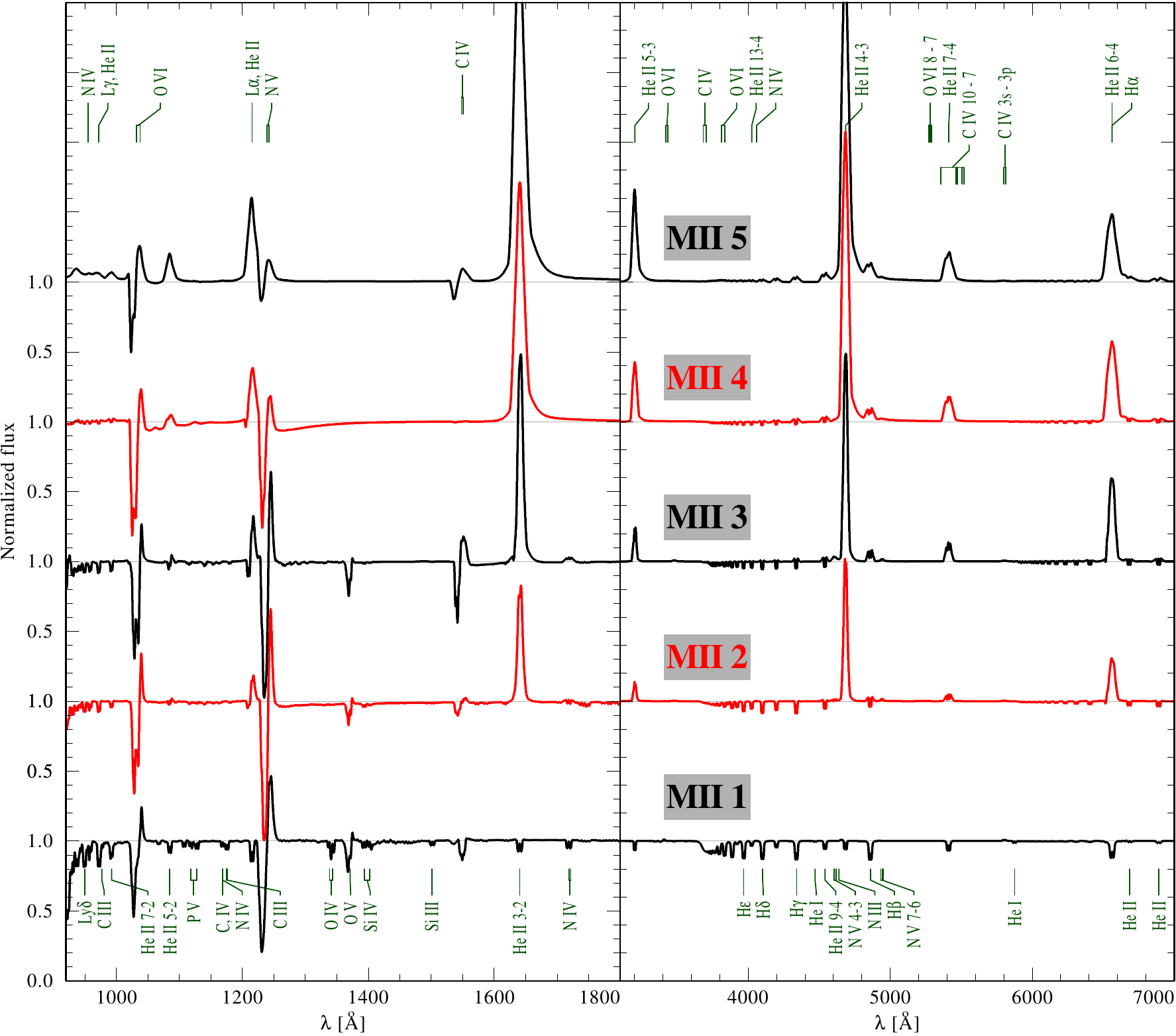}
\caption{Same as Fig.\,\ref{fig:spec_comb_01} but for the $50\,M_\odot$ track provided by 
\citet{Marchant2016} for a metallicity of $0.05\,Z_\odot$.
} 
\label{fig:spec_comb_005}
\end{figure*}
%---------------------------------------------------------------

%---------------------------------------------------------------
\begin{figure*}[htbp]
\centering
\includegraphics[width=0.8\hsize]{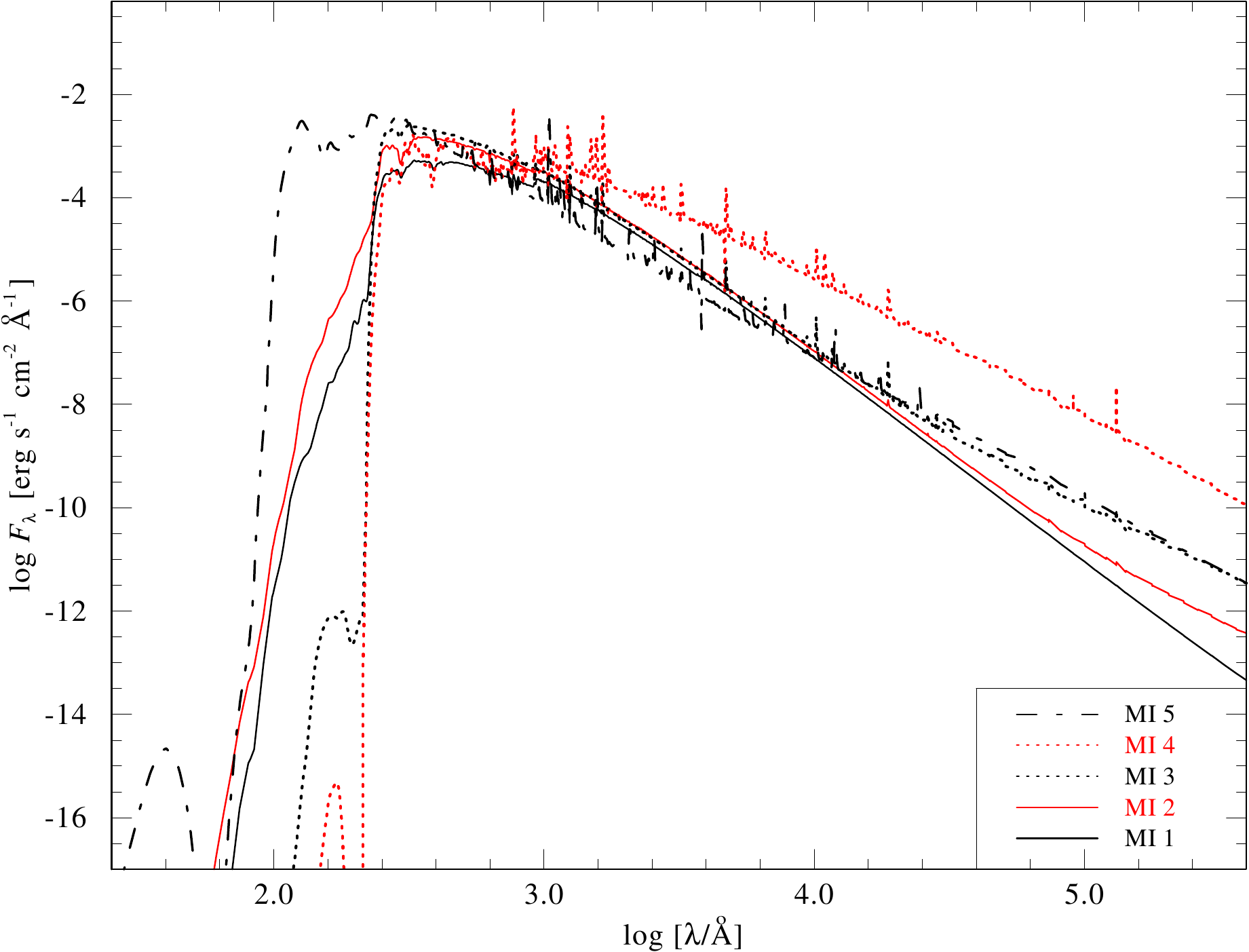}
\caption{Sequence of model SEDs calculated at 
characteristic points of the stellar evolution track provided by 
\citet{Marchant2016} for a metallicity of $0.1\,Z_\odot$. The model SEDs are convolved
with a Gaussian profile of 8\,\AA\ to increase lucidity.
} 
\label{fig:sed_01}
\end{figure*}
%---------------------------------------------------------------

%---------------------------------------------------------------
\begin{figure*}[htbp]
\centering
\includegraphics[width=0.8\hsize]{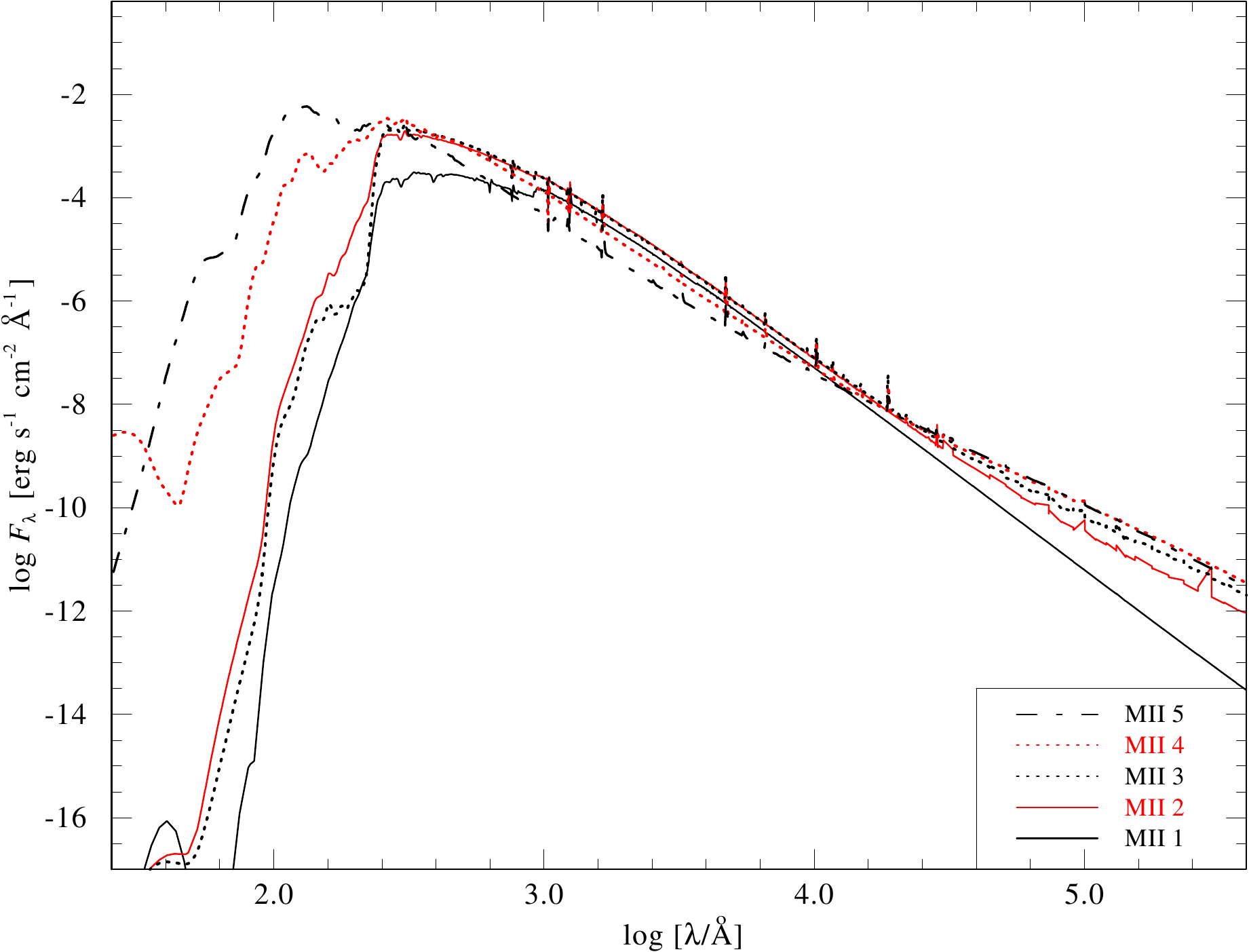}
\caption{Same as Fig.\,\ref{fig:sed_01} but for a metallicity of $0.05\,Z_\odot$.
} 
\label{fig:sed_005}
\end{figure*}
%---------------------------------------------------------------

%---------------------------------------------------------------
\begin{figure*}[htbp]
\centering
\includegraphics[width=0.8\hsize]{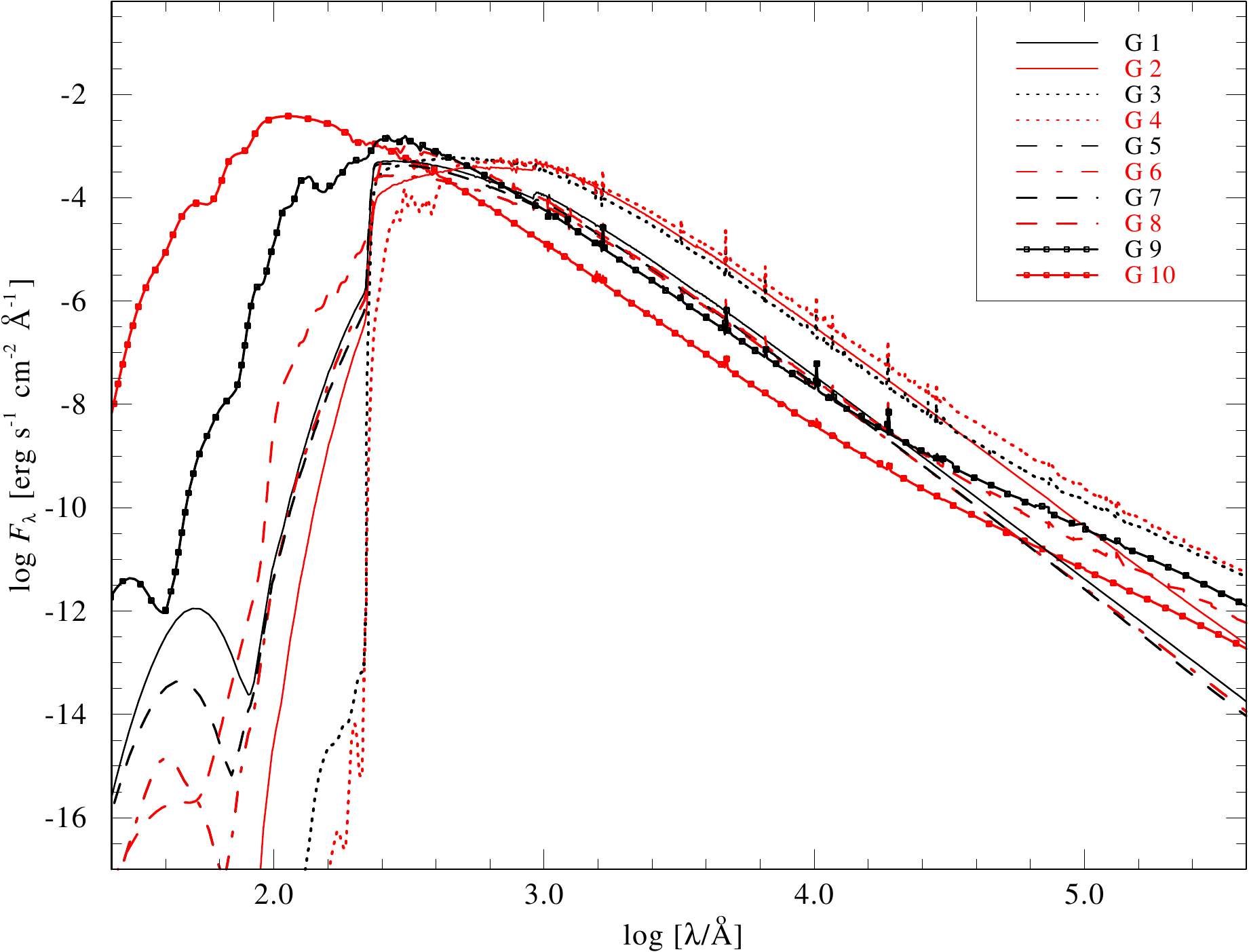}
\caption{Same as Fig.\,\ref{fig:sed_005} but for the BPASS evolution models \citep{Eldridge2016}.
} 
\label{fig:sed_jj}
\end{figure*}
%---------------------------------------------------------------

%---------------------------------------------------------------
\clearpage
\begin{figure*}
  \centering  
  \includegraphics[width=0.9\textwidth,page=1]{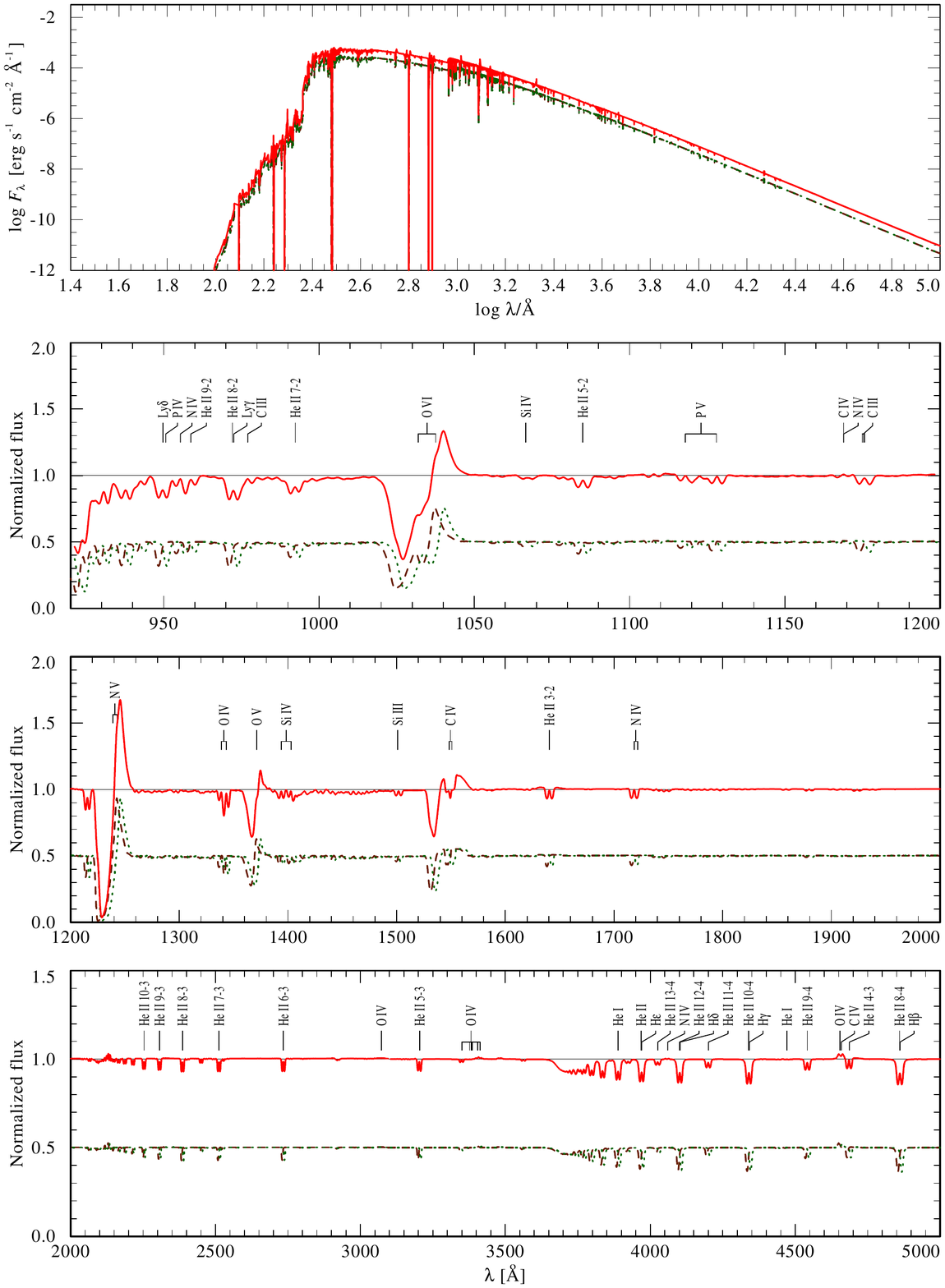} 
  \caption{Synthetic binary spectrum (red straight line) of model MI\,1 for the $60\,M_\odot$ track calculated by 
  \citet{Marchant2016} for a metallicity of $0.1\,Z_\odot$ (see Fig.\,\ref{fig:hrd_pablo} and Table\,\ref{table:parameters-pablo-01Zsun}). The composite spectrum is the sum of the primary spectrum (brown dashed line) and the secondary spectrum (green dotted line). 
  The continuum level is indicated by a thin black line.}
  \label{fig:ms_01Zsun}
\end{figure*}
\clearpage
\setcounter{figure}{\value{figure}-1}
\begin{figure*}
  \centering
  \includegraphics[width=0.92\textwidth,page=2]{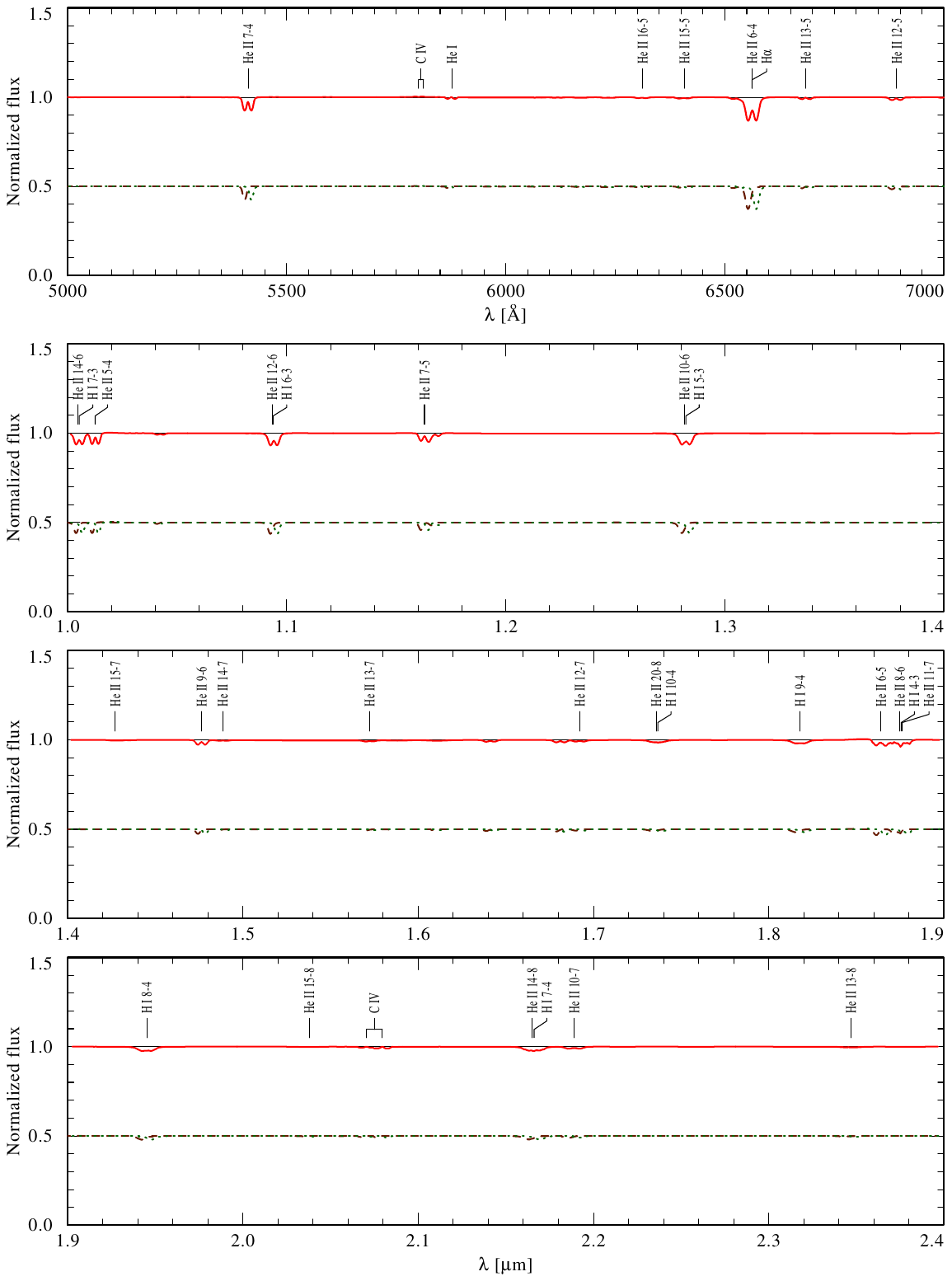}
  \caption{continued.}
\end{figure*}

\clearpage
\begin{figure*}
  \centering  
  \includegraphics[width=0.92\textwidth,page=1]{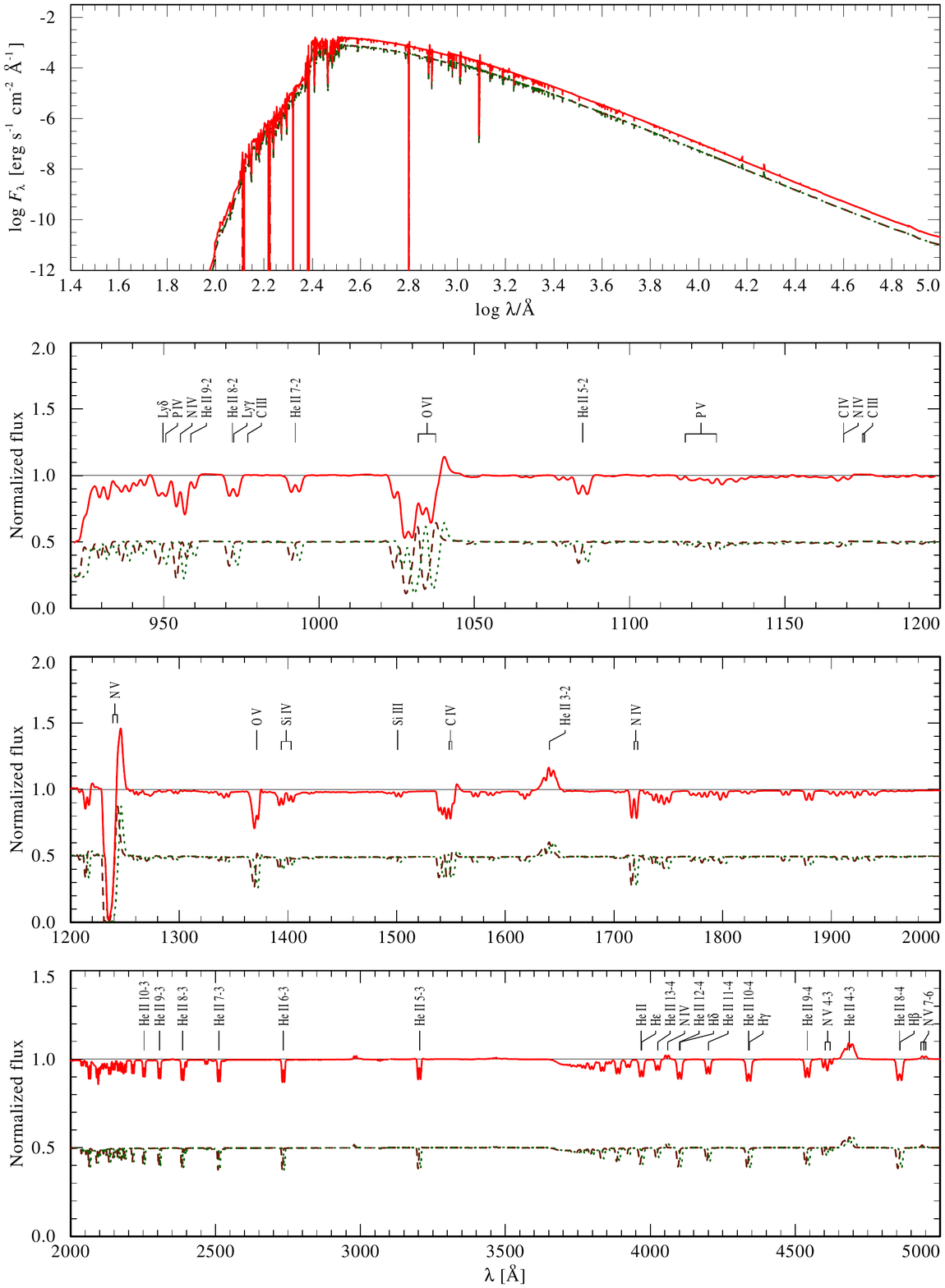} 
  \caption{Same as Fig.\,\ref{fig:ms_01Zsun} but for model MI\,2.}
  \label{fig:pmt_01Zsun}
\end{figure*}
\clearpage
\setcounter{figure}{\value{figure}-1}
\begin{figure*}
  \centering
  \includegraphics[width=0.92\textwidth,page=2]{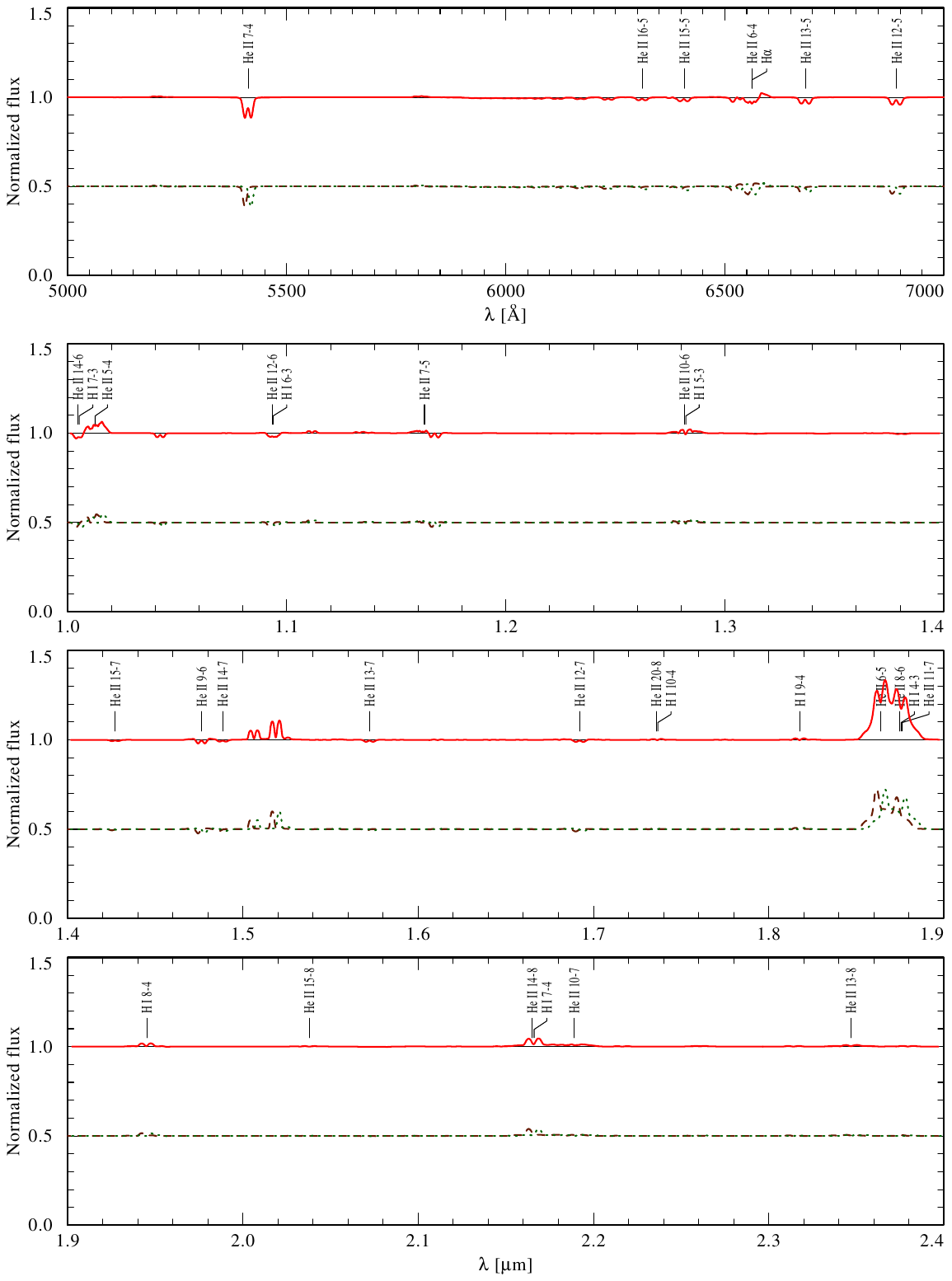}
  \caption{continued.}
\end{figure*}

\clearpage
\begin{figure*}
  \centering  
  \includegraphics[width=0.92\textwidth,page=1]{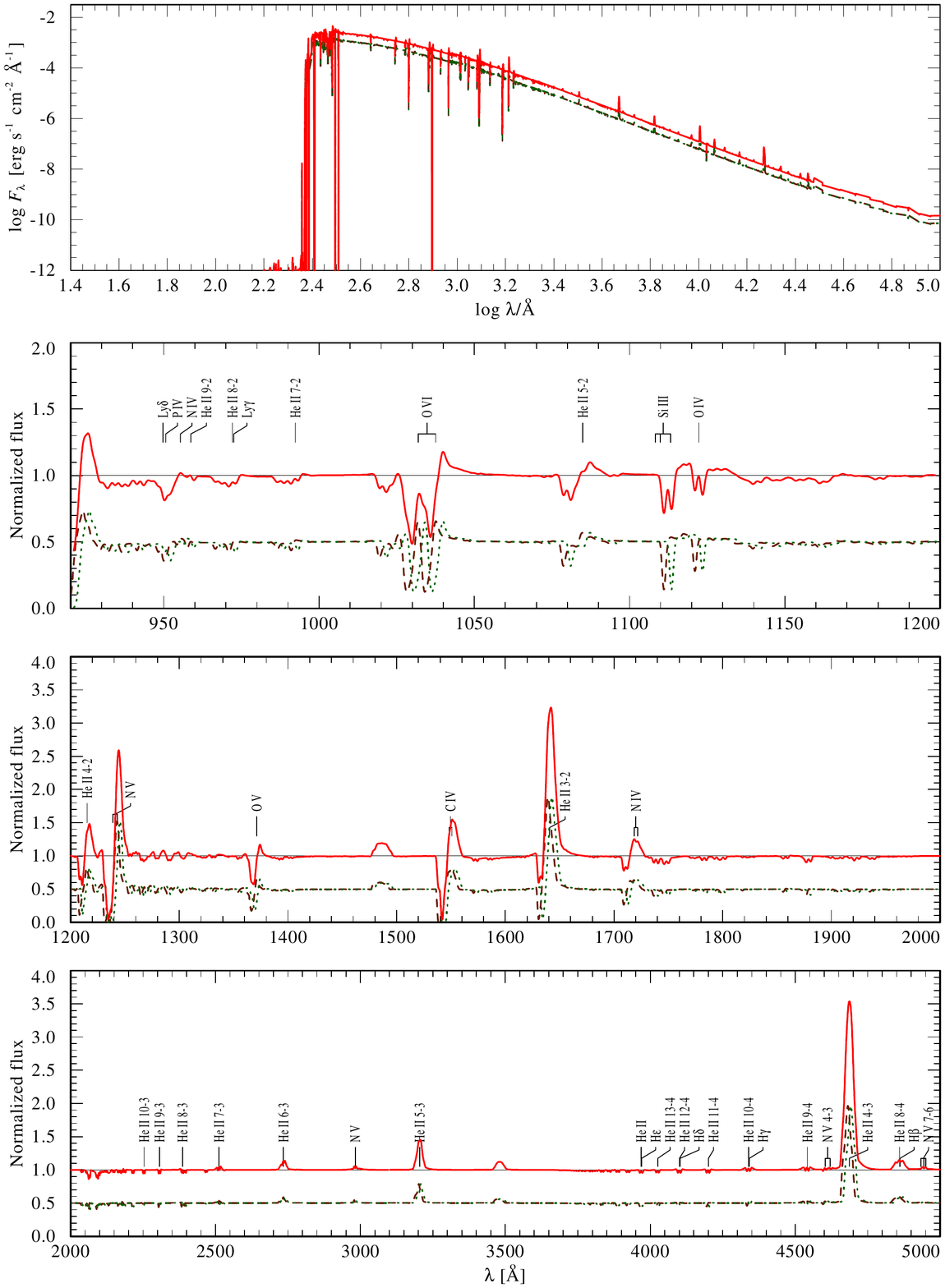} 
  \caption{Same as Fig.\,\ref{fig:ms_01Zsun} but for model MI\,3.}
  \label{fig:tp_01Zsun}
\end{figure*}
\clearpage
\setcounter{figure}{\value{figure}-1}
\begin{figure*}
  \centering
  \includegraphics[width=0.92\textwidth,page=2]{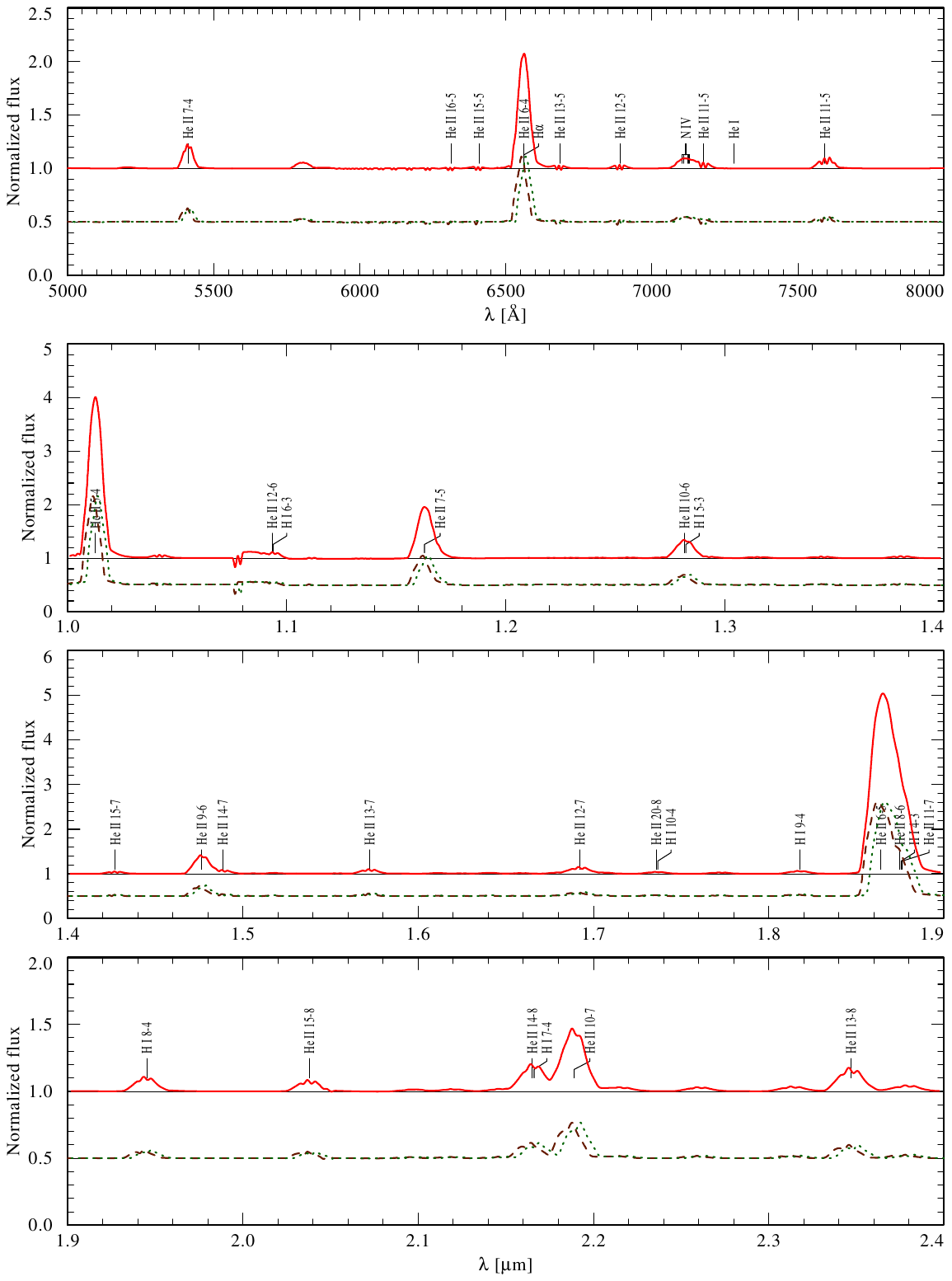}
  \caption{continued.}
\end{figure*}

\clearpage
\begin{figure*}
  \centering  
  \includegraphics[width=0.92\textwidth,page=1]{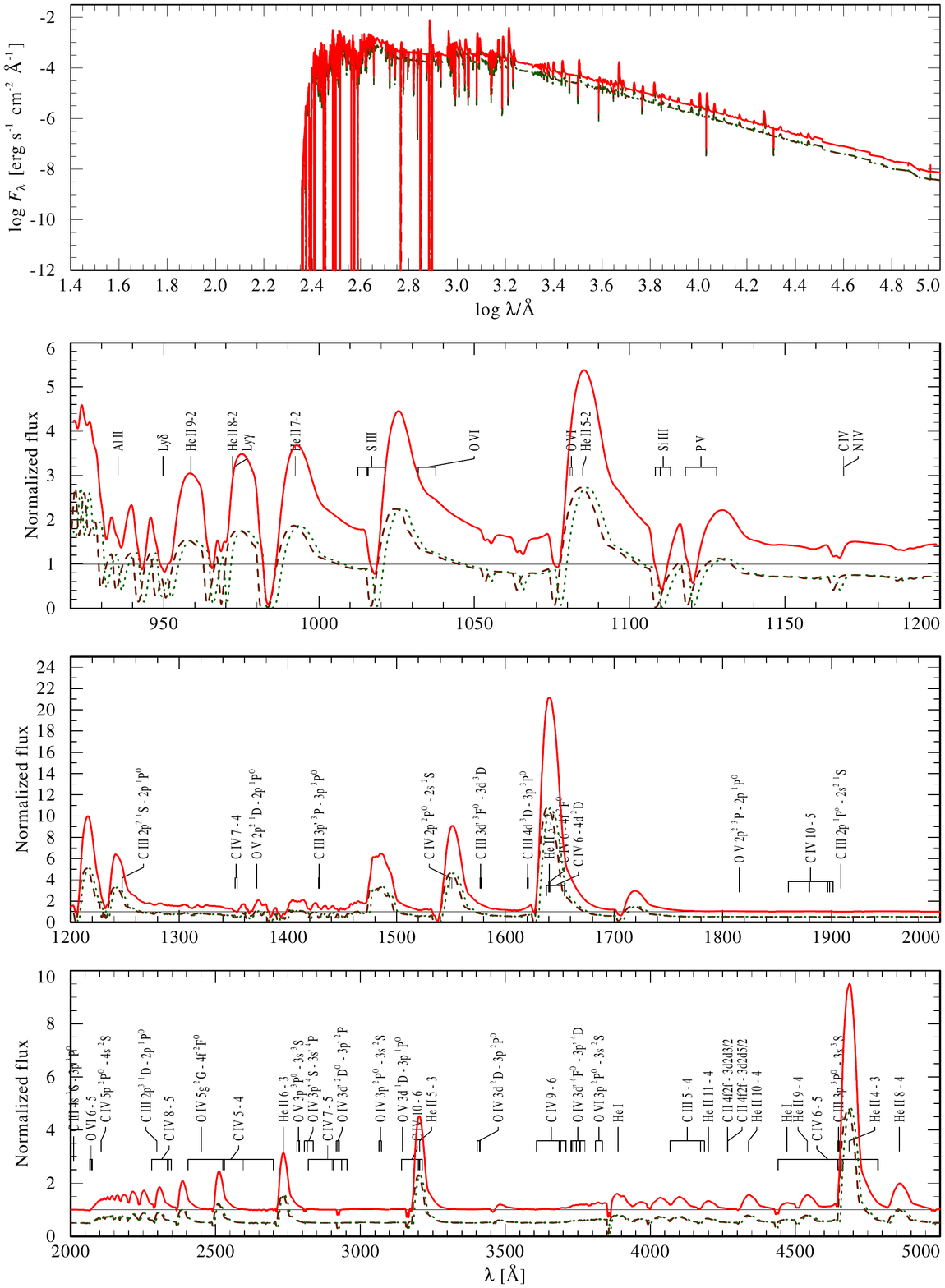} 
  \caption{Same as Fig.\,\ref{fig:ms_01Zsun} but for model MI\,4.}
  \label{fig:100kK_01Zsun}
\end{figure*}
\clearpage
\setcounter{figure}{\value{figure}-1}
\begin{figure*}
  \centering
  \includegraphics[width=0.92\textwidth,page=2]{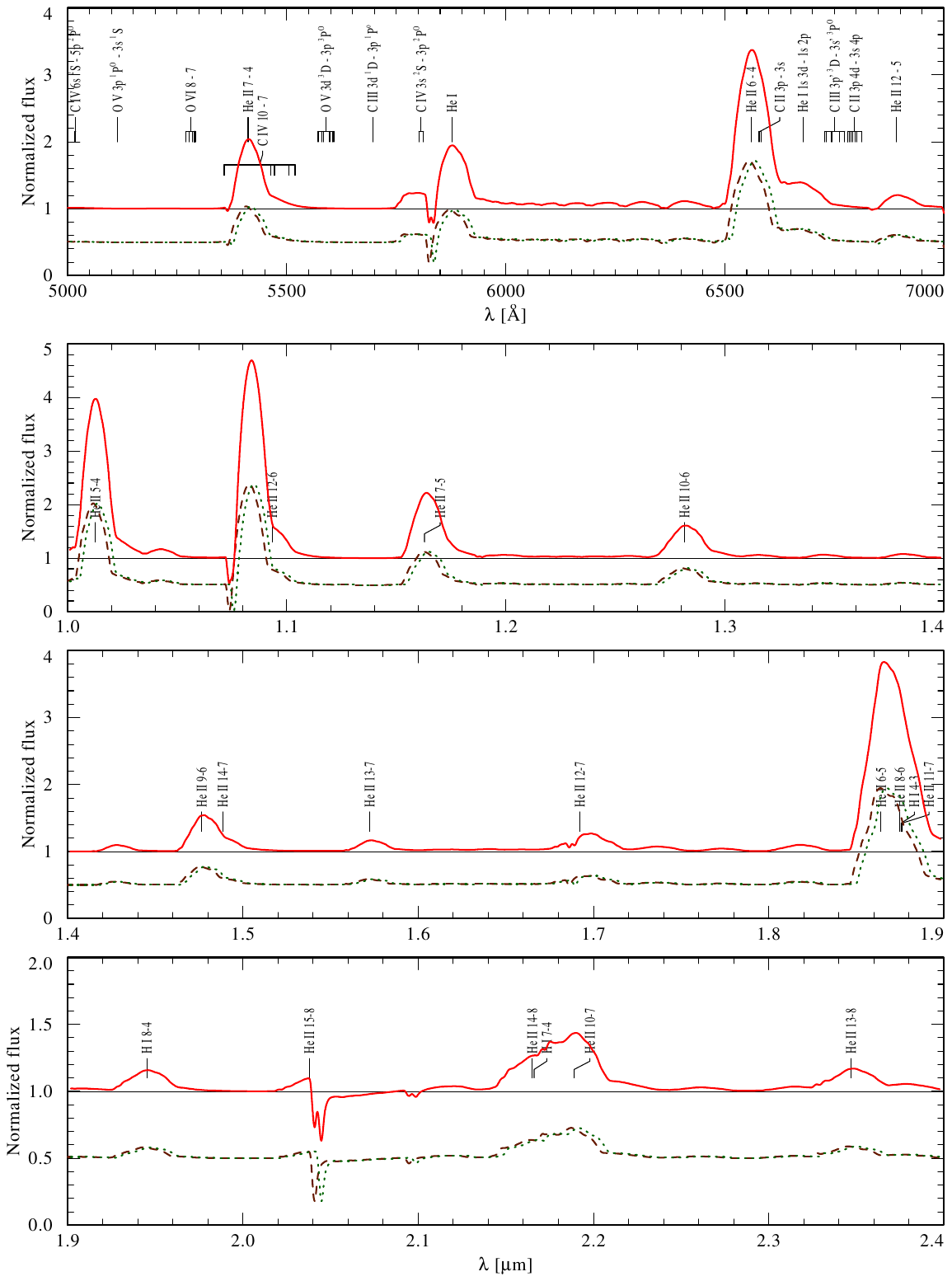}
  \caption{continued.}
\end{figure*}

\clearpage
\begin{figure*}
  \centering  
  \includegraphics[width=0.92\textwidth,page=1]{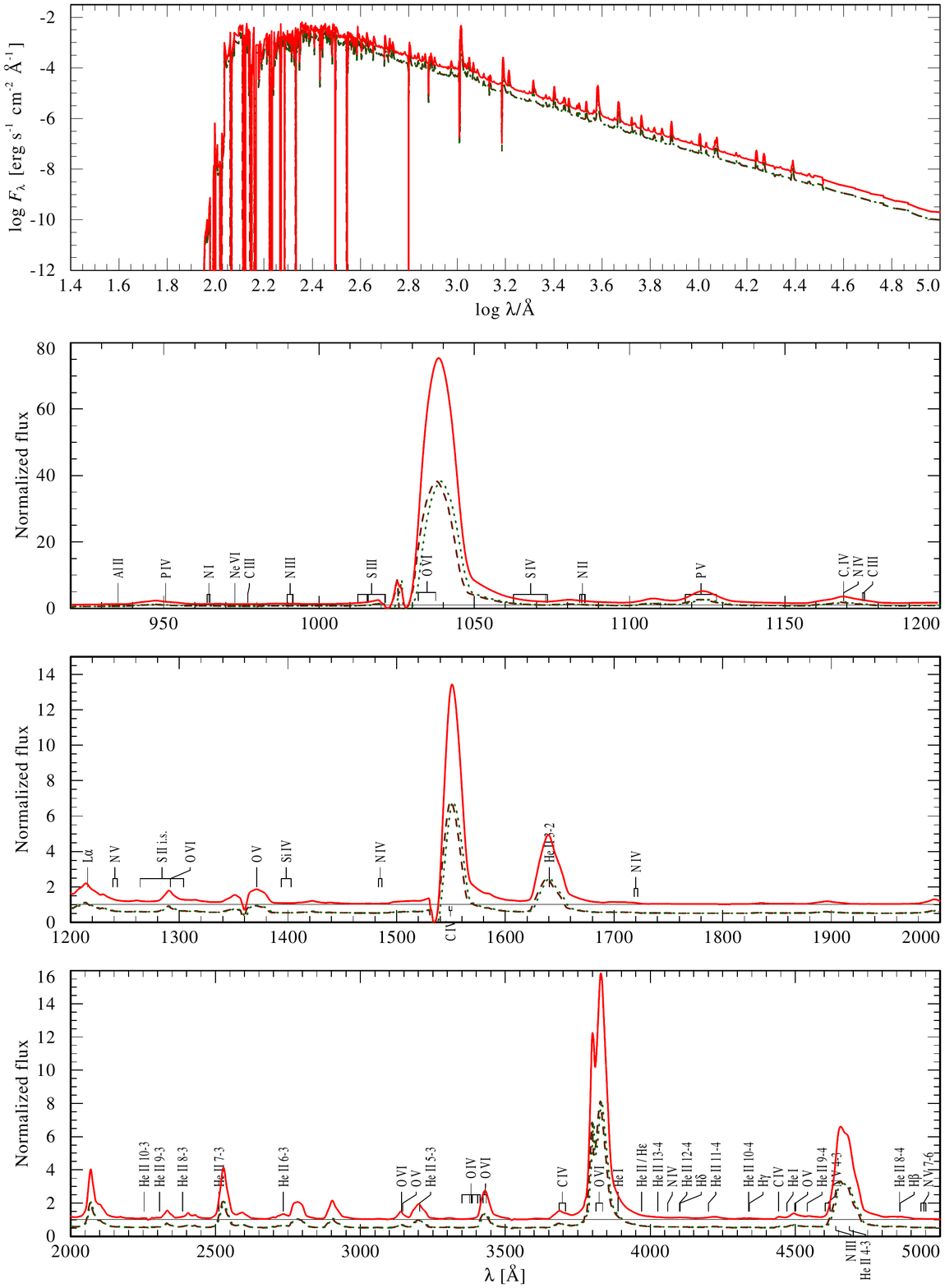} 
  \caption{Same as Fig.\,\ref{fig:ms_01Zsun} but for model MI\,5.}
  \label{fig:150kK_01Zsun}
\end{figure*}
\clearpage
\setcounter{figure}{\value{figure}-1}
\begin{figure*}
  \centering
  \includegraphics[width=0.92\textwidth,page=2]{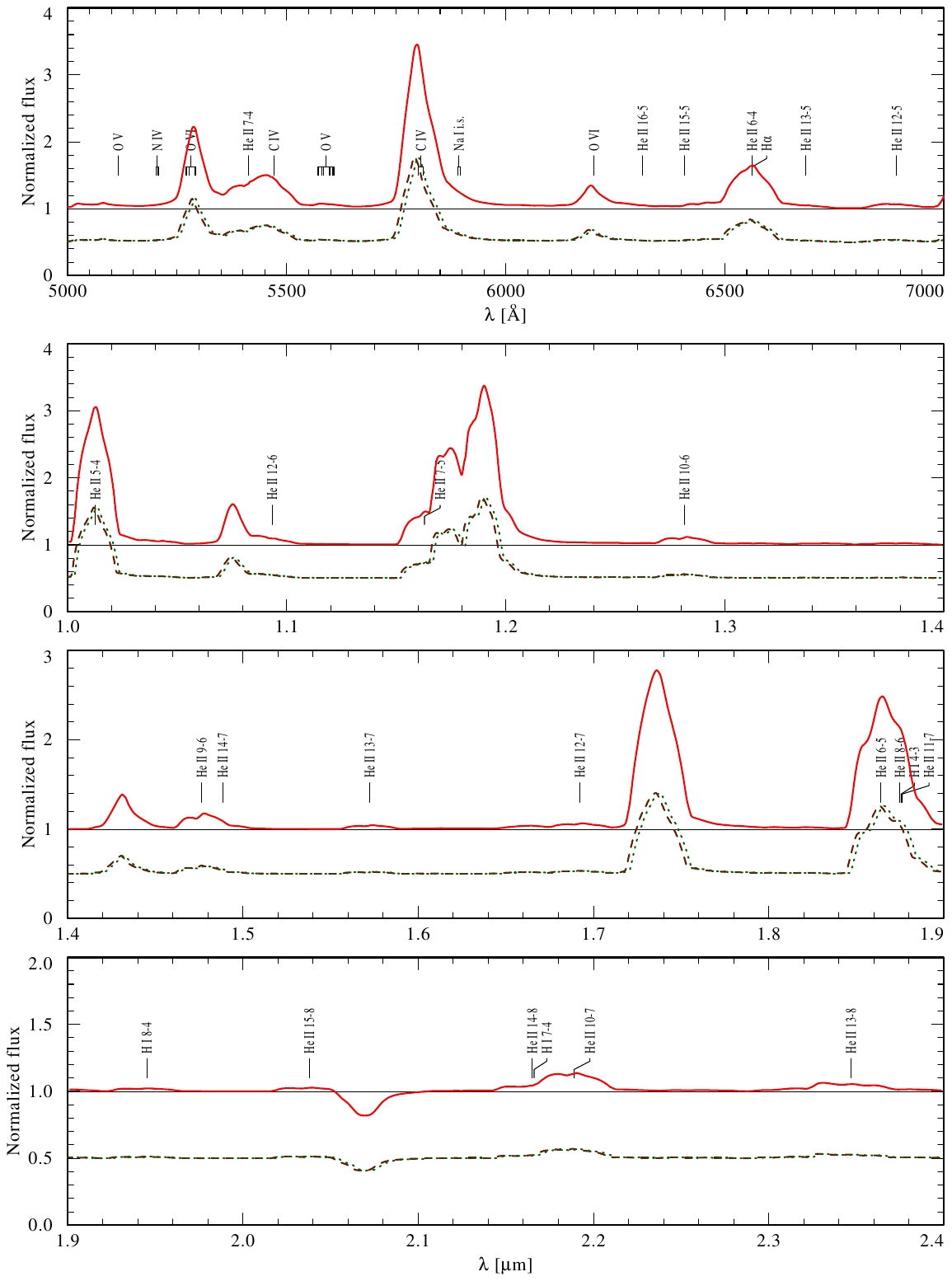}
  \caption{continued.}
\end{figure*}
%---------------------------------------------------------------

%---------------------------------------------------------------
\clearpage
\begin{figure*}
  \centering  
  \includegraphics[width=0.9\textwidth,page=1]{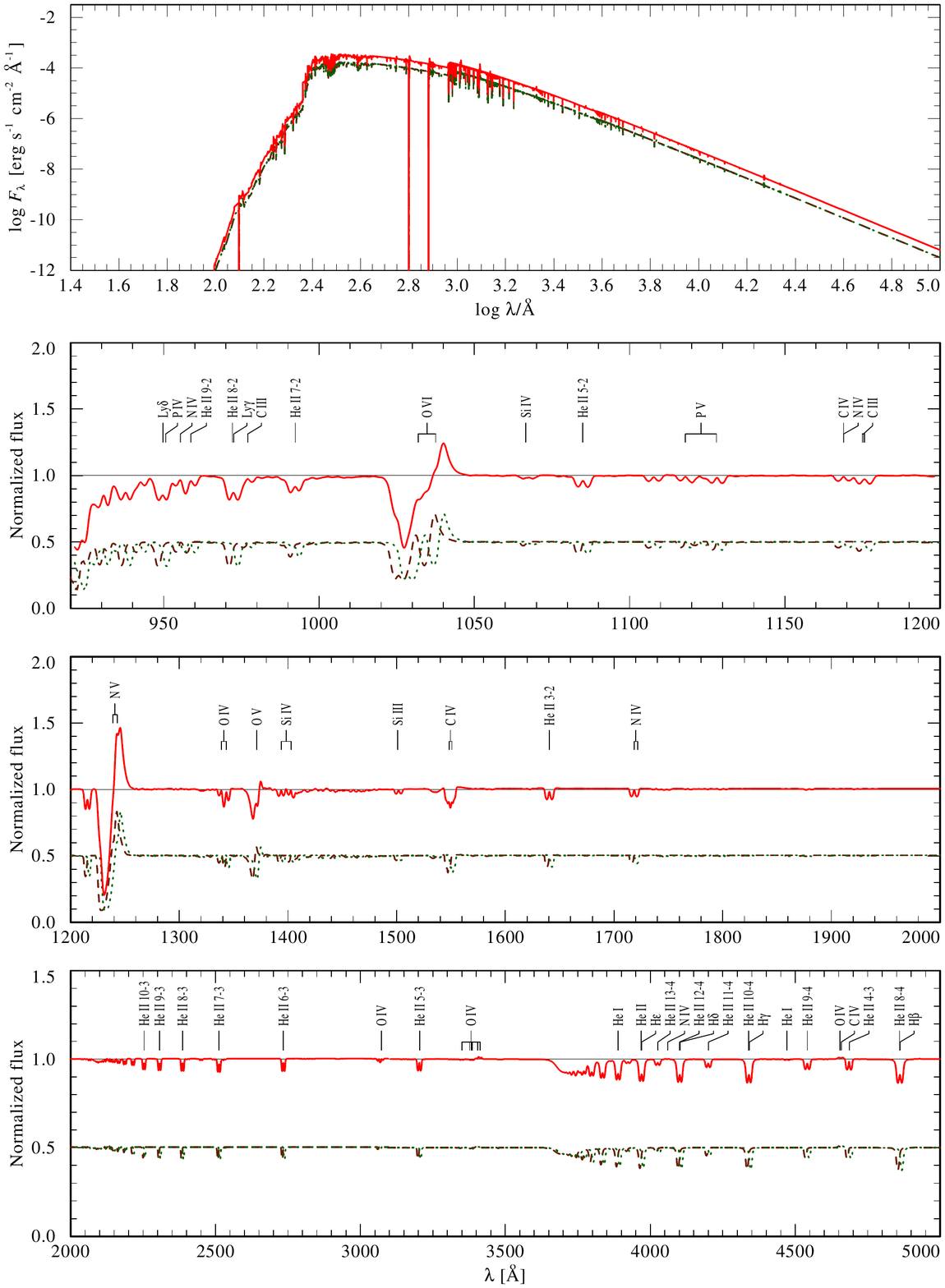} 
  \caption{Synthetic binary spectrum (red straight line) of model MII\,1 for the $50\,M_\odot$ track calculated by 
  \citet{Marchant2016} for a metallicity of $0.05\,Z_\odot$ (see Fig.\,\ref{fig:hrd_pablo} and Table\,\ref{table:parameters-pablo-005Zsun}). The composite spectrum is the sum of the primary spectrum (brown dashed line) and the secondary spectrum (green dotted line).
  The continuum level is indicated by a thin black line.}
  \label{fig:ms_005Zsun}
\end{figure*}
\clearpage
\setcounter{figure}{\value{figure}-1}
\begin{figure*}
  \centering
  \includegraphics[width=0.92\textwidth,page=2]{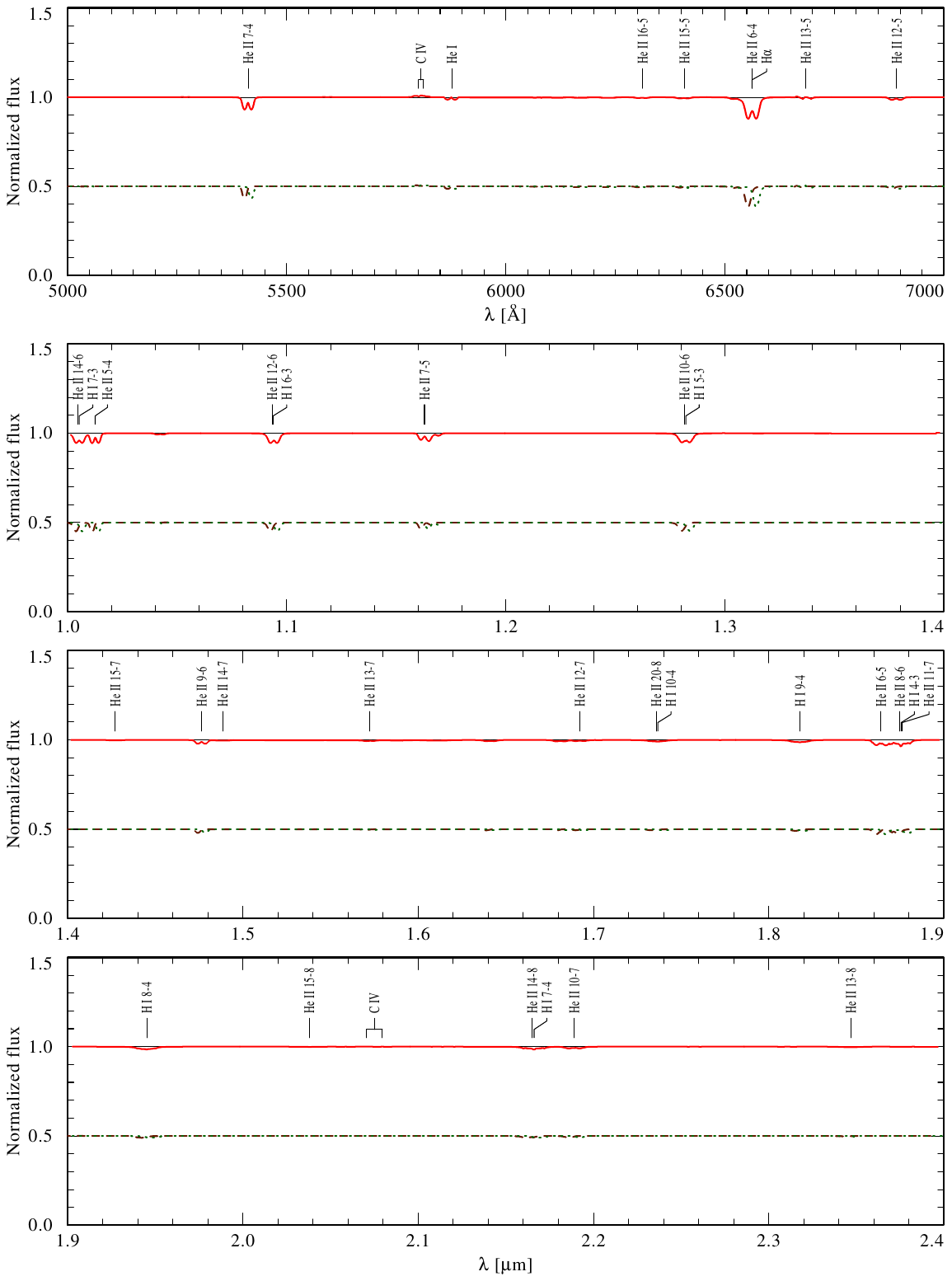}
  \caption{continued.}
\end{figure*}

\clearpage
\begin{figure*}
  \centering  
  \includegraphics[width=0.92\textwidth,page=1]{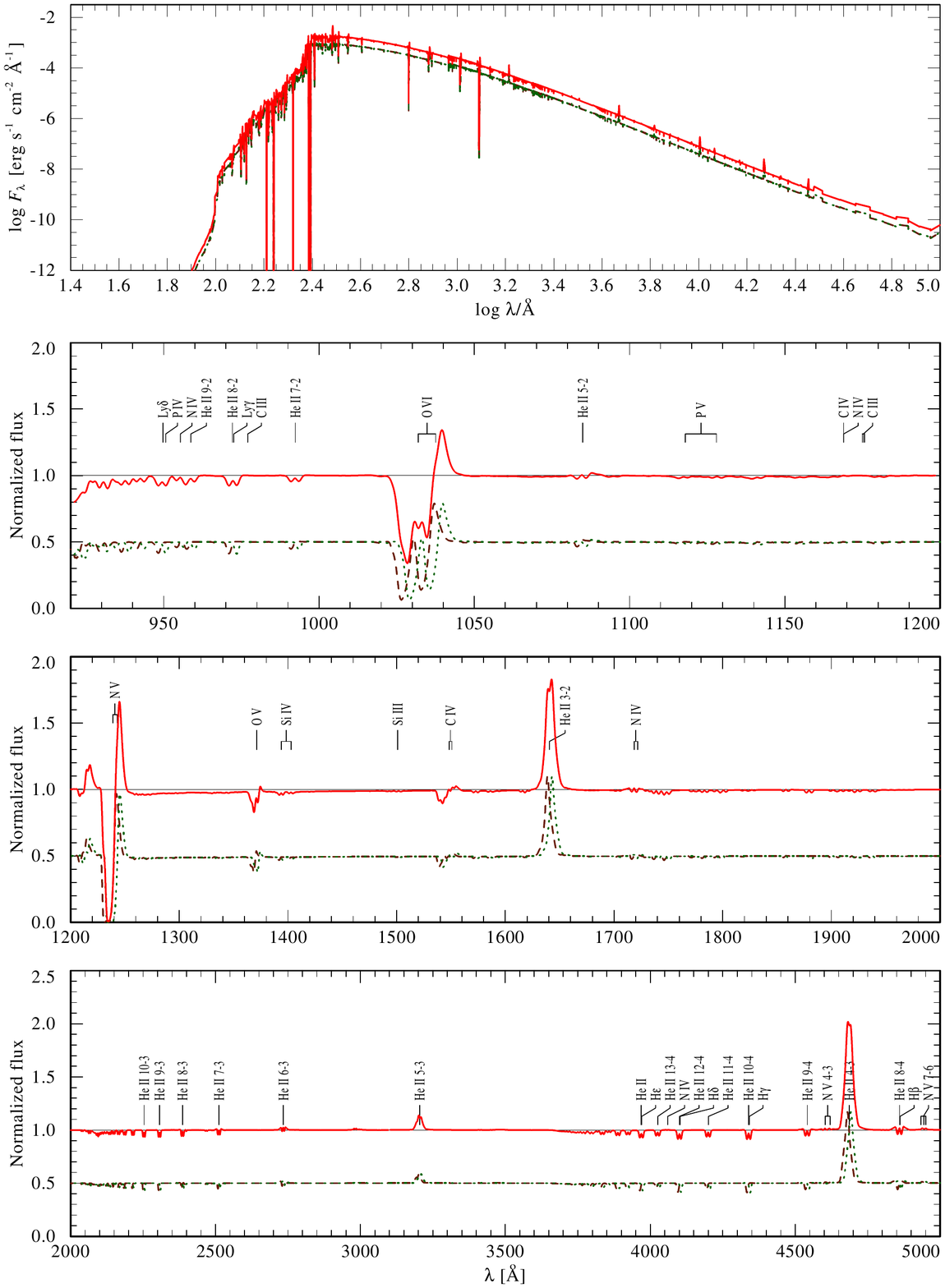} 
  \caption{Same as Fig.\,\ref{fig:ms_005Zsun} but for model MI\,2.}
  \label{fig:pmt_005Zsun}
\end{figure*}
\clearpage
\setcounter{figure}{\value{figure}-1}
\begin{figure*}
  \centering
  \includegraphics[width=0.92\textwidth,page=2]{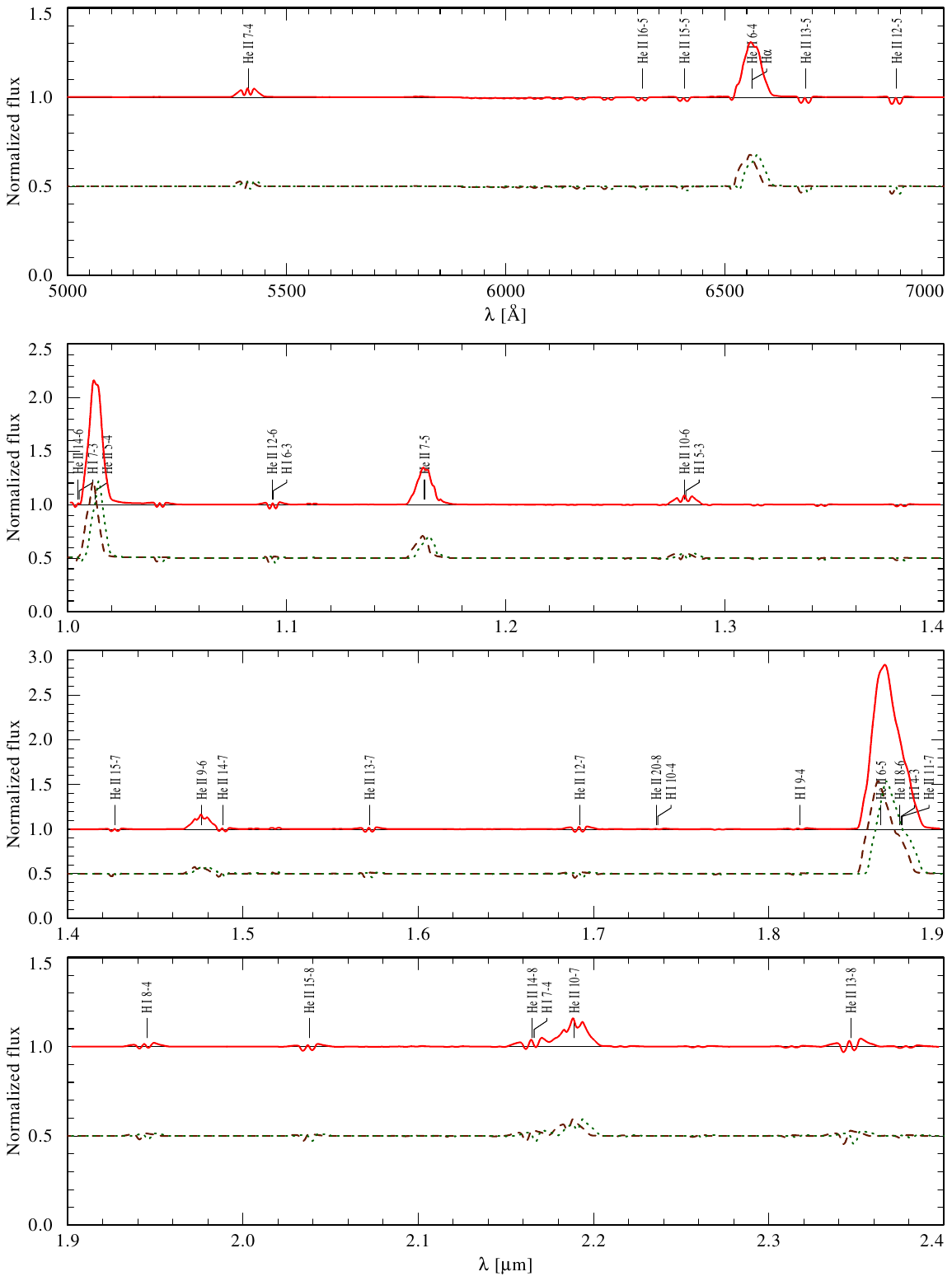}
  \caption{continued.}
\end{figure*}

\clearpage
\begin{figure*}
  \centering  
  \includegraphics[width=0.92\textwidth,page=1]{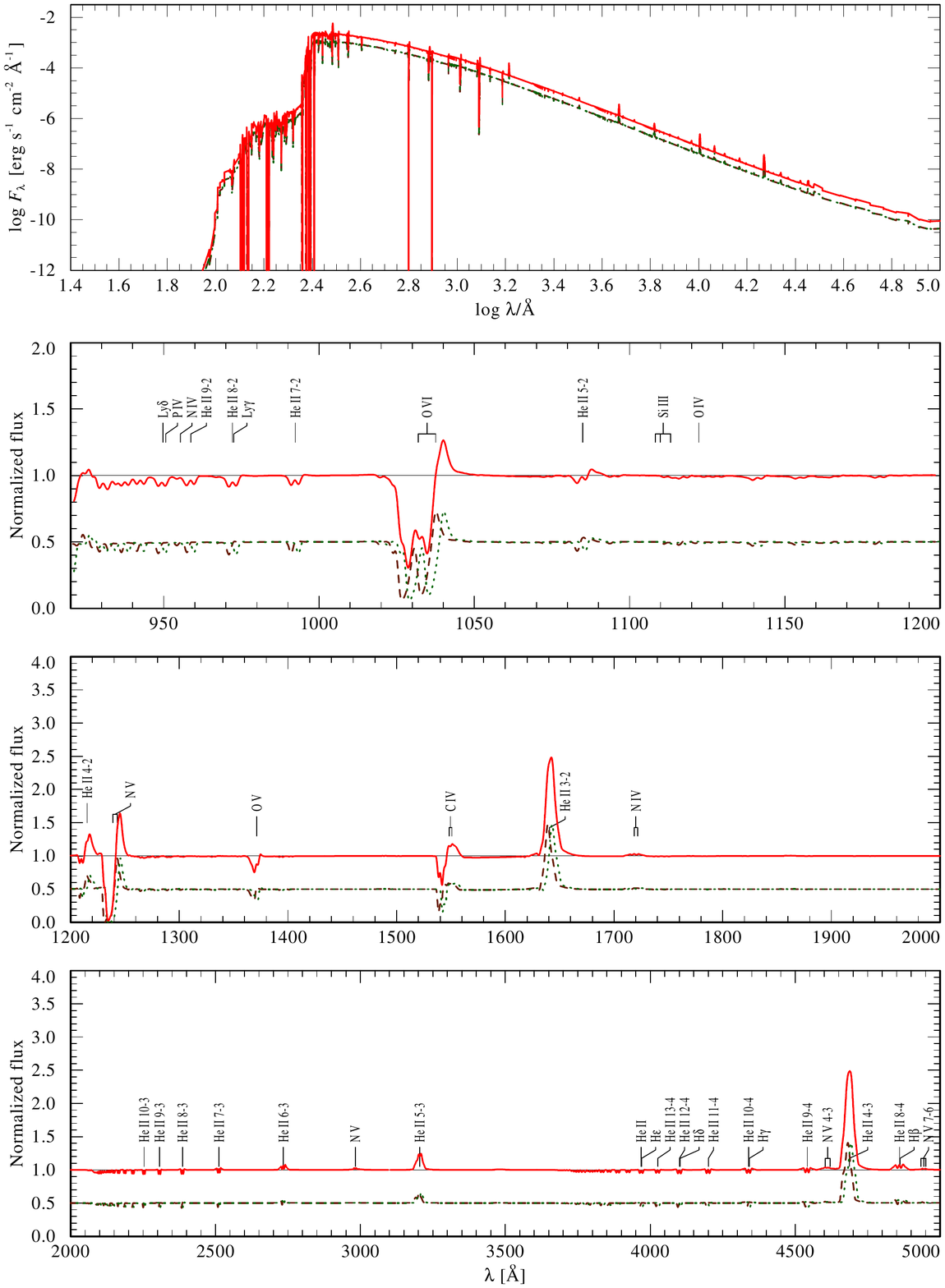} 
  \caption{Same as Fig.\,\ref{fig:ms_005Zsun} but for model MI\,3.}
  \label{fig:tp_005Zsun}
\end{figure*}
\clearpage
\setcounter{figure}{\value{figure}-1}
\begin{figure*}
  \centering
  \includegraphics[width=0.92\textwidth,page=2]{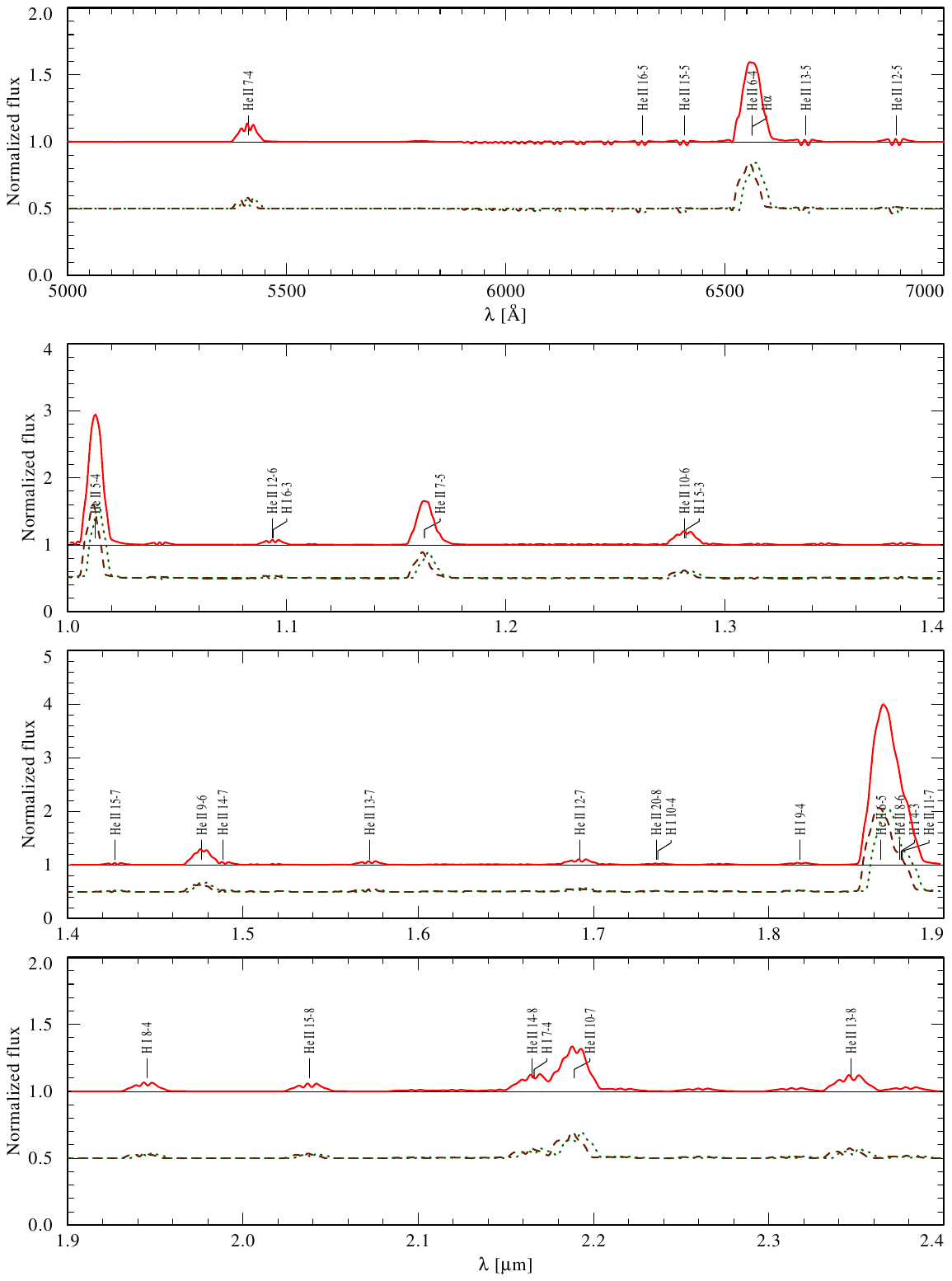}
  \caption{continued.}
\end{figure*}

\clearpage
\begin{figure*}
  \centering  
  \includegraphics[width=0.92\textwidth,page=1]{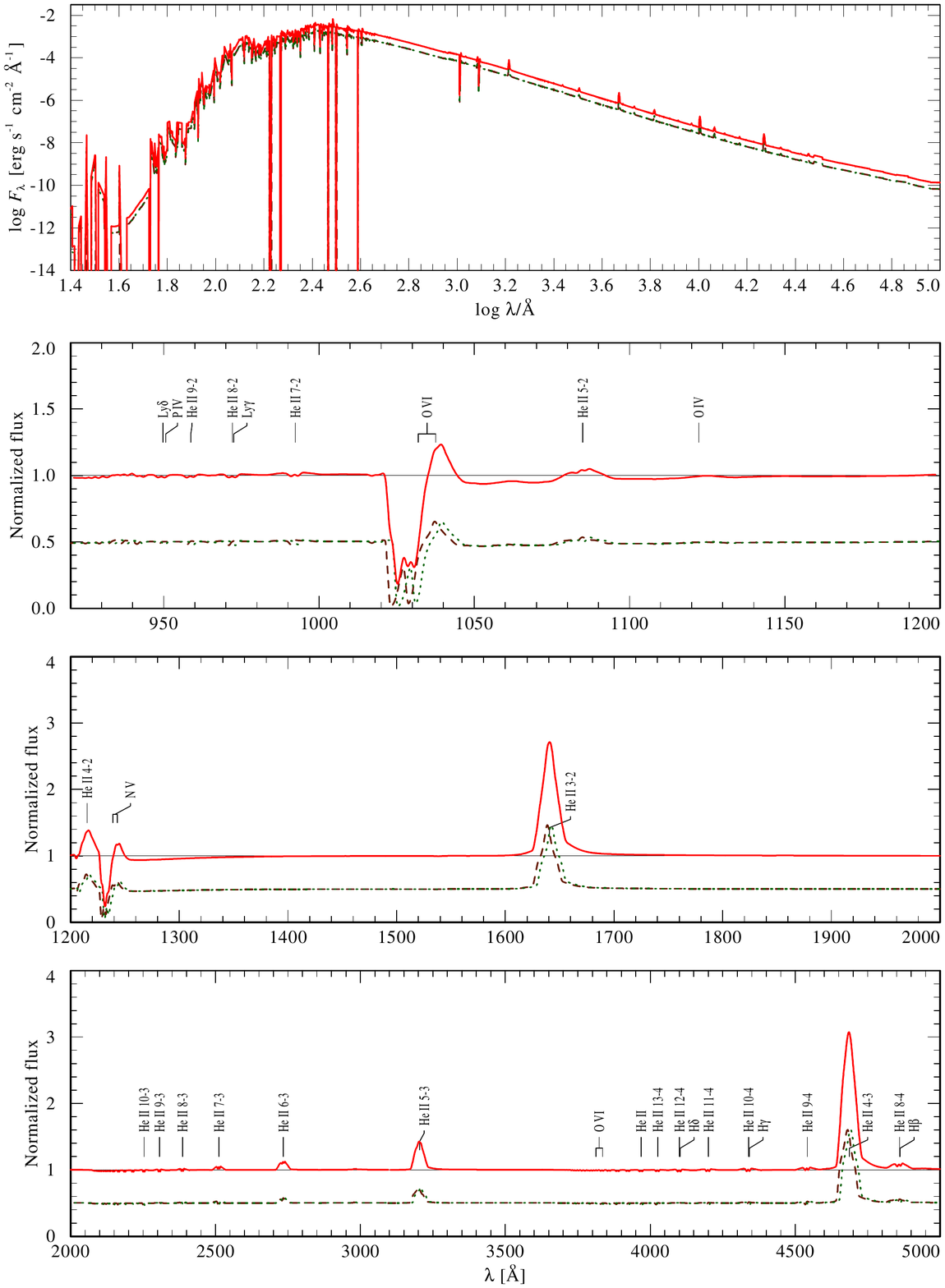} 
  \caption{Same as Fig.\,\ref{fig:ms_005Zsun} but for model MI\,4.}
  \label{fig:100kK_005Zsun}
\end{figure*}
\clearpage
\setcounter{figure}{\value{figure}-1}
\begin{figure*}
  \centering
  \includegraphics[width=0.92\textwidth,page=2]{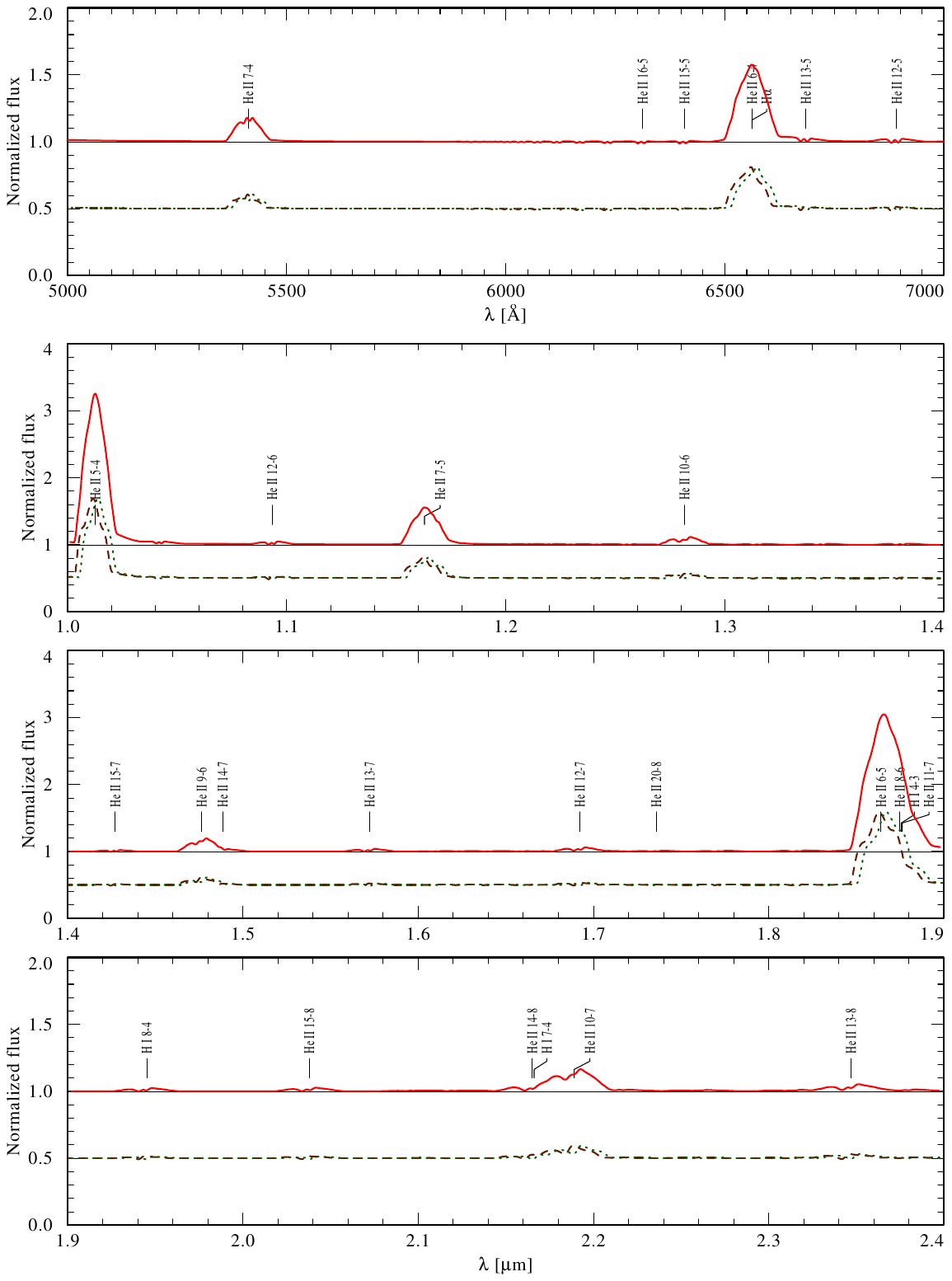}
  \caption{continued.}
\end{figure*}

\clearpage
\begin{figure*}
  \centering  
  \includegraphics[width=0.92\textwidth,page=1]{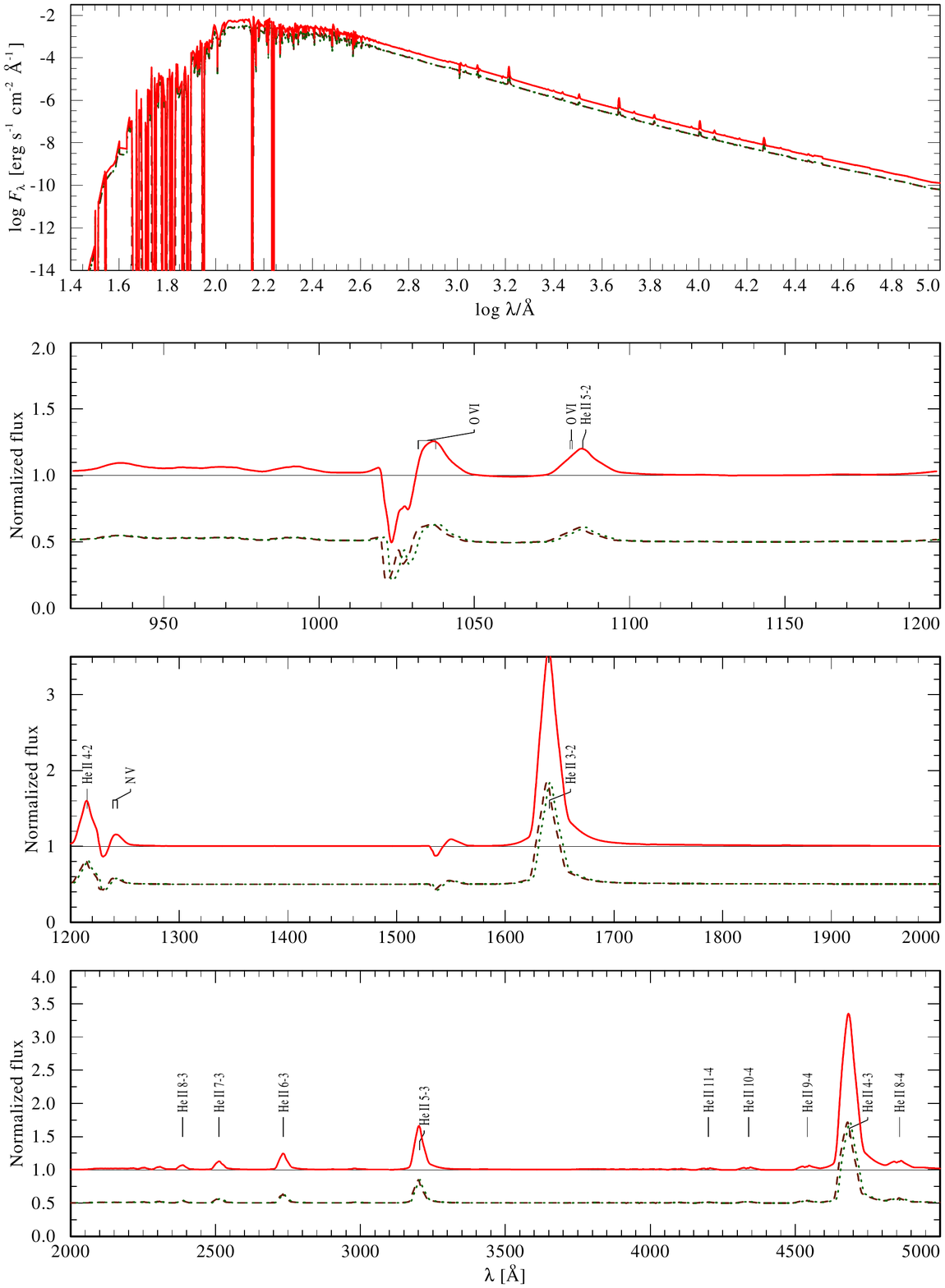} 
  \caption{Same as Fig.\,\ref{fig:ms_005Zsun} but for model MI\,5.}
  \label{fig:150kK_005Zsun}
\end{figure*}
\clearpage
\setcounter{figure}{\value{figure}-1}
\begin{figure*}
  \centering
  \includegraphics[width=0.92\textwidth,page=2]{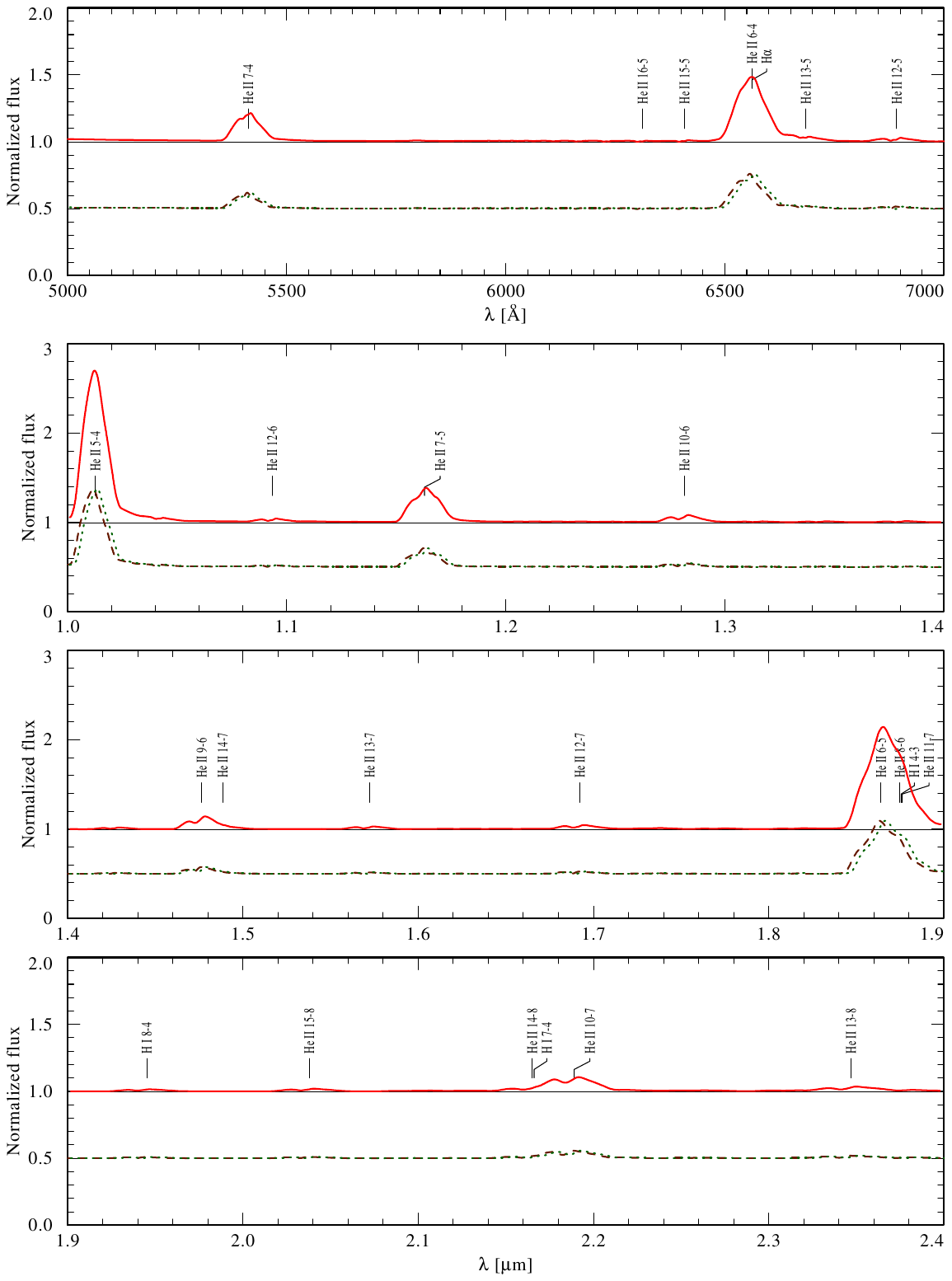}
  \caption{continued.}
\end{figure*}
%---------------------------------------------------------------

%---------------------------------------------------------------
\clearpage
\begin{figure*}
  \centering  
  \includegraphics[width=0.88\textwidth,page=1]{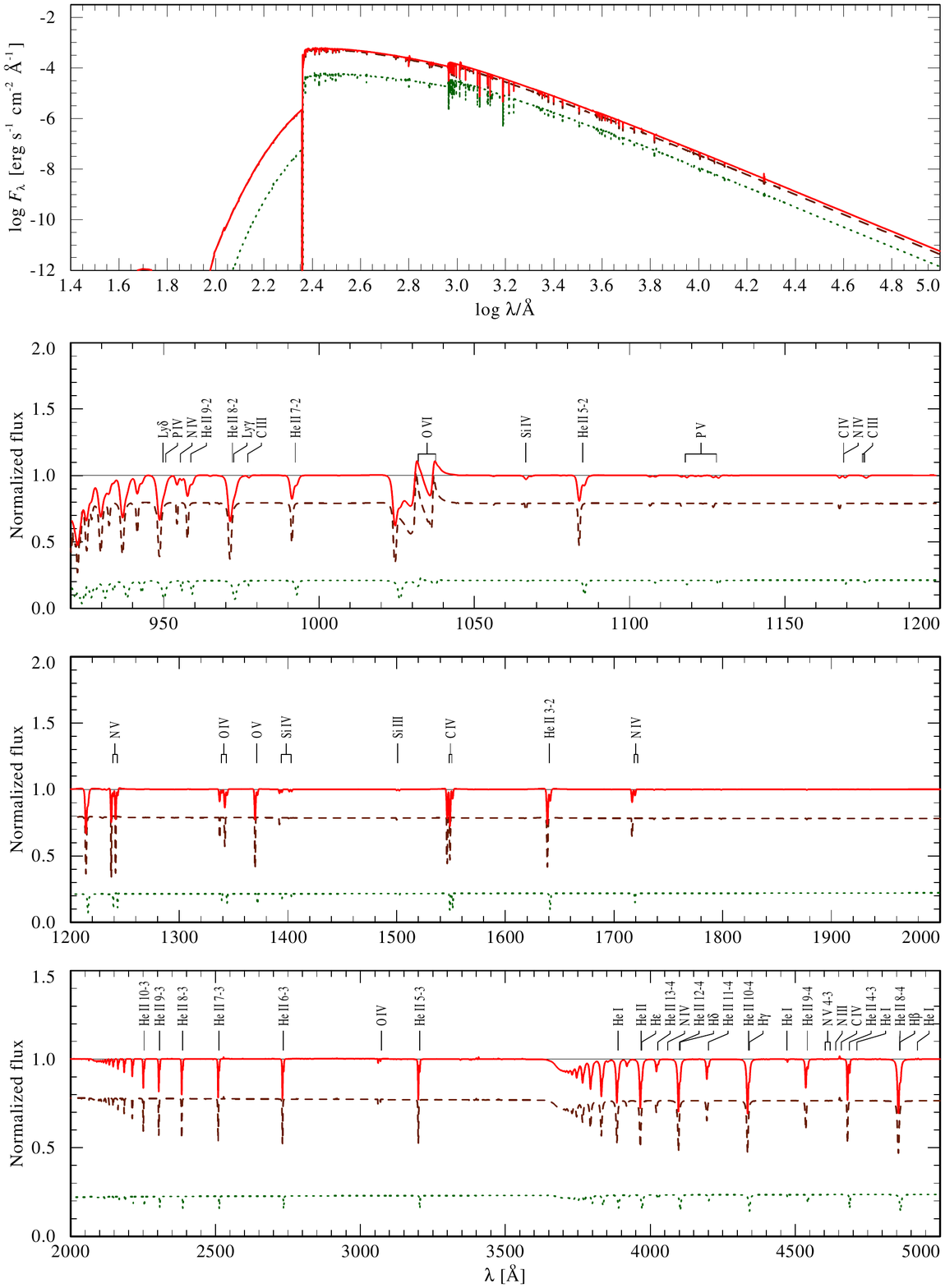} 
  \caption{Synthetic binary spectrum (red straight line) of model E\,1 for the BPASS evolution track calculate by \citet{Eldridge2016} (see Fig.\,\ref{fig:hrd_JJ} and Table\,\ref{table:parameters-JJ}). The composite spectrum is the sum of the primary spectrum (brown dashed line) and the secondary spectrum (green dotted line). The offsets between the spectra refers to the wavelength dependent light ratio of the components. The continuum level is indicated by a thin black line.}
  \label{fig:ms_jj}
\end{figure*}
\clearpage
\setcounter{figure}{\value{figure}-1}
\begin{figure*}
  \centering
  \includegraphics[width=0.92\textwidth,page=2]{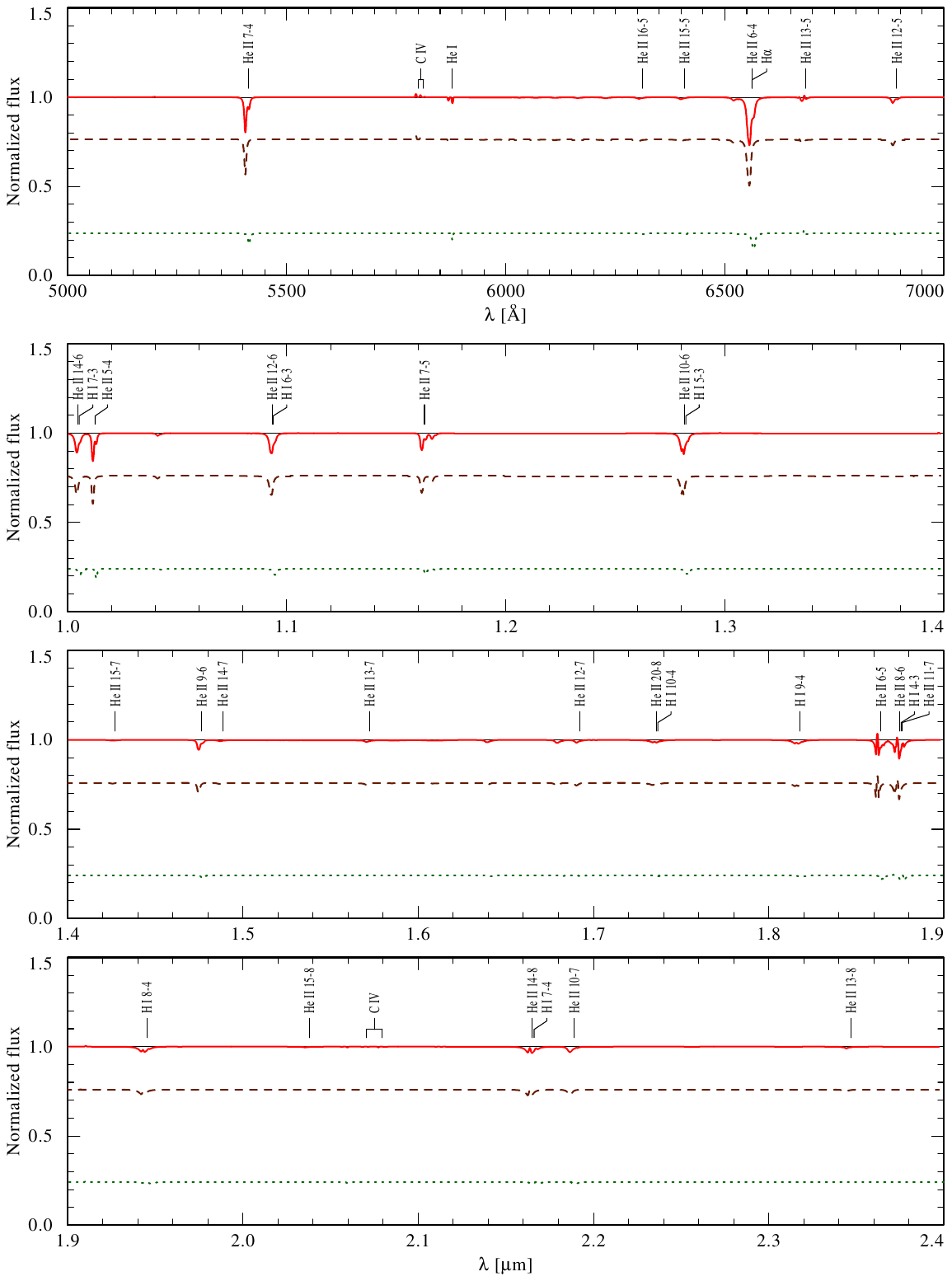}
  \caption{continued.}
\end{figure*}

\clearpage
\begin{figure*}
  \centering  
  \includegraphics[width=0.92\textwidth,page=1]{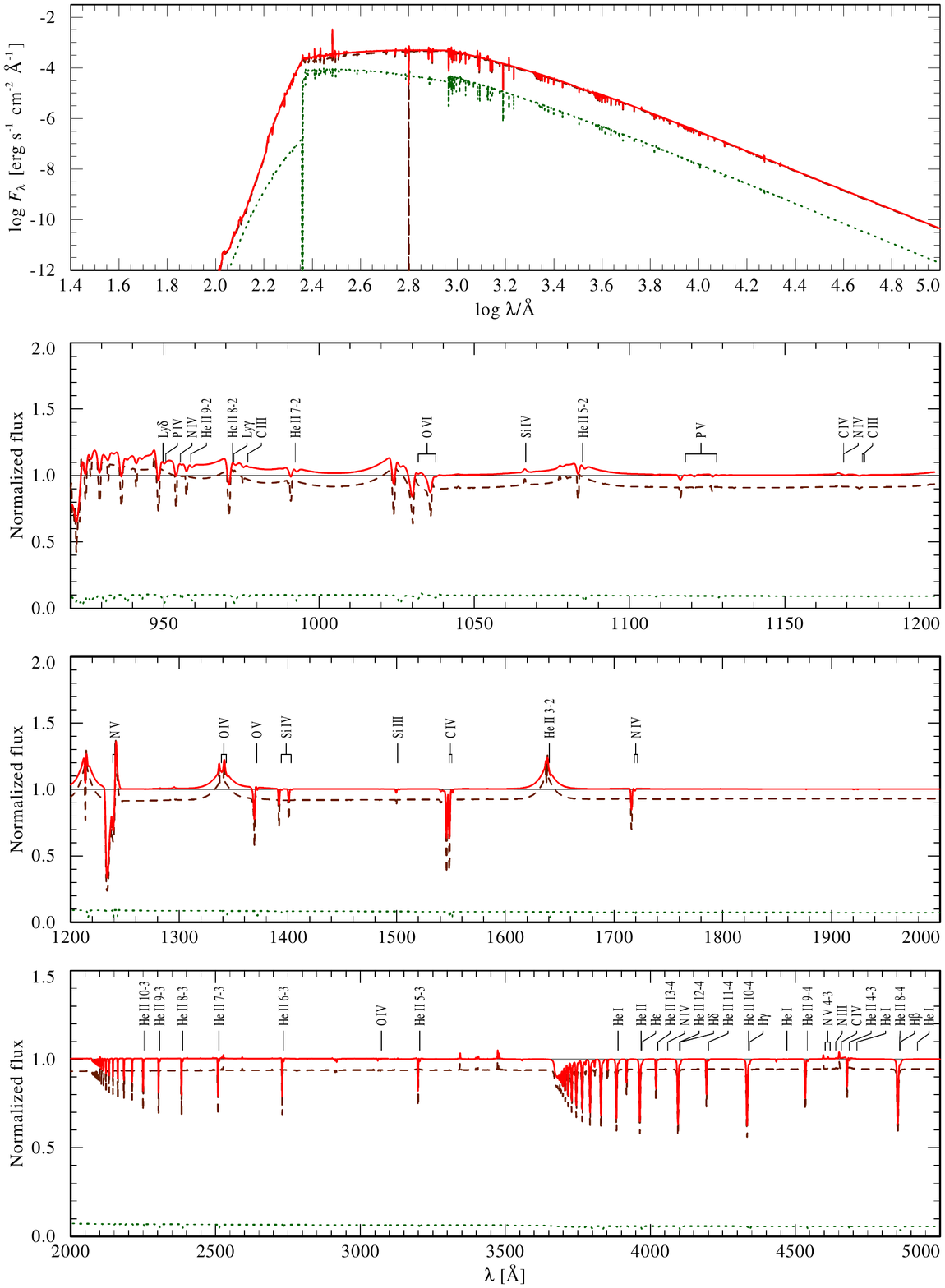} 
  \caption{Same as Fig.\,\ref{fig:ms_jj} but for model E\,2.}
  \label{fig:pre_rlof_jj}
\end{figure*}
\clearpage
\setcounter{figure}{\value{figure}-1}
\begin{figure*}
  \centering
  \includegraphics[width=0.92\textwidth,page=2]{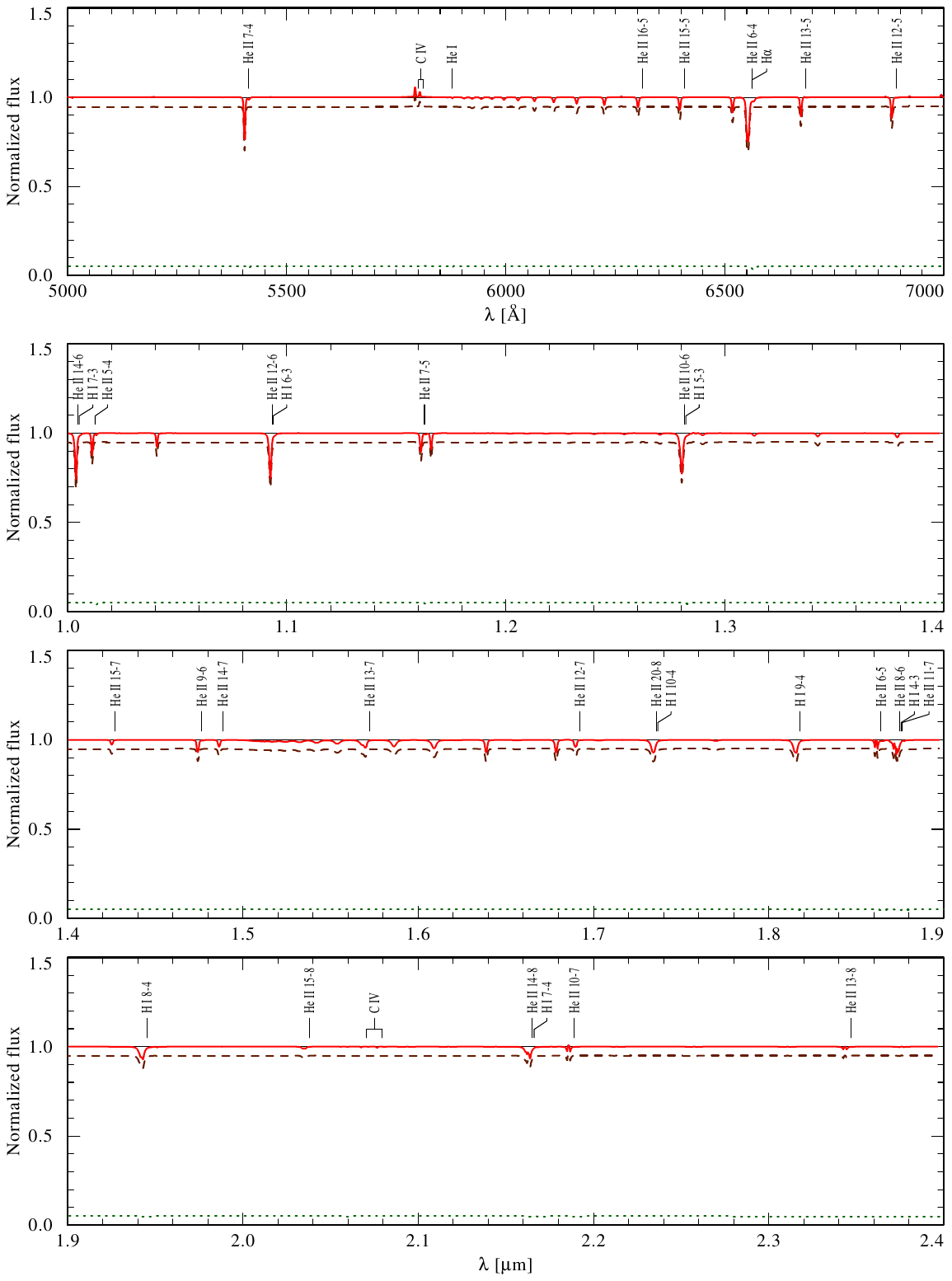}
  \caption{continued.}
\end{figure*}

\clearpage
\begin{figure*}
  \centering  
  \includegraphics[width=0.92\textwidth,page=1]{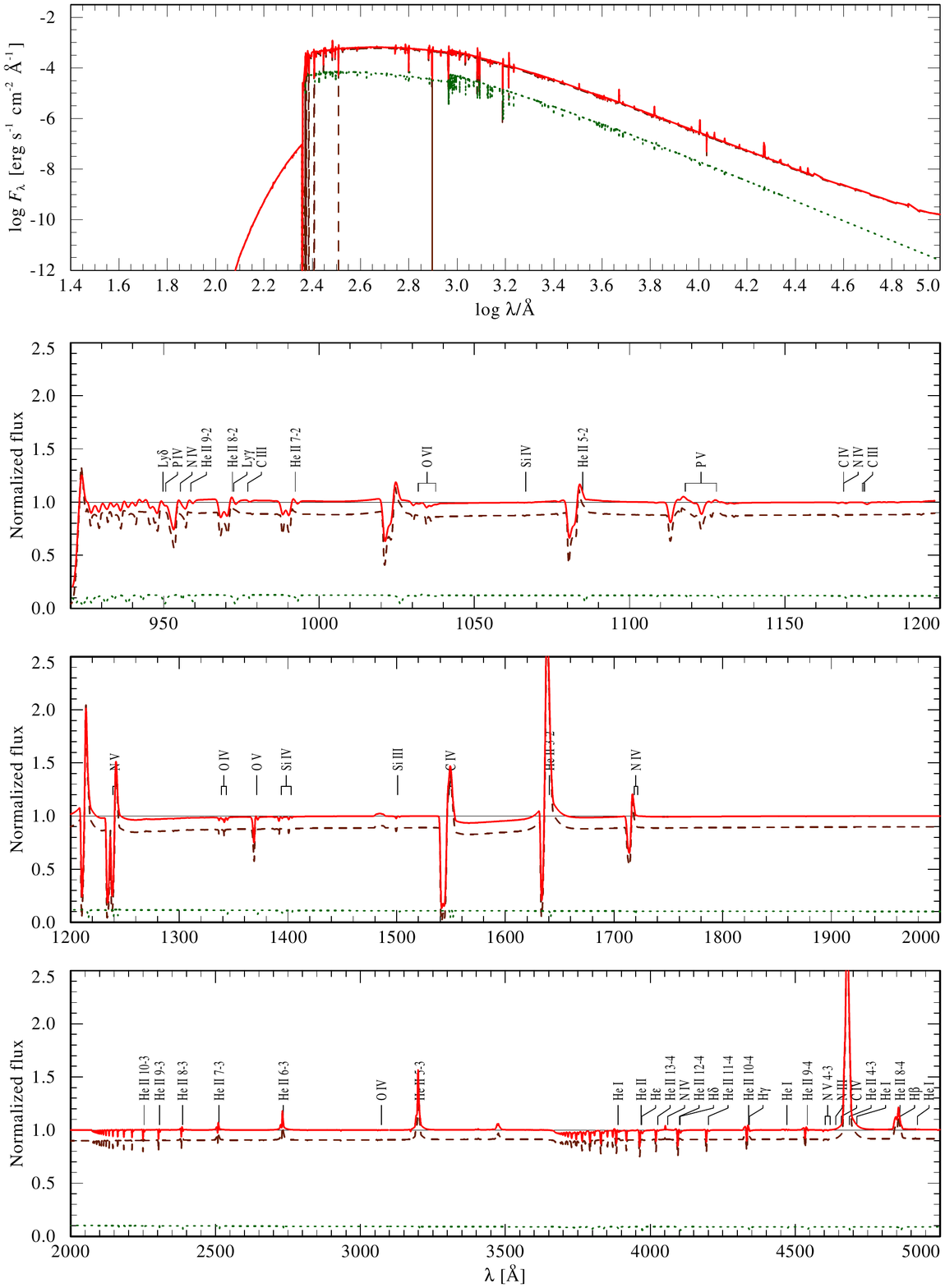} 
  \caption{Same as Fig.\,\ref{fig:ms_jj} but for model E\,3.}
  \label{fig:wn_1_jj}
\end{figure*}
\clearpage
\setcounter{figure}{\value{figure}-1}
\begin{figure*}
  \centering
  \includegraphics[width=0.92\textwidth,page=2]{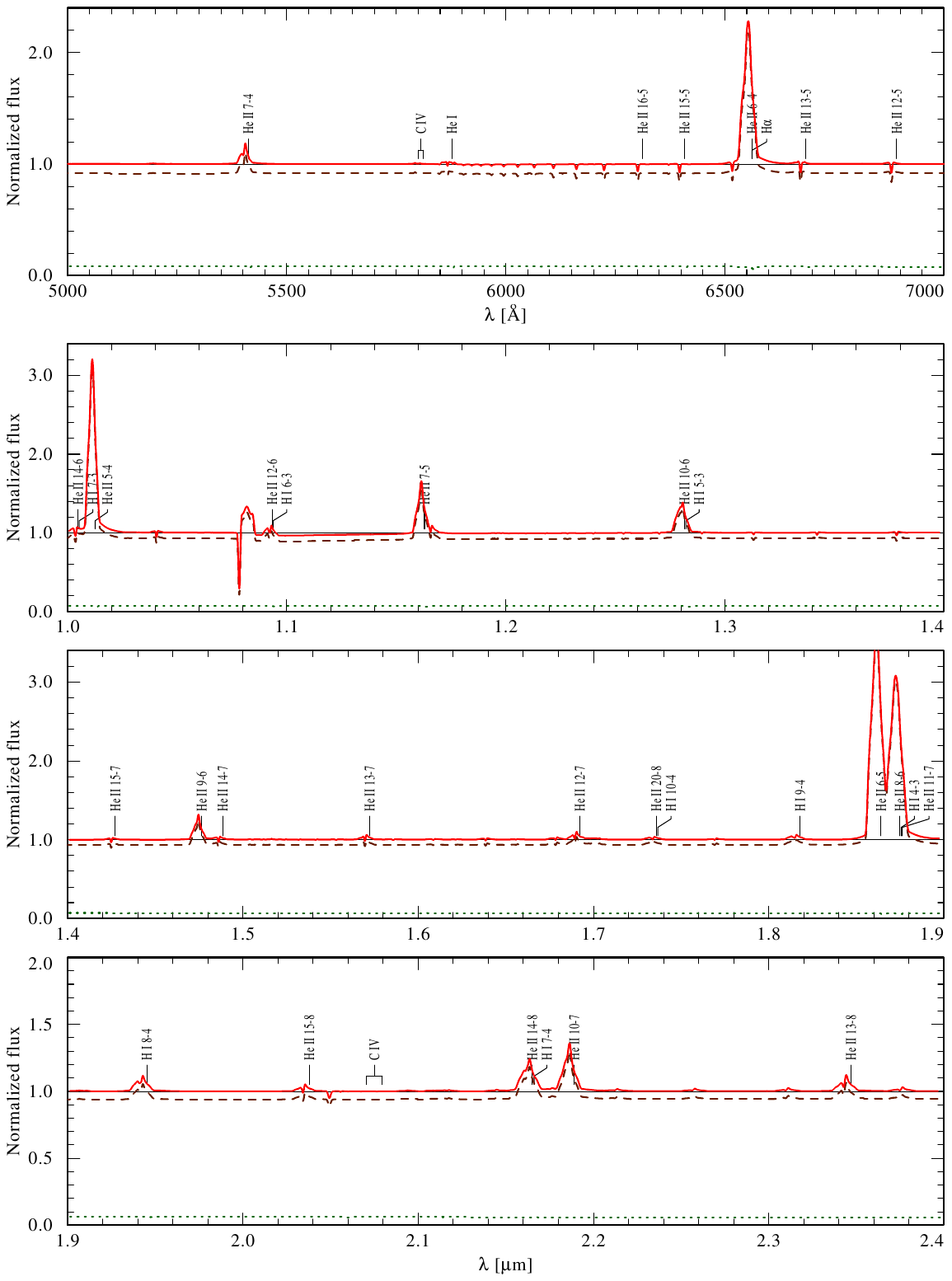}
  \caption{continued.}
\end{figure*}

\clearpage
\begin{figure*}
  \centering  
  \includegraphics[width=0.92\textwidth,page=1]{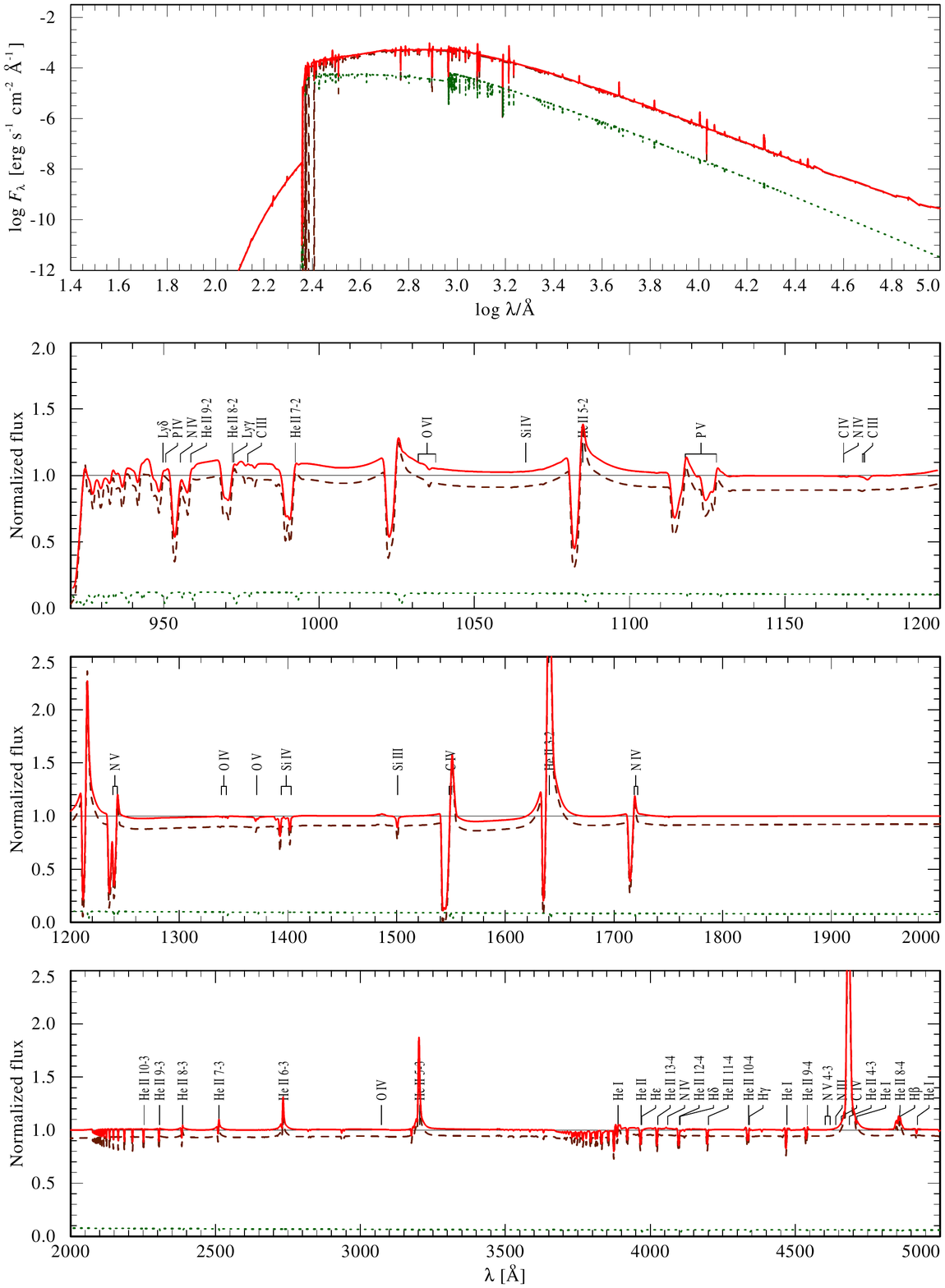} 
  \caption{Same as Fig.\,\ref{fig:ms_jj} but for model E\,4.}
  \label{fig:wn_2_jj}
\end{figure*}
\clearpage
\setcounter{figure}{\value{figure}-1}
\begin{figure*}
  \centering
  \includegraphics[width=0.92\textwidth,page=2]{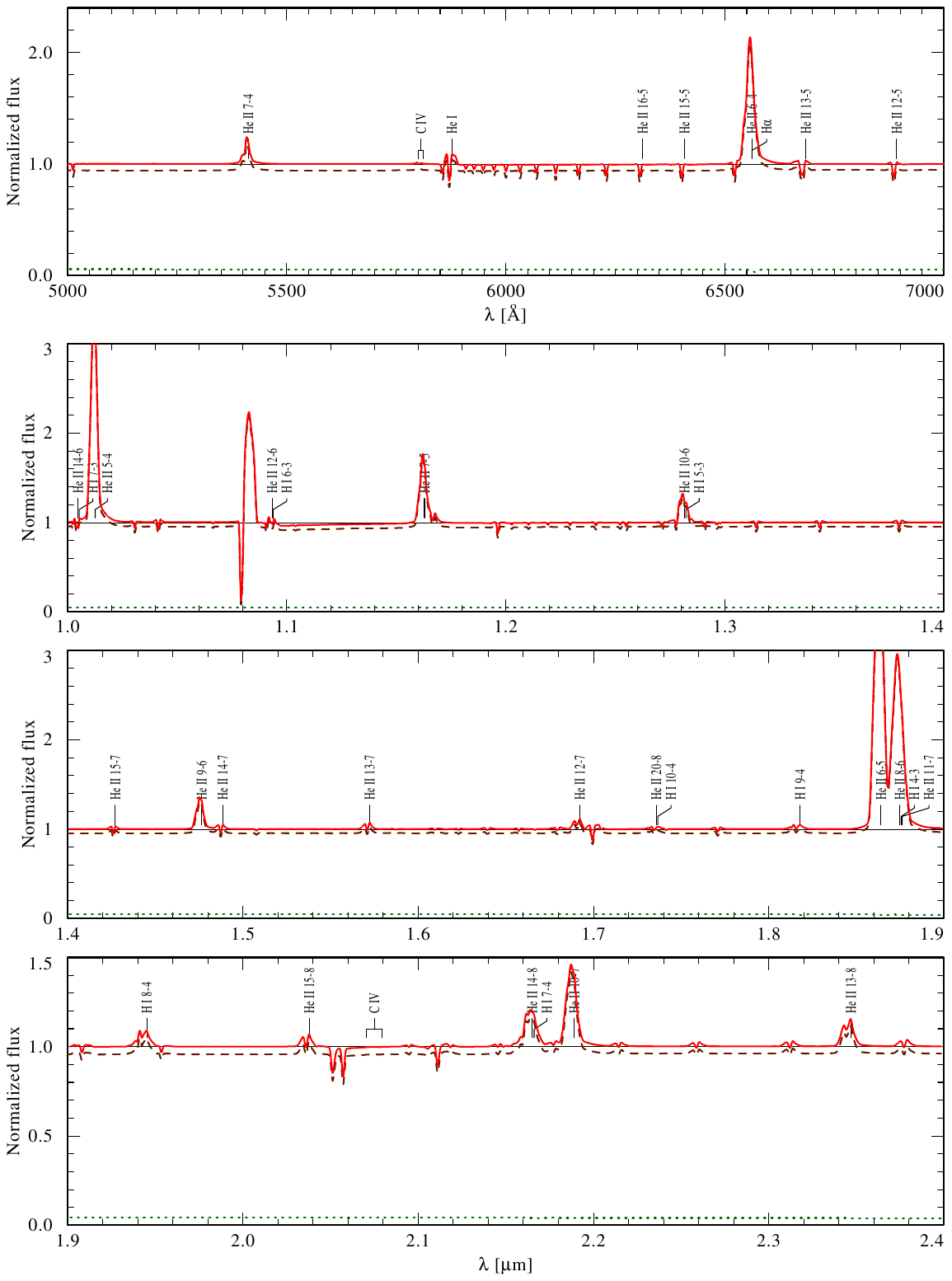}
  \caption{continued.}
\end{figure*}

%---------------------------------------------------------------

\clearpage
\begin{figure*}
  \centering  
  \includegraphics[width=0.91\textwidth,page=1]{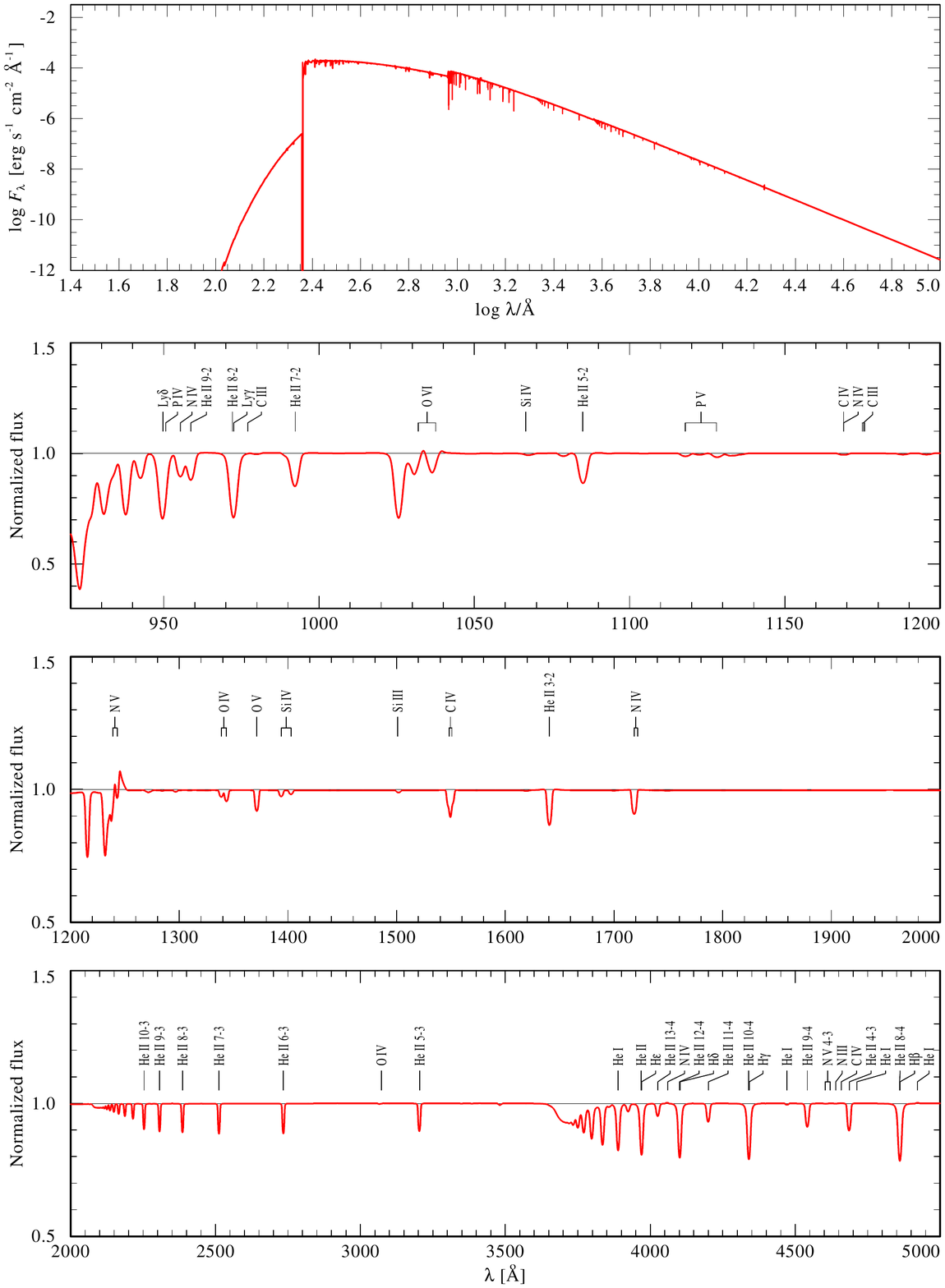} 
  \caption{Synthetic spectrum (red straight line) of model E\,5 for the BPASS evolution track (see Fig.\,\ref{fig:hrd_JJ} and Table\,\ref{table:parameters-JJ}), referring to the secondary after the primary collapsed into a BH.}
  \label{fig:ms_sec_jj}
\end{figure*}
\clearpage
\setcounter{figure}{\value{figure}-1}
\begin{figure*}
  \centering
  \includegraphics[width=0.92\textwidth,page=2]{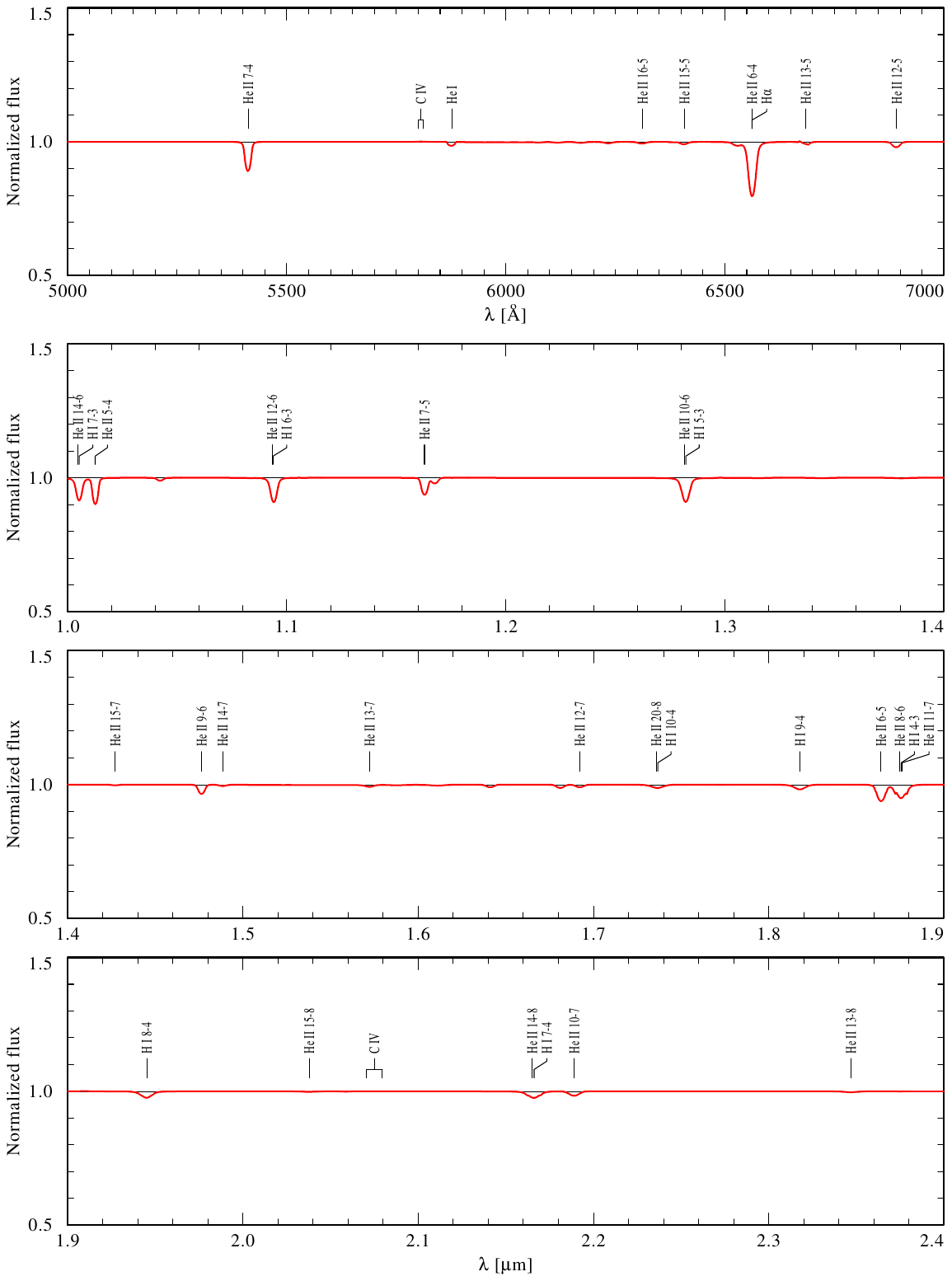}
  \caption{continued.}
\end{figure*}

\clearpage
\begin{figure*}
  \centering  
  \includegraphics[width=0.92\textwidth,page=1]{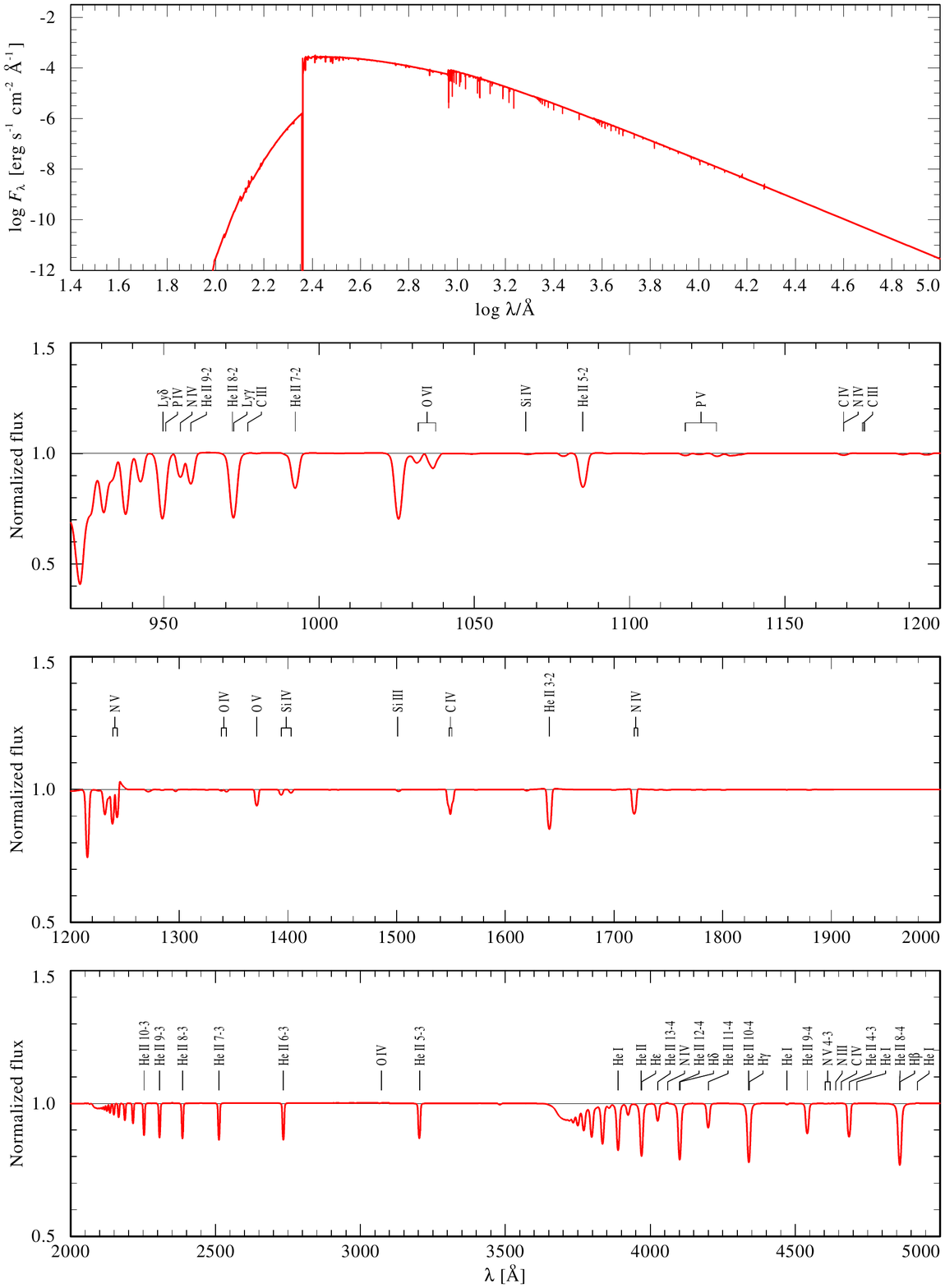} 
  \caption{Same as Fig.\,\ref{fig:ms_sec_jj} but for model E\,6.}
  \label{fig:pre_wn_sec_jj}
\end{figure*}
\clearpage
\setcounter{figure}{\value{figure}-1}
\begin{figure*}
  \centering
  \includegraphics[width=0.92\textwidth,page=2]{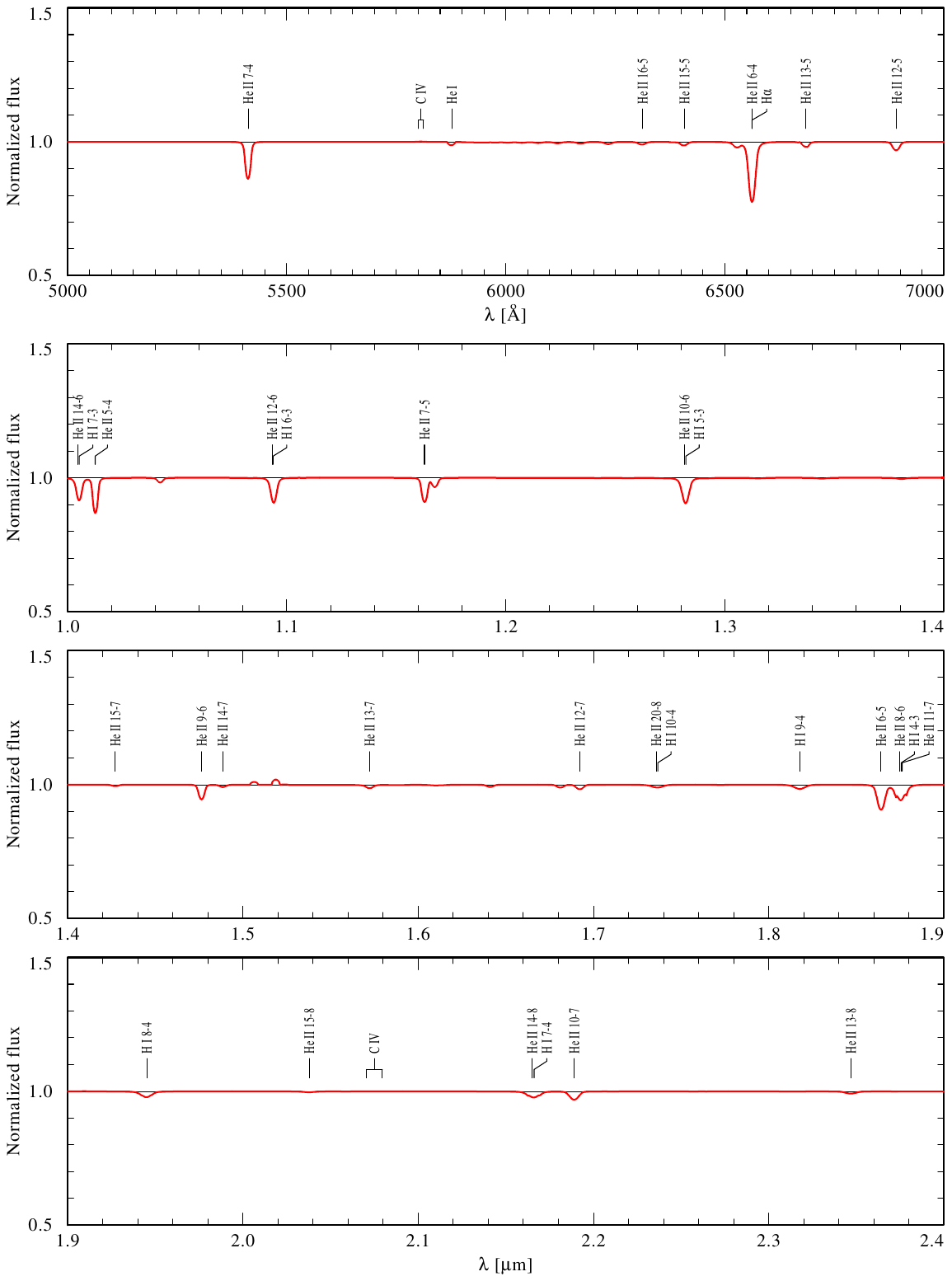}
  \caption{continued.}
\end{figure*}

\clearpage
\begin{figure*}
  \centering  
  \includegraphics[width=0.92\textwidth,page=1]{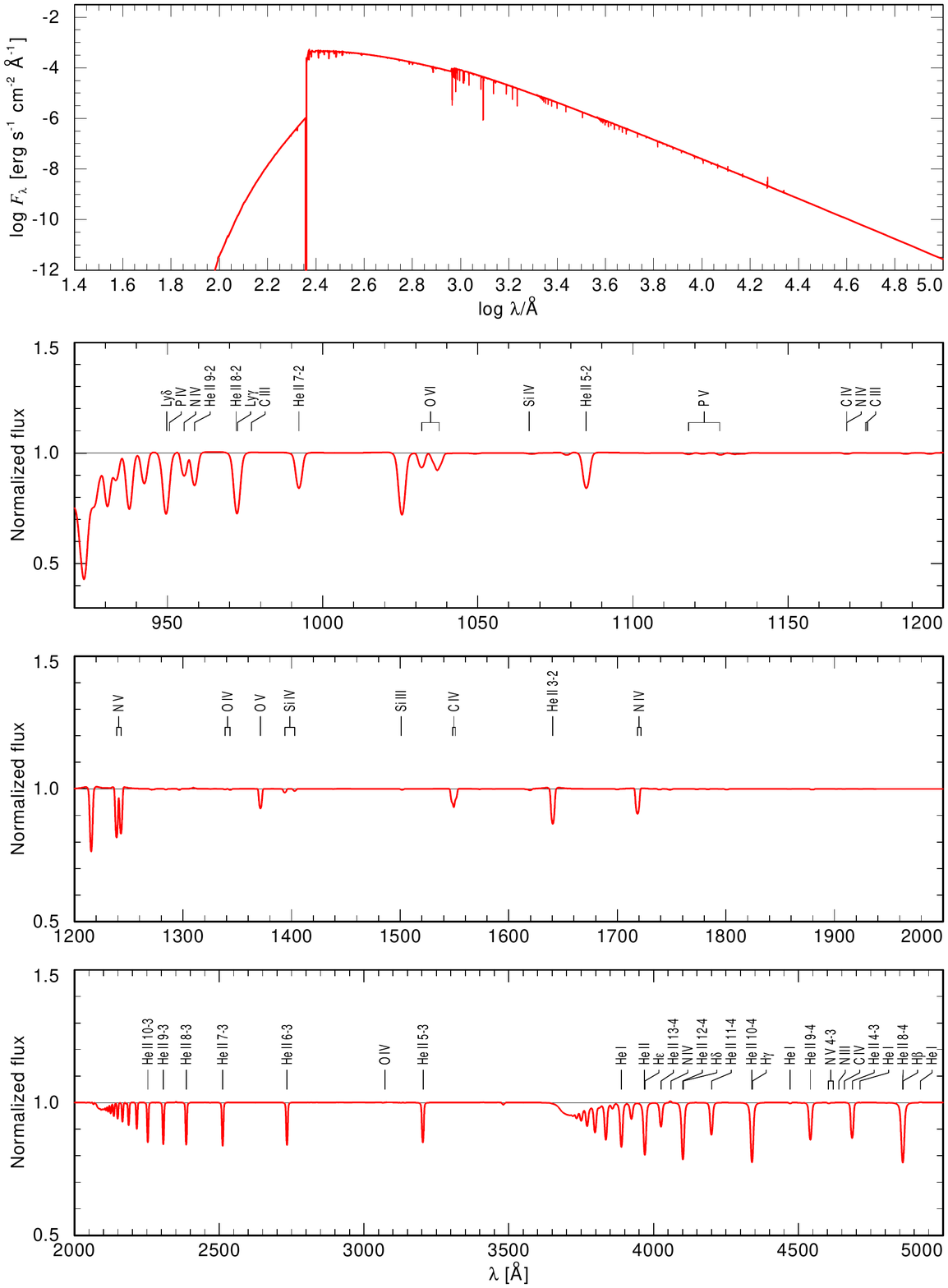} 
  \caption{Same as Fig.\,\ref{fig:ms_sec_jj} but for model E\,7.}
  \label{fig:wn_1_sec_jj}
\end{figure*}
\clearpage
\setcounter{figure}{\value{figure}-1}
\begin{figure*}
  \centering
  \includegraphics[width=0.92\textwidth,page=2]{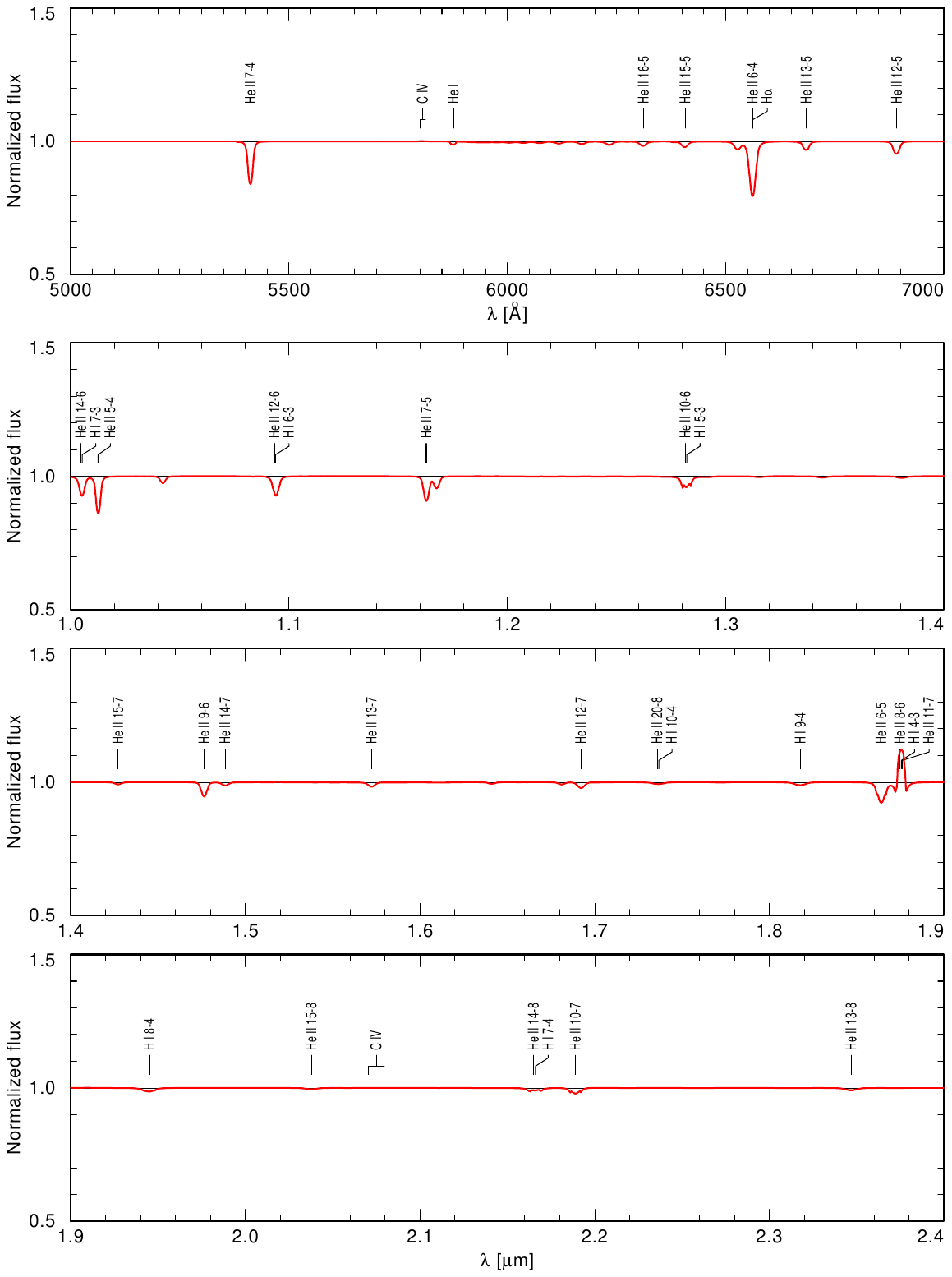}
  \caption{continued.}
\end{figure*}

\clearpage
\begin{figure*}
  \centering  
  \includegraphics[width=0.92\textwidth,page=1]{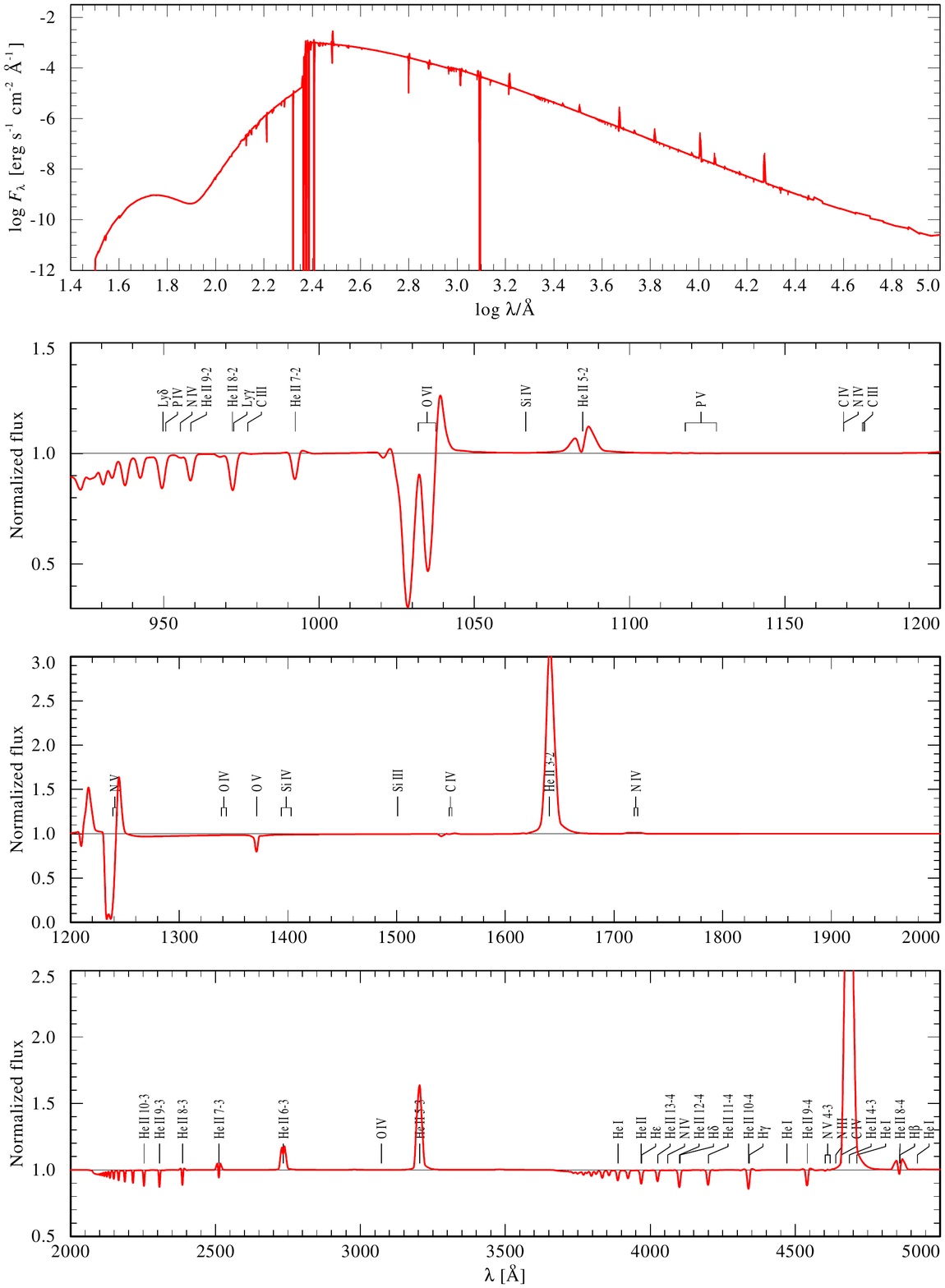} 
  \caption{Same as Fig.\,\ref{fig:ms_sec_jj} but for model E\,8.}
  \label{fig:wn_2_sec_jj}
\end{figure*}
\clearpage
\setcounter{figure}{\value{figure}-1}
\begin{figure*}
  \centering
  \includegraphics[width=0.92\textwidth,page=2]{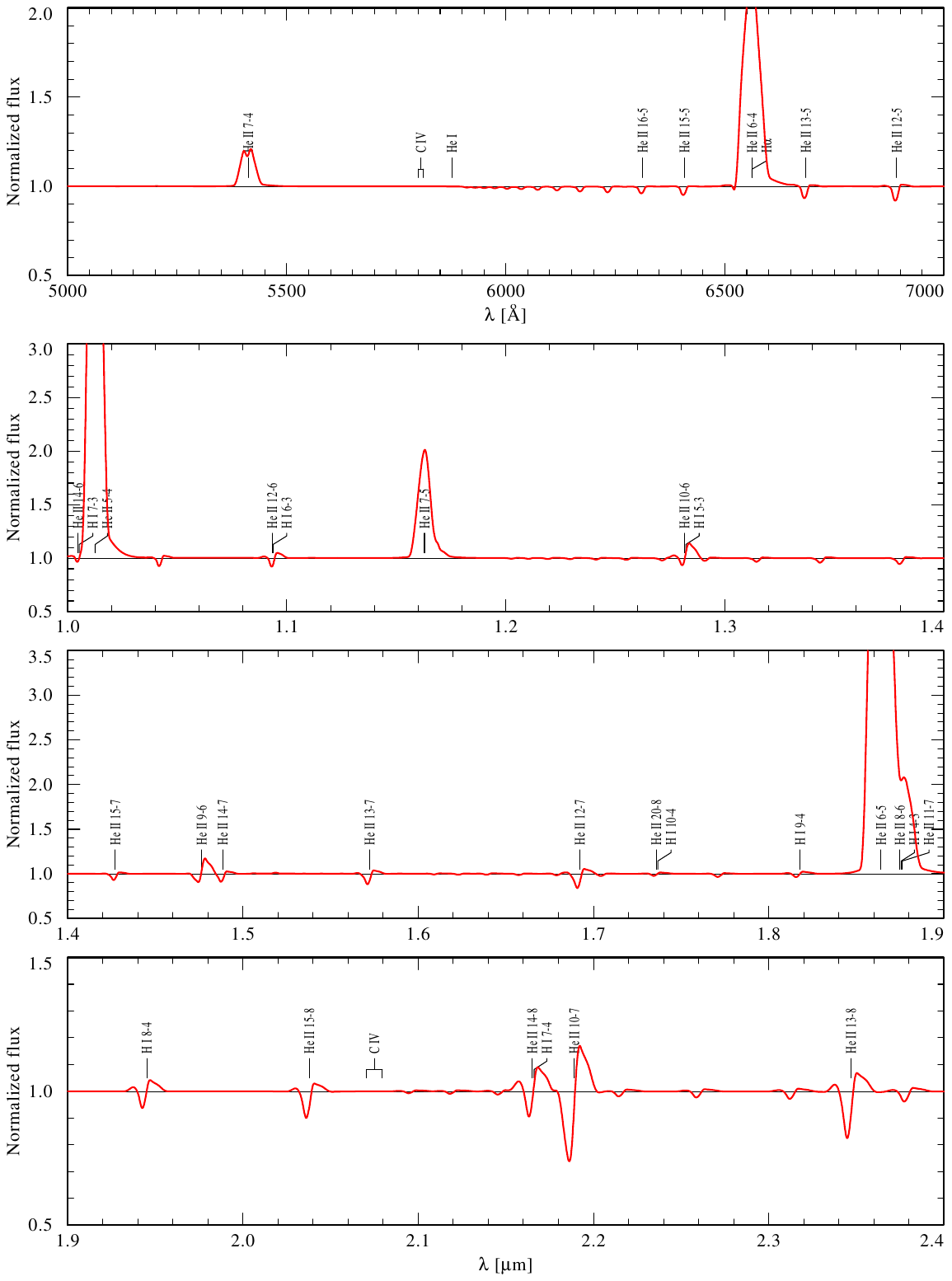}
  \caption{continued.}
\end{figure*}

\clearpage
\begin{figure*}
  \centering  
  \includegraphics[width=0.92\textwidth,page=1]{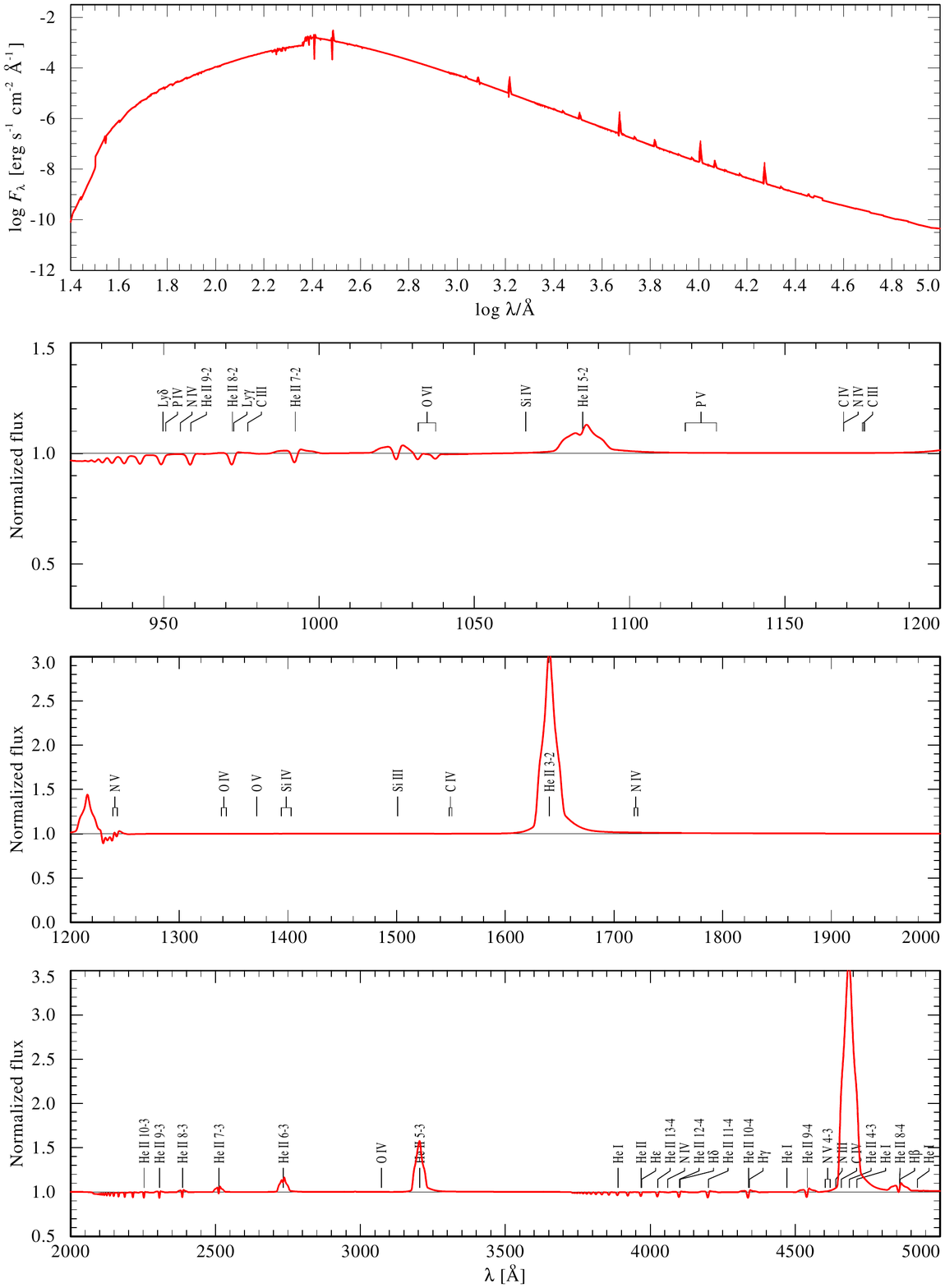} 
  \caption{Same as Fig.\,\ref{fig:ms_sec_jj} but for model E\,9.}
  \label{fig:wn_3_sec_jj}
\end{figure*}
\clearpage
\setcounter{figure}{\value{figure}-1}
\begin{figure*}
  \centering
  \includegraphics[width=0.92\textwidth,page=2]{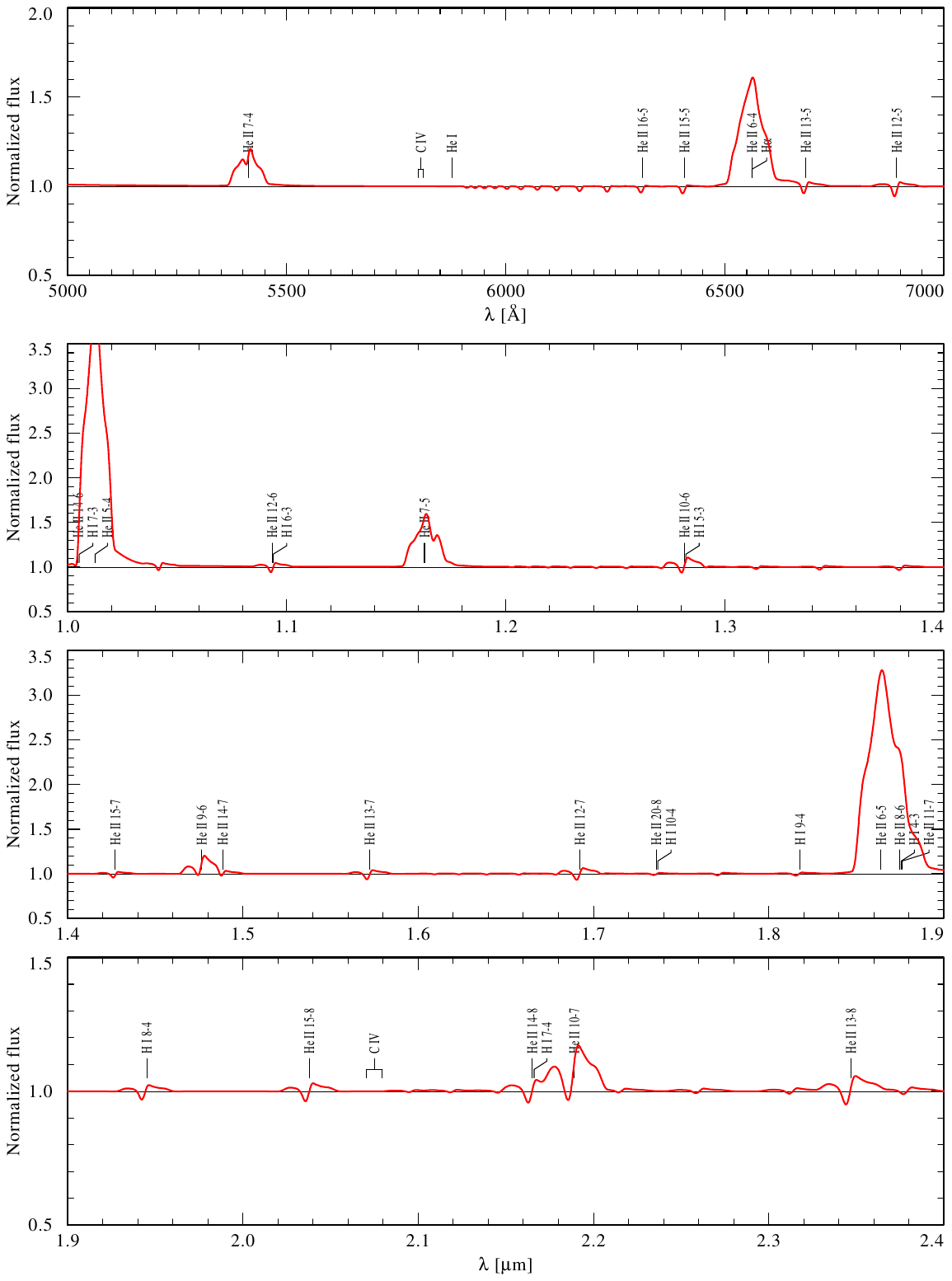}
  \caption{continued.}
\end{figure*}

\clearpage
\begin{figure*}
  \centering  
  \includegraphics[width=0.92\textwidth,page=1]{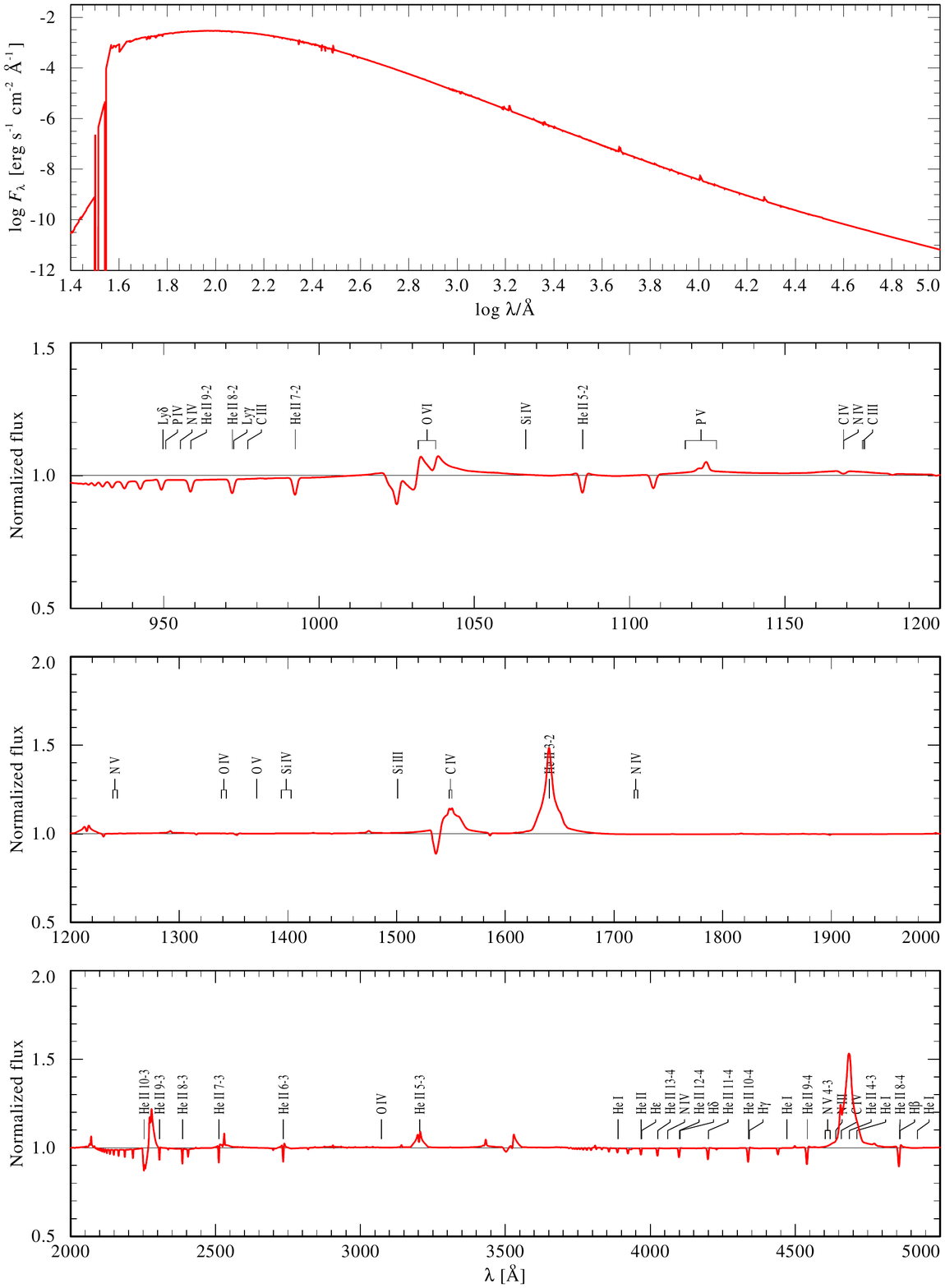} 
  \caption{Same as Fig.\,\ref{fig:ms_sec_jj} but for model E\,10.}
  \label{fig:wn_4_sec_jj}
\end{figure*}
\clearpage
\setcounter{figure}{\value{figure}-1}
\begin{figure*}
  \centering
  \includegraphics[width=0.92\textwidth,page=2]{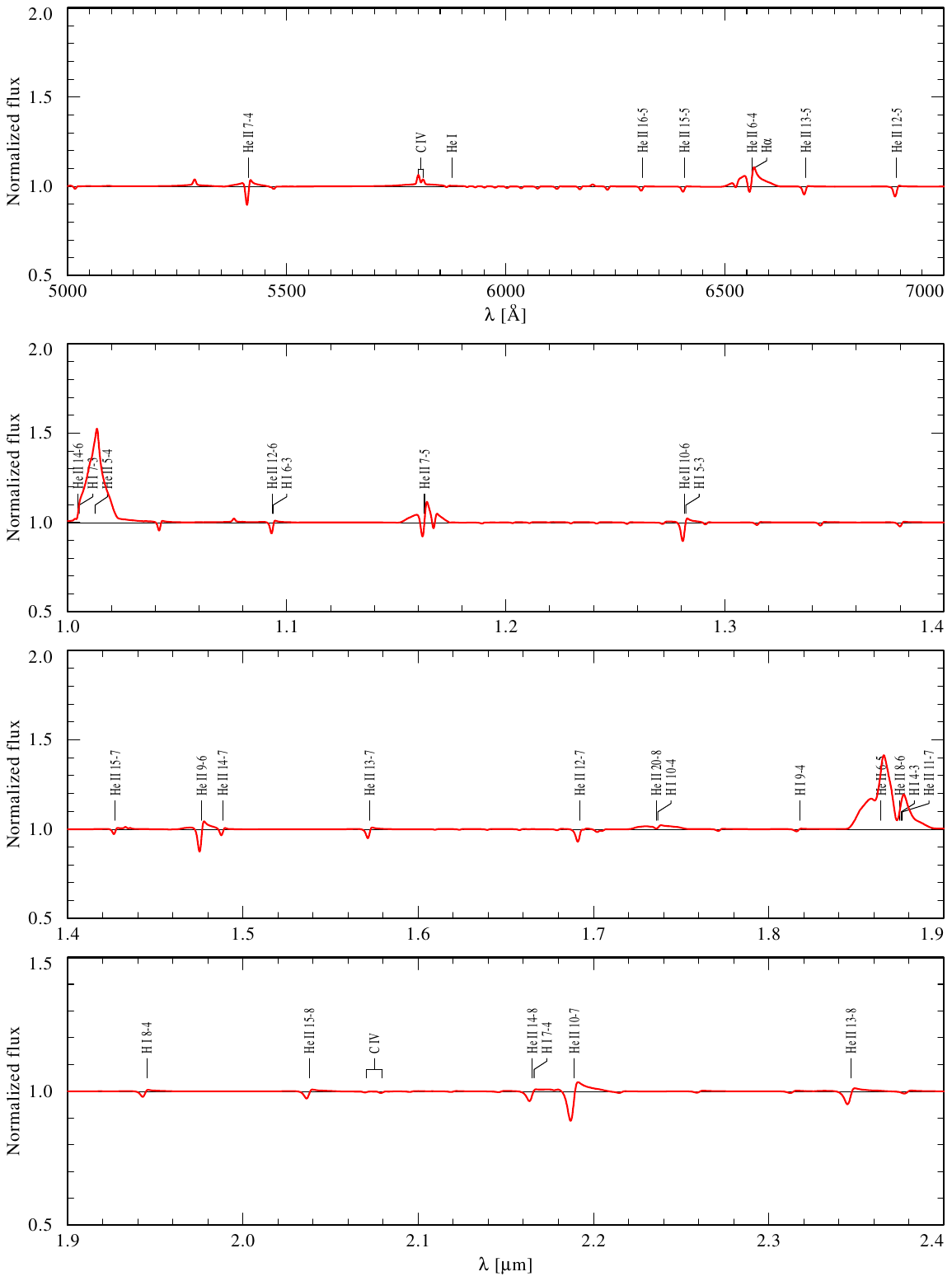}
  \caption{continued.}
\end{figure*}
%---------------------------------------------------------------

\end{appendix} 

\end{document}